\documentclass[longauth]{aa} 
\usepackage{graphicx}
\usepackage{tabularx}
\usepackage{multirow}
\usepackage[colorlinks=true, linkcolor=blue, citecolor=blue]{hyperref}
\usepackage{gensymb}
\usepackage{color}
\usepackage{txfonts}
\usepackage{natbib}
\usepackage{amsmath}
\usepackage{ulem}
\usepackage[flushleft]{threeparttable}
\usepackage[dvipsnames]{xcolor}
\usepackage{rotating}
\begin{document}

\title{The hidden side of cosmic star formation at $z$ $>$ 3}

\subtitle{Bridging optically dark and Lyman-break galaxies with GOODS-ALMA
}
\author{ M.-Y.~Xiao\inst{\ref{inst1},\ref{inst2},\ref{inst3}}\thanks{E-mail: my.xiao@smail.nju.edu.cn}
\and D. Elbaz\inst{\ref{inst2}} 
\and C. G\'omez-Guijarro\inst{\ref{inst2}} 
\and L.~Leroy\inst{\ref{inst2}} 
\and L.-J. Bing\inst{\ref{inst4}} 
\and E. Daddi\inst{\ref{inst2}} 
\and B. Magnelli\inst{\ref{inst2}} 
\and M. Franco\inst{\ref{inst5}}
\and L. Zhou\inst{\ref{inst1},\ref{inst3}}
\and M. Dickinson\inst{\ref{inst6}}
\and T. Wang\inst{\ref{inst1},\ref{inst3}}
\and W. Rujopakarn\inst{\ref{inst7},\ref{inst8},\ref{inst9}}
\and G. E. Magdis\inst{\ref{inst10},\ref{inst11},\ref{inst12}}
\and E. Treister\inst{\ref{inst13}}
\and H. Inami\inst{\ref{inst14}}
\and R. Demarco\inst{\ref{inst15}}
\and M. T. Sargent\inst{\ref{inst16},\ref{inst17}}
\and X. Shu\inst{\ref{inst18}}
\and J. S. Kartaltepe\inst{\ref{inst19}}
\and D. M. Alexander\inst{\ref{inst20}}
\and M. B\'ethermin\inst{\ref{inst4}} 
\and F. Bournaud\inst{\ref{inst2}} 
\and L. Ciesla\inst{\ref{inst4}} 
\and H. C. Ferguson\inst{\ref{inst21}}
\and S. L. Finkelstein\inst{\ref{inst5}}
\and M. Giavalisco\inst{\ref{inst22}}
\and Q.-S. Gu\inst{\ref{inst1},\ref{inst3}}
\and D. Iono\inst{\ref{inst23},\ref{inst24}}
\and S. Juneau\inst{\ref{inst6}}
\and G. Lagache\inst{\ref{inst4}} 
\and R. Leiton\inst{\ref{inst15}}
\and H. Messias\inst{\ref{inst25},\ref{inst26}}
\and K. Motohara\inst{\ref{inst27}}
\and J. Mullaney\inst{\ref{inst28}}
\and N. Nagar\inst{\ref{inst15},\ref{inst29}}
\and M. Pannella\inst{\ref{inst30},\ref{inst31}}
\and C. Papovich\inst{\ref{inst32},\ref{inst33}}
\and A. Pope\inst{\ref{inst25}}
\and C. Schreiber\inst{\ref{inst34}}
\and J. Silverman\inst{\ref{inst9}}
}

\institute{School of Astronomy and Space Science, Nanjing University, Nanjing 210093,P. R. China\label{inst1}
\and AIM, CEA, CNRS, Universit\'e Paris-Saclay, Universit\'e Paris Diderot, Sorbonne Paris Cit\'e, F-91191 Gif-sur-Yvette, France\label{inst2}
\and Key Laboratory of Modern Astronomy and Astrophysics (Nanjing University), Ministry of Education, Nanjing 210093, China\label{inst3}
\and Aix Marseille Universit\'e, CNRS, LAM, Laboratoire d’Astrophysique de Marseille, Marseille, France\label{inst4}
\and Department of Astronomy, The University of Texas at Austin, 2515 Speedway Blvd Stop C1400, Austin, TX 78712, USA\label{inst5}
\and Community Science and Data Center/NSF’s NOIRLab., 950 N. Cherry Ave., Tucson, AZ 85719, USA\label{inst6}
\and Department of Physics, Faculty of Science, Chulalongkorn University, 254 Phayathai Road, Pathumwan, Bangkok 10330, Thailand\label{inst7}
\and National Astronomical Research Institute of Thailand (Public Organization), Don Kaeo, Mae Rim, Chiang Mai 50180, Thailand\label{inst8}
\and Kavli IPMU (WPI), UTIAS, The University of Tokyo, Kashiwa, Chiba 277-8583, Japan\label{inst9}
\and Cosmic Dawn Center (DAWN), Denmark\label{inst10}
\and DTU-Space, Technical University of Denmark, Elektrovej 327, 2800, Kgs. Lyngby, Denmark\label{inst11}
\and Niels Bohr Institute, University of Copenhagen, Jagtvej 128, 2200, Copenhagen N, Denmark\label{inst12}
\and Instituto de Astrof\'isica, Facultad de F\'isica, Pontificia Universidad Cat\'olica de Chile, Casilla 306, Santiago 22, Chile\label{inst13}
\and Hiroshima Astrophysical Science Center, Hiroshima University, 1-3-1 Kagamiyama, Higashi-Hiroshima, Hiroshima 739-8526, Japan\label{inst14}
\and Departamento de Astronom\'ia, Facultad de Ciencias F\'sicas y Matem\'aticas, Universidad de Concepci\'on, Concepci\'on, Chile\label{inst15}
\and Astronomy Centre, Department of Physics and Astronomy, University of Sussex, Brighton BN1 9QH, UK\label{inst16}
\and International Space Science Institute (ISSI), Hallerstrasse 6, CH-3012 Bern, Switzerland\label{inst17}
\and Department of Physics, Anhui Normal University, Wuhu, Anhui 241000, PR China\label{inst18}
\and Laboratory for Multiwavelength Astrophysics, School of Physics and Astronomy, Rochester Institute of Technology, 84 Lomb Memorial Drive, Rochester, NY 14623, USA\label{inst19}
\and Centre for Extragalactic Astronomy, Department of Physics, Durham University, Durham DH1 3LE, UK\label{inst20}
\and Space Telescope Science Institute, 3700 San Martin Drive, Baltimore, MD 21218, USA\label{inst21}
\and Astronomy Department, University of Massachusetts, Amherst, MA 01003, USA\label{inst22}
\and National Astronomical Observatory of Japan, 2-21-1 Osawa, Mitaka, Tokyo 181-8588, Japan\label{inst23}
\and Department of Astronomical Science, SOKENDAI (The Graduate University for Advanced Studies), Mitaka, Tokyo 181-8588, Japan\label{inst24}
\and Joint ALMA Observatory, Alonso de C\'{o}rdova 3107, Vitacura 763-0355, Santiago, Chile\label{inst25}
\and European Southern Observatory, Alonso de C\'{o}rdova 3107, Vitacura, Casilla 19001, 19 Santiago, Chile\label{inst26}
\and Institute of Astronomy, Graduate School of Science, The University of Tokyo, 2-21-1 Osawa, Mitaka, Tokyo 181-0015, Japan\label{inst27}
\and Department of Physics and Astronomy, University of Sheffield, Sheffield S3 7RH, UK\label{inst28}
\and Department of Astronomy, Universidad de Concepci\'on, Casilla 160- C Concepci\'on, Chile\label{inst29}
\and Astronomy Unit, Department of Physics, University of Trieste, via Tiepolo 11, I-34131 Trieste, Italy\label{inst30}
\and Fakult\"at f\"ur Physik der Ludwig-Maximilians-Universit\"at, D-81679 M\"unchen, Germany\label{inst31}
\and Department of Physics and Astronomy, Texas A\&M University, College Station, TX, 77843-4242, USA\label{inst32}
\and George P. and Cynthia Woods Mitchell Institute for Fundamental Physics and Astronomy, Texas A\&M University, College Station, TX, 77843-4242, SA\label{inst33}
\and Astrophysics, Department of Physics, Keble Road, Oxford OX1 3RH, UK\label{inst34}
}
\date{Received October 3, 2022; accepted xxx}

 
  \abstract
 {Our current understanding of the cosmic star formation history at $z>3$ is primarily based on UV-selected galaxies (Lyman-break galaxies, i.e., LBGs). 
Recent studies of $H$-dropouts (HST-dark galaxies) have revealed that we may be missing a large proportion of star formation that is taking place in massive galaxies at $z>3$. In this work, we extend the $H$-dropout criterion to lower masses to select optically dark or faint galaxies (OFGs) at high redshifts in order to complete the census between LBGs and $H$-dropouts. Our criterion  ($H>$ 26.5 mag \& [4.5] $<$ 25 mag) combined with a de-blending technique is designed to select not only extremely dust-obscured massive galaxies but also normal star-forming galaxies (typically E(B-V) $>$ 0.4) with lower stellar masses at high redshifts. In addition, with this criterion, our sample is not contaminated by massive passive or old galaxies. In total, we identified 27 OFGs at $z_{\rm phot}$ > 3 (with a median of $z_{\rm med}=4.1$) in the GOODS-ALMA field, covering a wide distribution of stellar masses with log($M_{\star}$/$M_{\odot}$) = $9.4-11.1$ (with a median of log($M_{\rm \star med}$/$M_{\odot}$) = 10.3).
We find that up to 75\% of the OFGs with log($M_{\star}$/$M_{\odot}$) = $9.5-10.5$ were neglected by previous LBGs and $H$-dropout selection techniques. After performing an optical-to-millimeter stacking analysis of the OFGs, we find that 
rather than being limited to a rare population of extreme starbursts, these OFGs represent a normal population of dusty star-forming galaxies at $z > 3$. The OFGs exhibit shorter gas depletion timescales, slightly lower gas fractions, and lower dust temperatures than the scaling relation of typical star-forming galaxies. Additionally, the total star formation rate ($\rm{SFR}_{\rm tot} = \rm{SFR}_{\rm IR}+\rm{SFR}_{\rm UV}$) of the stacked OFGs is much higher than the SFR$_{\rm UV}^{\rm corr}$ (SFR$_{\rm UV}$ corrected for dust extinction), with an average SFR$_{\rm tot}$/SFR$_{\rm UV}^{\rm corr}$ = $8\pm1$, which lies above ($\sim$0.3 dex) the 16-84th percentile range of typical star-forming galaxies at $3 \leq z \leq 6$.  All of the above suggests the presence of hidden dust regions in the OFGs that absorb all UV photons, which cannot be reproduced with dust extinction corrections. 
The effective radius of the average dust size measured by a circular Gaussian model fit in the $uv$ plane is $R_{\rm e(1.13 mm)}=1.01\pm0.05$ kpc. After excluding the five LBGs in the OFG sample, we investigated their contributions to the cosmic star formation rate density (SFRD). We found that the SFRD at $z>3$ contributed by massive OFGs (log($M_{\star}$/$M_{\odot}$) > 10.3) is at least two orders of magnitude higher than the one contributed by equivalently massive LBGs.  
Finally, we calculated the combined contribution of OFGs and LBGs to the cosmic SFRD at $z=4-5$ to be 4 $\times$ 10$^{-2}$ $M_{\odot}$ yr$^{-1}$Mpc$^{-3}$, which is about 0.15 dex (43\%) higher than the SFRD derived from UV-selected samples alone at the same redshift. This value could be even larger, as our calculations were performed in a very conservative way.}
  



   \maketitle

\section{Introduction}
Our current knowledge of the first two billion years of cosmic star formation history is based mainly on (i) UV-selected galaxies, such as Lyman-break galaxies \citep[LBGs; e.g., ][]{Giavalisco2004,Bouwens2012a,Bouwens2015,Bouwens2020,Oesch2014,Oesch2015,Oesch2018,Madau2014}, which are known to be biased against massive galaxies;
 and (ii) the most massive and extremely dusty starburst galaxies \citep[e.g.,][]{Walter2012,Marrone2018}, which are limited to a rare population and are not representative of the most common galaxies typically on the star-formation main sequence \citep[SFMS; e.g.,][]{Elbaz2007,Elbaz2011,Noeske2007, Magdis2010,Whitaker2012,Whitaker2014,Speagle2014,Schreiber2015,Lee2015,Leslie2020}. 
Recent Atacama Large Millimeter and Submillimeter Array (ALMA) and \textit{Spitzer} observations have identified a more abundant and less extreme population of obscured galaxies at $z>3$ (e.g., $H$-dropouts in \citealt{Wang2019}; HST-dark galaxies in \citealt{Zhou2020}, optically dark/faint galaxies in \citealt{Carlos2021}), revealing that a significant population of high-$z$ optically dark/faint galaxies have been missed, and they may dominate the massive end of the stellar mass function. The contribution of these optically dark/faint galaxies to the cosmic star formation rate density (SFRD) at $z>3$ could be substantial, corresponding up to $\sim$10-25\% of the SFRD from LBGs, or even up to $\sim$ 40\%, depending on the methodology \citep[e.g.,][]{Wang2019, Williams2019, Gruppioni2020, Fudamoto2021,Talia2021,Enia2022,Shu2021,Barrufet2022}.

Optically dark/faint galaxies have generally been completely undetected or tentatively detected with very low significance even in the deepest HST/WFC3 images (typical 5$\sigma$ depth of $H>27$ mag), but brighter at longer wavelengths such as \textit{Spitzer}/IRAC 3.6 and 4.5$\mu$m, \citep[e.g.,][]{Franco2018,Yamaguchi2019,Zhou2020,Smail2021,Carlos2021}. In GOODS-ALMA 1.0, \cite{Franco2020a} reported six optically dark galaxies (i.e., HST-dark galaxies) out of 35 galaxies detected above 3.5$\sigma$ at 1.13 mm. With the ALMA spectroscopic follow-up, \cite{Zhou2020} further analyzed these six optically dark galaxies in detail and found that four ($\sim$70\%) could be associated with a $z\sim3.5$ overdensity (corresponding to OFG1, 2, 25, and 27 in the southwest region of Fig.~\ref{Fig:sky}). Afterward, in the deeper GOODS-ALMA 2.0, \cite{Carlos2021} updated the sample with 13 optically dark/faint galaxies (including six in the GOODS-ALMA 1.0), among a total of 88 sources detected above 3.5$\sigma$ at 1.13 mm. So far, we do not have a unified and clear definition of optically dark/faint galaxies.  The six optically dark galaxies in GOODS-ALMA 1.0 have no optical counterparts in the deepest $H$-band based on the CANDELS catalog down to $H$ = 28.16 AB (5$\sigma$ limiting depth in CANDELS-deep field). However, two of them show $H$-band magnitudes of approximately 25 mag and 27 mag following a de-blending process \citep{Zhou2020}. The remaining seven sources were classified as optically dark/faint galaxies because they are currently undetected or very faint in the $H$-band of the deepest fields and other shorter wavelength bands \citep{Carlos2021}. 


Therefore, the purpose of our work is to first make a clear definition of the selection of optically dark/faint galaxies. Furthermore, by systematically studying optically dark/faint galaxies in the GOODS-ALMA field, we aim to obtain a more complete picture of the cosmic star formation history in the $z>3$ Universe. 
In this work, 
our sample includes not only sources detected by ALMA 1.13 mm, but also those that are currently undetected (i.e., no millimeter counterparts in the GOODS-ALMA 2.0 catalog) to obtain a somewhat complete sample of optically dark/faint galaxies. By stacking their optical to millimeter emission, we can, however, investigate the differences between the optically dark/faint galaxies detected by ALMA 1.13 mm and those that remain undetected.
 


 This paper is organized as follows. In $\S$\ref{Sect::Data}, we describe the GOODS-ALMA survey and the multiwavelength data used. In $\S$\ref{Selection}, we present our selection criterion for optically dark/faint galaxies at $z>3$. 
 In $\S$\ref{individual}, we study the properties of individual sources in our sample, such as the redshift, stellar mass, star formation rate (SFR), molecular gas mass, and dust temperature.  In $\S$\ref{stack}, we present and discuss the properties of optically dark/faint galaxies mainly based on our optical to millimeter stacking analysis. In $\S$\ref{SFRD}, we calculate the cosmic SFRD contributed by optically dark/faint galaxies and discuss the level of the incompleteness of our understanding of dust-obscured star formation in the $z > 3$ Universe. Finally, we summarize our main conclusions in $\S$\ref{Sec:summary}.

Throughout this paper, we adopt a Chabrier initial mass function \citep[IMF;][]{Chabrier:2003} to estimate SFR and stellar mass. We assume cosmological parameters of $H_{0}$ = 70 km s$^{-1}$ Mpc$^{-1}$, $\Omega_{M}$ = 0.3, and $\Omega_{\Lambda}$ = 0.7.
When necessary, data from the literature have been converted with a conversion factor of $M_{\star}$ \citep[][IMF]{Salpeter1955} = 1.7  $\times$ $M_{\star}$ \citep[][IMF]{Chabrier:2003}. All magnitudes are in the AB system \citep{Oke1983}, such that $m_{\rm AB} = 23.9 - 2.5$ $\times$ log(S$_{\nu}$ [$\mu$Jy]). 

\section{Data}\label{Sect::Data}
\subsection{GOODS-ALMA survey}\label{Sect::ALMAobs}
GOODS-ALMA is an ALMA 1.13 mm survey in the deepest part of the Great Observatories Origins Deep Survey South field \citep[GOODS-South;][]{Dickinson2003,Giavalisco2004}. It covers a continuous area of 72.42 arcmin$^2$ (effective area of $\sim$69 arcmin$^2$ if the shallower areas at the edges are trimmed off) with ALMA band 6 receivers,
centered at $\alpha$\,=\,3$^{\rm h}$ 32$^{\rm m}$ 30.0$^{\rm s} $, $\delta$\,=\,-27$\degree$ 48$\arcmin$ 00$\arcsec$ (J2000). The observations were obtained from Cycle 3 and Cycle 5, with two different array configurations to include both small and large spatial scales. The ALMA Cycle 3 observations (high-resolution dataset; Project ID: 2015.1.00543.S; PI: D. Elbaz) were conducted between August and September 2016 in the C40-5 array configuration with a total on-source exposure time of approximately 14.06 h, providing a high-resolution image with the longest baseline of 1808 m. The ALMA Cycle 5 observations (low-resolution dataset; Project ID: 2017.1.00755.S; PI: D. Elbaz) were performed between July 2018 and March 2019 with the C43-2 array configuration with a total on-source exposure time of 14.39 h, providing a low-resolution image with the longest baseline of 360.5 m.

\begin{figure}
\centering
\includegraphics[scale=0.38]{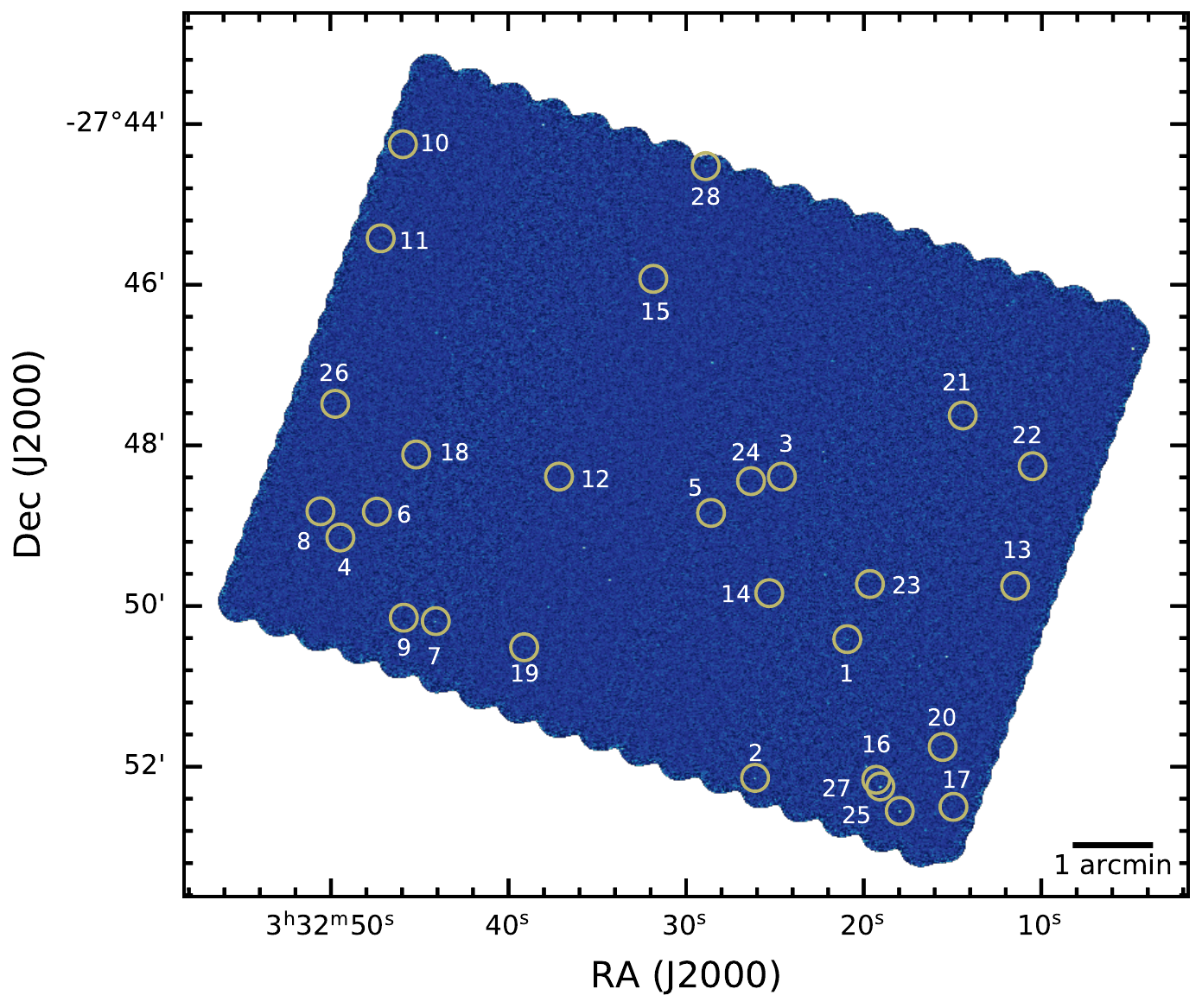}
\caption{Sky distributions of the optically dark/faint galaxies (OFGs) in the GOODS-ALMA 2.0 map at 1.13 mm. North is up and east is to the left.
\label{Fig:sky}}
\end{figure}

The calibration was processed using the Common Astronomy Software Application package \citep[CASA;][]{McMullin2007} with the standard pipeline. We systematically inspected calibrated visibilities and added a few additional flags to the original calibration scripts. The calibrated visibilities were then time- and frequency-averaged over 120 s and 8 channels, respectively, to reduce the computational time for subsequent continuum imaging. Given the excellent coverage of the $uv$ plane and the absence of very bright sources, we used the task \texttt{TCLEAN} in CASA version 5.6.1-8 to produce a dirty map with 0.05$\arcsec$ pixels and a natural weighting scheme to avoid potential biases from the \texttt{CLEAN} algorithm. The resulting high- and low-resolution 1.13 mm continuum maps have similar root mean square (rms) sensitivities, that is, $\sigma$ $\simeq$ 89.0 and 95.2\,$\mu$Jy beam$^{-1}$, with spatial resolutions of full width at half maximum (FWHM) $\simeq$ 0\farcs251 $\times$ 0\farcs232 and 1\farcs330 $\times$ 0\farcs935, respectively. To improve the sensitivity, we concatenated these two data configurations in the $uv$ plane with visibility weights proportional to 1:1. The combined map achieves an rms sensitivity of $\sigma$ $\simeq$ 68.4\,$\mu$Jy beam$^{-1}$ with a spatial resolution of 0\farcs447 $\times$ 0\farcs418 (see Fig.~\ref{Fig:sky}). For more details on the same data reduction, we refer to \cite{Franco2018} for the high-resolution dataset (GOODS-ALMA 1.0) and \cite{Carlos2021} for the low-resolution dataset and the combined dataset (GOODS-ALMA 2.0). 

\subsection{Multiwavelength images}\label{Sect::add_data}

Here we list the multiwavelength data we used for ultraviolet (UV) to mid-infrared (MIR) and MIR to millimeter (mm) spectral energy distribution (SED) fitting, as well as those used for the stacking analysis (see $\S$\ref{Sect::redshift}, $\S$\ref{Sect::LIR}, and $\S$\ref{Sect::stacked SED}):
($i$) X-ray data:  $Chandra$ 7 Ms (0.5-7.0 keV, 0.5-2.0 keV, and 2-7 keV bands) images in the $Chandra$ Deep Field-South (CDF-S) field \citep{Luo2017};
($ii$) UV, optical (OPT), and near-infrared (NIR) data:  HST/ACS (F435W, F606W, F775W, F814W, F850LP) and HST/WFC3 (F105W, F125W, F140W, F160W) images from the $Hubble$ Legacy Fields Program \citep[HLF v2.0;][]{Whitaker2019}, 
VLT/VIMOS ($U$, $R$; \citealt{Nonino2009}) images, VLT/ISAAC ($H$, $J$, $K_{\rm s}$; \citealt{Retzlaff2010}) images, VLT/HAWK-I ($K_{\rm s}$; \citealt{Fontana2014}) images, \textit{Magellan}/FourStar ($H_{\rm s}$, $H_{\rm l}$, $J_{\rm 1}$, $J_{\rm 2}$, $J_{\rm 3}$, $K_{\rm s}$) images from ZFOURGE (\citealt{Straatman2016}), and CFHT/WIRCAM ($J$, $K_{\rm s}$; \citealt{Hsieh2012}) images;
($iii$) MIR data:  deepest \textit{Spitzer}/IRAC images from the GREATS program \citep[3.6\,$\mu$m, 4.5\,$\mu$m, 5.8\,$\mu$m, 8\,$\mu$m;][]{Stefanon2021}, which were obtained by combining programs of the IUDF (PI: I. Labb{\'e}), IGOODS (PI: P. Oesch), GOODS (PI: M. Dickinson), ERS (PI: G. Fazio), S-CANDELS (PI: G. Fazio), SEDS (PI: G. Fazio), UDF2 (PI: R. Bouwens), and GREATS (PI: I. Labb{\'e}); 
($iv$) far-infrared (FIR) data: \textit{Spitzer}/MIPS (24\,$\mu$m) images from GOODS (PI: M. Dickinson). We used \textit{Herschel}/PACS images (100\,$\mu$m, 160\,$\mu$m; \citealt{Magnelli2013}) combined from the PEP \citep{Lutz2011} and GOODS-\textit{Herschel} \citep{Elbaz2011} programs and \textit{Herschel}/SPIRE (250\,$\mu$m, 350\,$\mu$m, 500\,$\mu$m; \citealt{Elbaz2011}) images;
($v$) millimeter data:  1.13 mm map of GOODS-ALMA 2.0, which is a combination of high- and low-resolution 1.13 mm continuum maps (see details in $\S$\ref{Sect::ALMAobs});
and ($vi$) radio data: radio image at 3 GHz (10 cm) from the VLA (PI: W. Rujopakarn, private communication), which covers the entire GOODS-ALMA field \citep[][and in prep.]{Rujopakarn2016}. 
 In Table \ref{table:filter}, we summarize the multiwavelength dataset used in this work from UV to mm. 
\begin{table}
\caption{Broad and intermediate bands (UV to mm) in this work.}            
\centering                          
\begin{tabularx}{0.49\textwidth}{l XXX}  
\hline\hline                
 Telescope/Camera      &  Filter & $\lambda_{\rm c}$ ($\mu$m)& Ref. \\ 
\hline             
VLT/VIMOS&$U$&0.3759 &(a)\\
&$R$&0.6481\\
HST/ACS&$F435W$&0.4347 &(b)\\
&$F606W$&0.6033\\
&$F775W$&0.7730\\
&$F814W$&0.8143\\
&$F850LP$&0.9085\\
HST/WFC3&$F105W$&1.0644 &(b)\\
&$F125W$&1.2561\\
&$F140W$&1.4064\\
&$F160W$&1.5463\\
CFHT/WIRCAM &$J$&1.2554 & (c)\\
&$K_{\rm s}$&2.1630\\
VLT/ISAAC&$J$&1.2423& (d)\\
&$H$&1.6560\\
&$K_{\rm s}$&2.1709\\
VLT/HAWK-I &$K_{\rm s}$&2.1586& (e)\\
\textit{Magellan}/FourStar&$J_{\rm 1}$&1.0552& (f)\\
&$J_{\rm 2}$&1.1472\\
&$J_{\rm 3}$&1.2819\\
&$H_{\rm s}$&1.5564\\
&$H_{\rm l}$&1.7038\\
&$K_{\rm s}$&2.1599\\
\textit{Spitzer}/IRAC&CH1&3.5763& (g)\\
&CH2&4.5289\\
&CH3&5.7875  \\
&CH4&8.0449\\
\hline  
\textit{Spitzer}/MIPS&MIPS&24 & (h)\\
\textit{Herschel}/PACS& blue&70 & (i)\\
& green&100\\
& red&160\\
\textit{Herschel}/SPIRE&PSW&250\\
&PMW&350\\
&PLW&500\\
ALMA&band 6&1130&(j)\\
\hline                               
\end{tabularx}
\begin{tablenotes}
\item \textbf{Notes.} These bands are used for the UV-to-MIR and MIR-to-mm SED fitting. From left to right: Telescope and/or instrument, the filter name, the central wavelength of the filters, and the reference for the survey, including the images and catalogs we used. $^{(a)}$\citealt{Nonino2009}. $^{(b)}$HLF program \citep{Whitaker2019}. $^{(c)}$\citealt{Hsieh2012}. $^{(d)}$\citealt{Retzlaff2010}. $^{(e)}$\citealt{Fontana2014}. $^{(f)}$ZFOURGE program \citep{Straatman2016}. $^{(g)}$Image: the GREATS program \citep{Stefanon2021}; catalog: this work (see Appendix \ref{IRAC catalog}). $^{(h)}$PI: M. Dickinson; catalog: \cite{Magnelli2011}. $^{(i)}$Image: \cite{Magnelli2013} and  \cite{Elbaz2011}; catalog: T. Wang (private communication) and \cite{Elbaz2011}. $^{(j)}$Image: this work (see $\S$\ref{Sect::ALMAobs}); catalog (and also image): \cite{Carlos2021}.
\end{tablenotes}
\label{table:filter}     
\end{table}

\subsection{The multiwavelength catalogs}\label{Sect::cat}
We list below the main catalogs we used in this work. Briefly, we combined catalogs from the HST \citep[HLF v2.0;][]{Whitaker2019} and IRAC to select our sample, and from \textit{Herschel} and ALMA \citep[GOODS-ALMA 2.0;][]{Carlos2021} for the MIR-to-mm SED fitting. In addition, we also used the GOODS-ALMA 2.0 catalog to classify our sources detected at ALMA 1.13 mm and the X-ray catalog \citep[CDF-S 7 Ms;][]{Luo2017} to help identify candidate X-ray active galactic nuclei (AGN). We note that here we have corrected for systematic and local astrometric offsets in different catalogs following \cite{Franco2020a} and \cite{Whitaker2019} to ensure a consistent astrometric reference frame.

  \begin{enumerate}
      \item The  HST $H$-band catalog: The HLF v2.0 in the GOODS-South region (HLF-GOODS-S) uses a deep noise-equalized combination of four HST bands (F850LP, F125W, F140W, F160W) for detection \citep{Whitaker2019}. The catalog includes all UV, optical, and NIR data (13 filters in total) taken by HST over 18 years across the field. The 5$\sigma$ point-source depth in the $H$-band is approximately $27.0-29.8$ mag. We note that the GOODS-South field includes the HUDF, which is much deeper than other parts of the HST data, but covers only a small portion of the field.
      
      \item IRAC catalog: We built the IRAC catalog in the GOODS-ALMA field using the deepest IRAC 3.6 and 4.5\,$\mu$m images from the GREATS program \citep[][]{Stefanon2021}. 
 The source detection was performed using \texttt{Source Extractor} \citep[SE version 2.25.0;][]{Bertin1996} on the background-subtracted 3.6 and 4.5\,$\mu$m images. 
 The total sources in the catalogs are 125,338 for 3.6 \,$\mu$m and 154,234 for 4.5\,$\mu$m  in the GOODS-S field ($\sim$150 arcmin$^2$). To ensure the purity of detections, we then cross-matched these two catalogs with a radius of 1.0$^{\prime\prime}$ ($\sim$0.5 FWHM). Furthermore, for sources in the GOODS-ALMA 2.0 catalog \citep{Carlos2021} that were detected at least in one IRAC band, we also considered them to be real sources and kept them in the final catalog.  Finally, we end up with 71,899 sources in the IRAC catalog, of which 5,127 are in the GOODS-ALMA region (see Appendix \ref{IRAC catalog} for more details).
     
     \item \textit{Spitzer}/MIPS 24\,$\mu$m and \textit{Herschel} catalog: We mainly used the catalog of Wang et al. (in prep.), which was built based on a state-of-the-art de-blending method, similar to that used in the `super-deblended' catalogs in GOODS-North \citep{Liu2018} and COSMOS \citep{Jin2018}. Briefly, Wang et al. used \textit{Spitzer}/MIPS 24\,$\mu$m detections as priors for source extraction on the PACS and SPIRE images. They performed super-deblending in one band each time, from shorter to longer wavelengths, and predicted fluxes at longer wavelengths based on the redshift and photometry information given by the shorter wavelengths. Then, the faint priors at longer wavelengths were removed, which helped break the blending degeneracies. 
For one source (OFG27 in our work) still affected by blending with problematic photometry from MIR to FIR, we used the catalog of \cite{Elbaz2011}.  
   
      \item GOODS-ALMA 2.0 catalog: This catalog contains 88 sources detected by ALMA at 1.13 mm in the GOODS-ALMA 2.0 field  \citep{Carlos2021}. These include 50\% of sources detected above 5$\sigma$ with a purity of 100\% and 50\% detected within the 3.5 and 5$\sigma$ range aided by priors. The median redshift and stellar mass of the 88 sources are $z_{\rm med}=2.46$ and log($M_{\rm \star med}$/$M_{\odot}$) = 10.56 (\citealt{Chabrier:2003} IMF), respectively.     
       
      \item X-ray catalog: We used the CDF-S 7 Ms catalog \citep{Luo2017}. It contains 1008 sources in the main source catalog, observed in three X-ray bands (0.5$-$7.0 keV, 0.5$-$2.0 keV, and 2$-$7 keV) and 47 lower-significance sources in a supplementary catalog. This catalog includes the candidate X-ray AGNs identified by \cite{Luo2017}, which we used in this work to search for X-ray AGN counterparts of our sources.
      
 \end{enumerate}

\section{Selection of optically dark/faint galaxies at  $z>3$}\label{Selection}
We aim to obtain a more complete picture of the cosmic star formation history, that is, to bridge the extreme population of optically dark galaxies \citep[e.g.,][]{Wang2019} with the most common population of lower-mass, less-attenuated galaxies, such as those selected using the LBG selection technique \citep[e.g.,][]{Bouwens2012a,Bouwens2015,Bouwens2020}. To reach this goal, we chose to select our sample with a less strict cut than the one used in \cite{Wang2019} for the $H$-band and 4.5\,$\mu$m magnitudes, which would allow for our sample to encompass lower-mass and less-attenuated galaxies, while still including extremely dust-obscured galaxies. Here, we call them optically dark/faint galaxies (hereafter, OFGs) and we select them with the criteria defined below (see $\S$\ref{Sect:method}). 



   \begin{figure*}
   \centering
   \includegraphics[width=8cm]{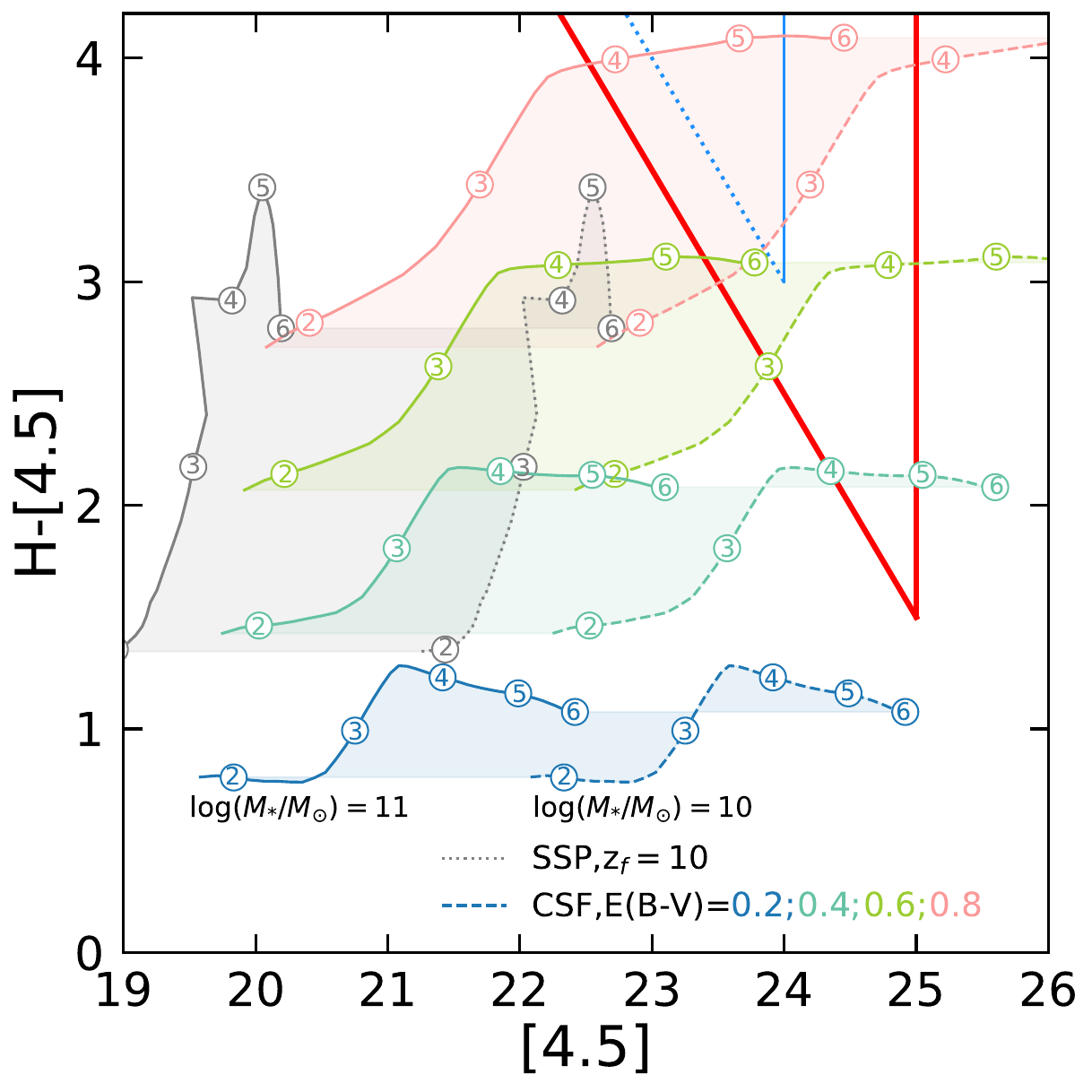}
    \includegraphics[width=8cm]{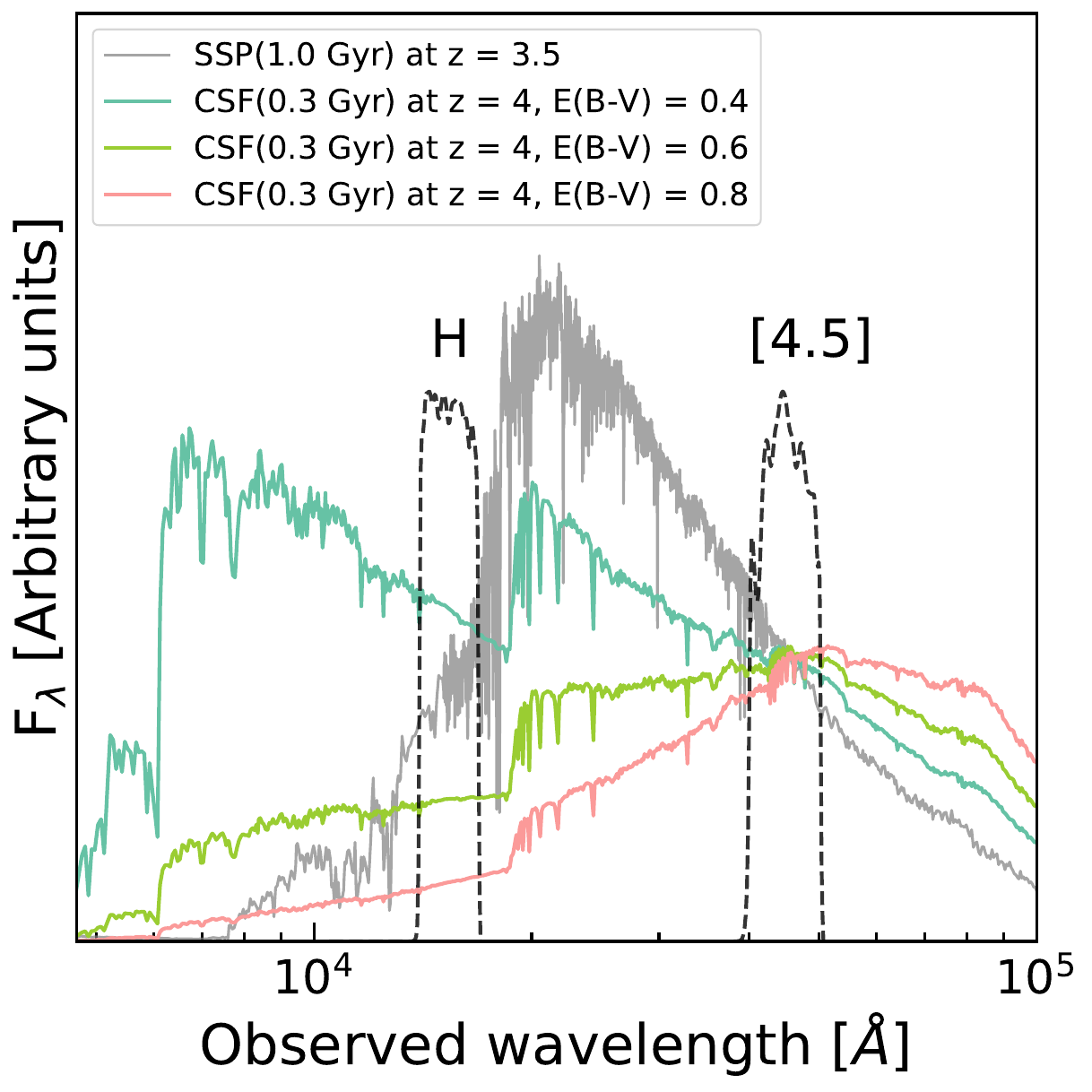}
   
      \caption{Selection criteria for OFGs at $z\gtrsim3$. $Left$: Color-magnitude diagram for the selection of OFGs. This diagram shows the evolutionary tracks of theoretical galaxy templates at $z=2-6$, with stellar mass log($M_{\rm \star}$/$M_{\odot}$) = $10-11$ and solar metallicity. The numbers in the circles indicate the redshift. The solid and dashed evolutionary tracks correspond to stellar masses with log($M_{\rm \star}$/$M_{\odot}$) = $11$ and log($M_{\rm \star}$/$M_{\odot}$) = $10$, respectively.  These templates are based on the BC03 models \citep{Bruzual2003}, including an instantaneous burst (i.e., simple stellar population; SSP) model formed at $z=$ 10 (grey area) and a non-evolving constant star formation (CSF) model with an age of 300 Myr with varying degrees of reddening (blue, green, olive, and red areas). The SSP model corresponds to a passive or old galaxy with an age of 1 Gyr at $z=3.5$. The blue triangular region shows the selection criteria for $H$-dropouts in \cite{Wang2019}, i.e., no $H$-band flux above 5$\sigma$ ($H>27$ mag; blue dashed line) and [4.5] < 24 mag (blue solid line). The red triangular region shows the selection criteria for OFGs in this work. 
$Right$: Different theoretical galaxy templates in HST/F160W ($H$-band) and IRAC 4.5\,$\mu$m filters. The templates include a passive galaxy with an age of 1 Gyr at $z=3.5$ and three star-forming galaxies  with different degrees of reddening at $z=4$. At a similar [4.5], the three star-forming galaxies  with different dust extinction values exhibit different $H$-band magnitudes. }
         \label{track}
   \end{figure*}
   \begin{figure*}
   \centering
   \includegraphics[width=12cm]{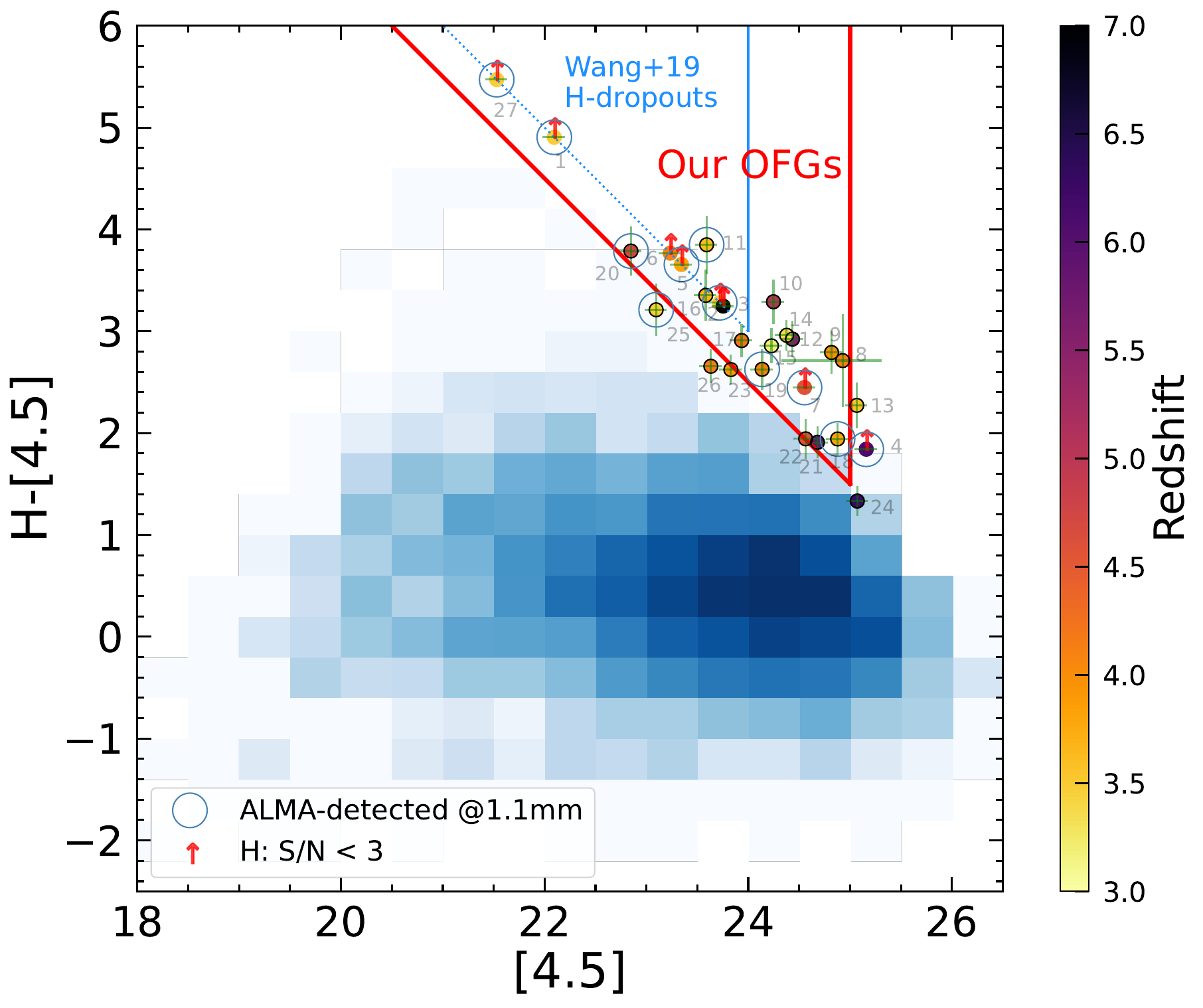}
      \caption{Color-magnitude diagram color-coded by photometric redshift. Our criteria (in red) for selecting OFGs are: $H>$ 26.5 mag \& [4.5] $<$ 25 mag. We note that we include sources outside the wedge whose 1$\sigma$ photometric uncertainties overlap the wedge, so we have points outside the red triangle. The arrows show our $H$-dropouts (S/N < 3), with the typical depth of H $=$ 27 mag (5$\sigma$) in the shallowest region of the HLF survey as their lower limits.  The blue shaded area describes the distribution of all the IRAC detected sources (see Appendix \ref{IRAC catalog} for more details) in the GOODS-ALMA field. The blue and red triangular regions are the same as in Fig.~\ref{track}.
              }
         \label{color_mag}
   \end{figure*}

\subsection{A selection criterion for optically dark/faint galaxies uncontaminated by massive passive galaxies}\label{Sect:method}
To start, we used a combination of $H$-band and IRAC 4.5\,$\mu$m photometry, where the $H$-band measures rest-frame UV light ($\lambda<$ 4000\,\AA\,) for galaxies at $z>3$, while the 4.5\,$\mu$m band probes the rest-frame $J$-band. Galaxies with red color in $H-$[4.5] 
are either quiescent or passive galaxies or dusty star-forming galaxies with significant dust extinction (UV attenuation). 

For optically dark galaxies at high redshifts, a common selection approach in the literature is to target $H$-dropouts, which are defined to be undetected in the $H$-band (i.e., absent in the $H$-band catalog) and bright in the IRAC band (e.g., [4.5] < 24 mag; \citealt{Wang2019}; see the blue triangular region in Fig.~\ref{track}), and/or extremely red in color (e.g., $H-$[4.5] > 4 in \citealt{Shu2021}). These methods can help to select extremely dust-obscured massive galaxies. 
However, the detection of an $H$-dropout obviously depends on the depth of the $H$-band. For example, in the HLF-GOODS-S field, the 5$\sigma$ depth in the $H$-band ranges from 27.0 mag to 29.8 mag \citep{Whitaker2019}. 
To avoid this imprecise selection and to extend the selection in view of bridging the heavily obscured star-forming galaxies with more common galaxies, we defined OFGs based on the following characteristics: 1) $H$ > 26.5 mag; 2) [4.5] < 25 mag (see the red triangular region in Fig.~\ref{track}). Instead of only selecting galaxies undetected in the $H$-band and/or extremely red in $H-$[4.5], we used the criterion of $H$ > 26.5 mag to select not only optically dark sources but also optically faint galaxies with less dust obscuration. 
The $H>$ 26.5 mag cut also helps to distinguish massive passive galaxies with stellar masses log($M_{\rm \star}$/$M_{\odot}$) > $10$ (the grey region in Fig.~\ref{track}; see more details afterward) from our selected OFGs (the red triangular region in Fig.~\ref{track}). In Fig.~\ref{track}, the faintest modeled passive galaxy in the $H$-band ($H=26.0$ mag) has a similar magnitude to the brightest OFGs (OFG25 and OFG26: $H=26.3\pm0.2$ mag), considering the 1$\sigma$ uncertainty in the flux measurements.  
The criterion of [4.5] < 25 mag can help to select not only massive galaxies, but also galaxies with intermediate stellar masses.

To verify the reliability of our selection criteria, we investigated the evolutionary tracks of theoretical galaxy templates at $z=2-6$ in the color-magnitude diagram ($H-$[4.5] vs. [4.5]; Fig.~\ref{track}). The templates are based on the BC03 models \citep{Bruzual2003}, including an instantaneous burst model (i.e., simple stellar population; SSP) formed at $z=$10 and a non-evolving constant star formation (CSF) model with an age of 300 Myr with varying degrees of reddening. The templates have stellar masses in the range of log($M_{\rm \star}$/$M_{\odot}$) = $10-11$, with the \cite{Calzetti2000} attenuation law and solar metallicity. The SSP model corresponds to passive or old galaxies with an age of 1 Gyr at $z$ = 3.5, while the CSF models represent star-forming galaxies with different dust obscurations. 

Our selection criteria for OFGs are shown in Fig.~\ref{track} as the red triangular region, which encompasses high redshift galaxies with lower stellar mass and less dust attenuation that were excluded by previous $H$-dropout selection criteria such as those of \cite{Wang2019} (the blue triangular region in Fig.~\ref{track}). 
For instance, the selected OFGs include those with E(B-V) = 0.4, log($M_{\rm \star}$/$M_{\odot}$) = $10$, and $z=4-5$, as well as those with E(B-V) = 0.6, log($M_{\rm \star}$/$M_{\odot}$) = $10$, and $z=3-4$. Similarly, extremely dust-obscured massive galaxies, such as those with E(B-V) = 0.8, log($M_{\rm \star}$/$M_{\odot}$) = $11$, and $z>4$ can also be selected by our criteria. We note that although a few OFGs (E(B-V) = 0.8 and log($M_{\rm \star}$/$M_{\odot}$) = $10$) at $z=2.5-3$ were selected, the total OFGs dominate at $z>3$. 
Overall, in our selection, the majority of OFGs have E(B-V) $>$ 0.4 and are at $z>3$. 
In addition, with these criteria, our sample is not contaminated by massive passive or old galaxies  (log($M_{\rm \star}$/$M_{\odot}$) > $10$; the grey region in Fig.~\ref{track}).  
Therefore, the selection of optically dark/faint galaxies at high redshifts with $H>$ 26.5 mag and [4.5] $<$ 25 mag is a reliable approach. In summary, the selected OFGs contain not only extremely dust-obscured massive galaxies at $z>4$, but also lower-mass and less-attenuated (typically E(B-V) $>$ 0.4) galaxies at $z>3$, without contamination from massive passive galaxies. 



\subsection{Sample selection}\label{Sect:selection}

We selected candidate OFGs by cross-matching our IRAC catalog at 3.6 and 4.5\,$\mu$m (see Appendix \ref{IRAC catalog} for details on the catalog construction) with the HLF catalog \citep[][]{Whitaker2019} in the GOODS-ALMA field. Candidates were required to have none or very faint HST counterparts ($H$ $>$ 26.5 mag) within a 0.6$^{\prime\prime}$ radius around the IRAC positions. The small radius of 0.6$^{\prime\prime}$ (roughly one-third of the point spread function size of IRAC 4.5 \,$\mu$m) was taken to select as many individual candidates as possible, while avoiding excessive contamination of our final sample by fake candidates.
After cross-matching, we had 88 candidates.

\subsection{Photometry}\label{Sect:deblend}
We visually inspected all the candidates and noted that blending was common in the IRAC images due to their relatively worse spatial resolution (e.g., $\sim$2$^{\prime\prime}$ at 4.5\,$\mu$m). To obtain photometric values in different bands without contamination from neighboring galaxies, we simultaneously de-blended sources in the multi-wavelength images (from UV to 8\,$\mu$m) by applying the de-blending code\footnote{qdeblend: \url{https://github.com/cschreib/qdeblend}} described in \cite{Schreiber2018b}. The de-blending method is briefly summarized in the following three steps. Step 1: for each OF candidate, to save computational time, we first cut the stacked HST image (four bands of HST/F105W, F125W, F140W, and F160W) into a 10$^{\prime\prime}$ $\times$ 10$^{\prime\prime}$ area around the IRAC position of the candidate. Then, we detected all sources in the clipped stacked HST image. 
      For some optically dark galaxies undetected even in the stacked HST image, we used positions of their $K_{\rm s}$-band, 4.5\,$\mu$m, or ALMA counterparts and modeled them as point sources.
Step 2: following \cite{Schreiber2018b}, we fitted all the sources detected in the clipped stacked HST image simultaneously with a single S\'ersic profile to obtain a best-fit deconvolved model (intrinsic light profile) for each source.
Step 3: these  models were then convolved with a point spread function (PSF) for each image at all wavelengths (up to 8\,$\mu$m). We then used the positions and the PSF-convolved models 
      as priors for all objects to fit the multi-wavelength images. The uncertainties of the fluxes were calculated by Monte Carlo simulations. 
We find that 60\% (53/88) of the sources in our candidate sample needed to be de-blended at 4.5\,$\mu$m.  
 
 \subsection{Incompleteness correction and final sample}\label{Sect::sample}

To identify high-$z$ OFGs, we used the simple selection techniques discussed in $\S$\ref{Sect:method}: (1) $H$ $>$ 26.5 mag 
and (2) [4.5] $<$ 25 mag. 
  Considering the 1$\sigma$ uncertainty of the flux measurements, we finally identified 26 individual OFGs in total in the GOODS-ALMA field (see Figs.~\ref{Fig:sky} and \ref{color_mag}).  

Here, we discuss the corrections for the incompleteness of our sample selection approach. Considering the criterion of no/very faint HST counterparts within the search radius of 0.6$^{\prime\prime}$ at the IRAC position, we may have missed some target sources simply due to random bright HST sources falling within the radius. 
Following the same method as in \cite{Lilly1999} \citep[also in][]{Wang2019}, at a given position, the probability of finding one or more random galaxies within a given radius is defined as: 
\begin{eqnarray}
p(n\ge1) = 1 - exp(-N \times \pi \times radius^2),
 \end{eqnarray}
 where $N$ represents the surface density of bright HST sources (H $<$ 26.5 mag) in our case. In the GOODS-ALMA field, $N$ is 0.05 galaxies arcsec$^{-2}$, so the derived $p(n\ge1) = 0.056$. It suggests that using this selection approach, we may have missed around 5\% of OFGs that we wrongly associated with a counterpart due to projection effects. That is, one source could have been missed due to the serendipitous presence of a bright $H$ detection within our 0.6$^{\prime\prime}$ radius search circle. 
 After comparing our sample with 13 ALMA-detected OFGs in the GOODS-ALMA 2.0 catalog, we confirmed that the missing galaxy is OFG27 \citep[A2GS7 in ][]{Carlos2021}. A random bright source (ID55970 in the HLF catalog) with $H$ = 24.5 mag is located at a distance of 0.33$^{\prime\prime}$  (< 0.6$^{\prime\prime}$ searching radius) from OFG27. This bright source has been confirmed not to be the $H$-band counterpart of OFG27 \citep[AGS17 in][]{Zhou2020}. Therefore, we included OFG27 in our catalog (see Table \ref{table:properties}) to correct the incompleteness of our selection approach. Also, we included OFG27 in the analysis of the main discussion presented in this paper.
 
 In addition, an extra IRAC 4.5$\mu$m dropout candidate was detected only in the longer wavelength images: JCMT/SCUBA-2 850$\mu$m and ALMA 870$\mu$m and 1.13 mm and 1.2 mm (OFG28; see Table \ref{table:properties}). Including this one, we have 28 OFGs in our final catalog (Table \ref{table:properties}). Considering that OFG28 is an IRAC 4.5$\mu$m dropout candidate with [4.5] $\gg$ 25 mag, in the following analysis, we focus only on the first 27 OFGs, which meet our criteria: H $>$ 26.5 mag \& [4.5] $<$ 25 mag.

\begin{table*}
\caption{Derived properties of the OFGs.}   
\tiny          
\centering
\begin{threeparttable} 
 
\begin{tabular}{l l l ccc c r c c c c}     
\hline\hline       
ID & RA & Dec & $H$ & [4.5] & $S_{\rm 1.13 mm}$& $z$ & log$(M_{\star})$ & log$(L_{\rm IR})$&SFR &Other ID\\  
& (deg) & (deg) &(mag)&(mag)& (mJy)&&log$(M_{\odot})$&log$(L_{\odot})$&($M_{\odot}$ yr$^{-1}$)&\\ 
(1)& (2) & (3) & (4)&(5)&(6)&(7)&(8)&(9)&(10)&(11)\\
\hline  
OFG1  &   53.087184$^\dagger$ & -27.840242$^\dagger$ &$(...)$ &  22.10$_{-0.10}^{+0.11}$  &$0.85\pm0.11$&3.47$^{o}$& 10.79$_{-0.16}^{+0.11}$ &12.41 $\pm$ 0.05&384 $\pm$ 48&AGS24, A2GS29\\                
OFG2  &   53.108810$^\dagger$ & -27.869037$^\dagger$ & $(...)$ &  23.72$_{-0.10}^{+0.11}$  &$1.24\pm0.10$& 3.47$^{o}$ & 10.76$_{-0.17}^{+0.11}$ &12.39 $\pm$ 0.03&365 $\pm$ 30& AGS11, A2GS15\\
OFG3  &   53.102536& -27.806531   & $(...)$ &  23.75$_{-0.10}^{+0.11}$  &$(...)$& 7.04$_{-1.19}^{+1.01}$         &  10.91$_{-0.14}^{+0.22}$ &$(...)$&$(...)$&$(...)$\\
OFG4  &   53.206064$^\dagger$ & -27.819142$^\dagger$   & $(...)$ &  25.16$_{-0.10}^{+0.11}$  &$0.70\pm0.13$& 6.13$_{-1.36}^{+1.02}$         &  10.60$_{-0.36}^{+0.16}$ &12.56 $\pm$ 0.07&537 $\pm$ 100& A2GS38\\
OFG5  &   53.119150$^\dagger$ & -27.814066$^\dagger$   & $(...)$ &  23.34$_{-0.10}^{+0.11}$  &$0.43\pm0.13$& 3.81$_{-0.94}^{+0.74}$ & 10.17$_{-0.27}^{+0.34}$ &12.26 $\pm$ 0.09&268 $\pm$ 61&A2GS87, GDS44539\\  
OFG6  &   53.197493& -27.813789   & $(...)$ &  23.23$_{-0.10}^{+0.11}$   & $(...)$& 4.10$_{-0.32}^{+0.31}$         &  10.93$_{-0.06}^{+0.06}$ &$(...)$&$(...)$&$(...)$\\
OFG7  &   53.183697$^\dagger$ & -27.836500$^\dagger$   & $(...)$ &  24.55$_{-0.10}^{+0.11}$  &$1.23\pm0.12$& 4.58$_{-0.42}^{+0.42}$ & 10.34$_{-0.09}^{+0.08}$ &12.86 $\pm$ 0.04&1070 $\pm$ 101&AGS25, A2GS17\\    
OFG8  &   53.210736& -27.813706   & 27.64$_{-0.24}^{+0.31}$ &  24.93$_{-0.40}^{+0.60}$   & $(...)$& 4.14$_{-0.44}^{+0.41}$        &  9.92$_{-0.27}^{+0.18}$&$(...)$&$(...)$&$(...)$  \\
OFG9  &   53.191204& -27.835791   & 27.61$_{-0.19}^{+0.22}$ &  24.82$_{-0.10}^{+0.11}$  & $(...)$& 3.99$_{-0.17}^{+0.18}$         &  9.59$_{-0.08}^{+0.08}$&$(...)$&$(...)$&$(...)$  \\
OFG10&  53.191348& -27.737554   & 27.54$_{-0.19}^{+0.23}$ &  24.25$_{-0.10}^{+0.11}$  & $(...)$& 5.04$_{-0.31}^{+0.34}$         &  10.37$_{-0.08}^{+0.06}$&$(...)$&$(...)$&$(...)$ \\
OFG11&  53.196569$^\dagger$ & -27.757065$^\dagger$   & 27.44$_{-0.27}^{+0.35}$ &  23.59$_{-0.10}^{+0.11}$ &$0.62\pm0.12$& 3.60$_{-0.27}^{+0.27}$ & 10.41$_{-0.07}^{+0.06}$ &12.41 $\pm$ 0.09&384 $\pm$ 90&A2GS40, GDS48885\\    
OFG12&  53.154787 & -27.806529   & 27.36$_{-0.15}^{+0.17}$ &  24.43$_{-0.10}^{+0.11}$  & $(...)$& 5.55$_{-0.28}^{+0.29}$         &  10.31$_{-0.06}^{+0.05}$&$(...)$&$(...)$&$(...)$ \\
OFG13&  53.047834 & -27.829186   & 27.34$_{-0.20}^{+0.24}$ &  25.07$_{-0.10}^{+0.11}$  & $(...)$& 3.56$_{-0.09}^{+0.09}$         &  9.50$_{-0.03}^{+0.05}$&$(...)$&$(...)$&$(...)$  \\
OFG14&  53.105489 & -27.830711   & 27.34$_{-0.11}^{+0.12}$ &  24.38$_{-0.10}^{+0.11}$  & $(...)$& 3.39$_{-0.57}^{+0.92}$         &  9.86$_{-0.22}^{+0.23}$&$(...)$&$(...)$&$(...)$\\
OFG15& 53.132675 & -27.765496   & 27.08$_{-0.14}^{+0.16}$ &  24.23$_{-0.10}^{+0.11}$  & $(...)$& 3.192$^{sp}$ &  9.96$_{-0.05}^{+0.04}$ &$(...)$&$(...)$&$(...)$ \\
OFG16&  53.080379 & -27.869420   & 26.93$_{-0.23}^{+0.29}$ &  23.58$_{-0.10}^{+0.11}$  & $(...)$& 3.69$_{-0.25}^{+0.27}$         &  10.38$_{-0.08}^{+0.08}$&$(...)$&$(...)$&$(...)$ \\
OFG17&  53.062276 & -27.875036   & 26.84$_{-0.13}^{+0.15}$ &  23.93$_{-0.10}^{+0.11}$  & $(...)$& 4.23$_{-0.21}^{+0.23}$         &  10.29$_{-0.05}^{+0.07}$&$(...)$&$(...)$&$(...)$ \\
OFG18&   53.188278$^\dagger$ & -27.801928$^\dagger$   & 26.82$_{-0.12}^{+0.14}$ &  24.88$_{-0.10}^{+0.11}$  &$0.37\pm0.11$& 3.81$_{-0.10}^{+0.11}$         &  9.44$_{-0.11}^{+0.03}$  &12.28 $\pm$ 0.11&286 $\pm$ 86& A2GS47\\
OFG19&   53.162978$^\dagger$ & -27.841940$^\dagger$   & 26.76$_{-0.17}^{+0.20}$ &  24.14$_{-0.10}^{+0.11}$  &$0.40\pm0.12$& 4.09$_{-0.32}^{+0.35}$         &  10.30$_{-0.09}^{+0.06}$ &12.26 $\pm$ 0.09&271 $\pm$ 59& A2GS82\\
OFG20&   53.064807$^\dagger$ & -27.862613$^\dagger$   & 26.64$_{-0.22}^{+0.27}$ &  22.85$_{-0.10}^{+0.11}$  &$0.54\pm0.10$& 4.74$_{-0.50}^{+0.42}$         &  10.88$_{-0.26}^{+0.16}$ &12.79 $\pm$ 0.04&913 $\pm$ 86& A2GS57\\
OFG21&   53.060144 & -27.793838   & 26.59$_{-0.12}^{+0.13}$ &  24.68$_{-0.10}^{+0.11}$  & $(...)$& 5.89$_{-0.38}^{+0.36}$         &  9.75$_{-0.13}^{+0.32}$&$(...)$&$(...)$&$(...)$  \\
OFG22&   53.043745 & -27.804347   & 26.51$_{-0.16}^{+0.19}$ &  24.56$_{-0.11}^{+0.12}$  & $(...)$& 4.56$_{-0.24}^{+0.26}$         &  9.97$_{-0.07}^{+0.07}$&$(...)$&$(...)$&$(...)$  \\
OFG23&   53.081890 & -27.828815   & 26.45$_{-0.10}^{+0.11}$ &  23.83$_{-0.10}^{+0.11}$  & $(...)$& 3.88$_{-0.18}^{+0.18}$         &  10.25$_{-0.04}^{+0.04}$&$(...)$&$(...)$&$(...)$ \\
OFG24&   53.109771 & -27.807466  & 26.40$_{-0.10}^{+0.11}$ &  25.07$_{-0.10}^{+0.11}$  & $(...)$& 6.27$_{-0.16}^{+0.16}$         &  9.87$_{-0.08}^{+0.10}$&$(...)$&$(...)$&$(...)$  \\
OFG25&   53.074868$^\dagger$ & -27.875889$^\dagger$   & 26.31$_{-0.24}^{+0.30}$ &  23.10$_{-0.10}^{+0.11}$  &$1.67\pm0.10$& 3.47$^{o}$ &  9.99$_{-0.11}^{+0.46}$ &12.92 $\pm$ 0.03&1227 $\pm$ 74& AGS15, A2GS10 \\
OFG26&   53.207252 & -27.791408   & 26.29$_{-0.19}^{+0.22}$ &  23.63$_{-0.10}^{+0.11}$ & $(...)$  & 4.16$_{-0.25}^{+0.25}$         &  10.33$_{-0.09}^{+0.08}$&$(...)$&$(...)$&$(...)$ \\
OFG27$^*$&  53.079416$^\dagger$ & -27.870820$^\dagger$ &   $(...)$ & 21.53$_{-0.10}^{+0.11}$  &$2.05\pm0.12$&3.467$^{sp}$& 11.11$_{-0.19}^{+0.15}$ &13.08 $\pm$ 0.02&1795 $\pm$ 90&AGS17, A2GS7\\   
\hline 
 OFG28$^{**}$& 53.120402$^\dagger$&-27.742111$^\dagger$& $(...)$ &   $(...)$& $0.95\pm0.12$ &$(...)$&$(...)$&$(...)$&$(...)$&A2GS33\\   
\hline 
                 
\end{tabular}
\begin{tablenotes}
\item \textbf{Note:} (1) Source ID; (2)(3) Right ascension and declination (J2000) of sources. Coordinates detected in the ALMA 1.13 mm image are marked with a "$\dagger$" exponent;  
(4)(5) $H$-band and IRAC 4.5\,$\mu$m AB magnitudes. These magnitudes are given for the best-fitting S\'ersic profile during de-blending procedure (see $\S$\ref{Sect:deblend}); (6) ALMA 1.13 mm flux density: obtained from the GOODS-ALMA 2.0 catalog \citep{Carlos2021}; (7) Photometric redshifts: determined with the \texttt{EAzY} code (see $\S$\ref{Sect::redshift}; spectroscopic redshifts expressed in three decimal places and flagged with a "sp" exponent). The spectroscopic redshifts of OFG15 and OFG27 are from \cite{Herenz2017} and \cite{Zhou2020}, respectively. OFG1, 2, 25, 27 were discovered in an overdensity region with the redshift peak to be 3.47 \citep[flagged with a "o" exponent;][]{Zhou2020}; (8) Stellar masses: determined with the \texttt{FAST++} (see $\S$\ref{Sect::redshift}); (9) Infrared luminosities: derived from \texttt{CIGALE}  for three sources (OFG2, OFG20, and OFG27) with a \textit{Herschel}  counterpart or from the IR template library \citep{Schreiber2018c} for the galaxies without a \textit{Herschel} counterpart (see $\S$\ref{Sect::LIR}); (10) SFR = SFR$_{\rm IR}$ + SFR$_{\rm UV}$ (see $\S$\ref{Sect::sfr}); 
(11) Source IDs in other work: AGS \citep[GOODS-ALMA 1.0 catalog;][]{Franco2018,Franco2020a}; A2GS \citep[GOODS-ALMA 2.0 catalog;][]{Carlos2021}; GDS \citep[$H$-dropouts catalog;][]{Wang2019}. 
$^*$The missing galaxy in our selection approach is due to incompleteness of the search radius of 0.6$^{\prime\prime}$. We add it to our catalog to correct this incompleteness. $^{**}$OFG28 is a candidate IRAC 4.5$\mu$m dropout, which is only detected in longer wavelength images, e.g., 850$\mu$m from the JCMT/SCUBA-2 and 870$\mu$m from the ALMA  \citep[ID68 in][]{Cowie2018}, 1.13 mm from the GOODS-ALMA 2.0 100\% pure source catalog \citep[A2GS33 in][]{Carlos2021}, and 1.2mm from the ALMA \citep[ID20 in][]{Yamaguchi2019}. We add it here to refine our OFG catalog in the GOODS-ALMA field. We note that OFG28 is not used in our analysis. We also note  that OFG9, OFG13, OFG18, OFG22, and OFG24 are identified as LBGs (see $\S$\ref{Sect::lyman break}). In this catalog, eight sources (OFG1-OFG7, and OFG27) are $H$-dropouts, which are not detected in the $H$-band (<3$\sigma$).

\end{tablenotes}
\label{table:properties} 
\end{threeparttable} 
\end{table*}

\subsection{Lyman-break galaxies and $H$-dropouts in final sample}\label{Sect::lyman break $H$-dropout}
To compare our sample with LBGs \citep[e.g.,][]{Bouwens2012a} and $H$-dropouts \cite[e.g.,][]{Wang2019}, we need to determine how many OFGs are LBGs or $H$-dropouts in our sample.

\subsubsection{Lyman-break galaxies}\label{Sect::lyman break}
LBGs are a UV-selected population of high-$z$ star-forming galaxies. 
To understand how many galaxies in our OFG catalog are missed by this UV-selected approach and to further know their contribution to the cosmic SFRD, we identify LBGs from our OFG catalog by employing the Lyman-break color criteria used in \cite{Bouwens2020} (also see similar methods in \citealt{Bouwens2012a,Bouwens2015}). The redshift range of our OFGs is $z = 3-7$ (see $\S$\ref{Sect::redshift}, Fig.~\ref{Fig:distribution}, and Table \ref{table:properties}). The Lyman-break color criteria are as follows:
\begin{eqnarray*}
z\sim3: (U_{336}-B_{435}>1)\wedge(B_{435}-V_{606}<1.2)\wedge \\
(i_{775}-Y_{105}<0.7)\wedge(\chi_{\rm UV_{\rm 225},UV_{\rm 275}}^2<2),
\end{eqnarray*}
\begin{eqnarray*}
z\sim4: (B_{\rm 435}-V_{\rm 606}>1)\wedge (i_{\rm 775}-J_{\rm 125}<1) \wedge \\
(B_{\rm 435}-V_{\rm 606} > 1.6(i_{\rm 775}-J_{\rm 125})+1),
\end{eqnarray*}
\begin{eqnarray*}
z\sim5: (V_{\rm 606}-i_{\rm 775}>1.2)\wedge (z_{\rm 850}-H_{\rm 160}<1.3) \wedge \\
(V_{\rm 606}-i_{\rm 775} > 0.8(z_{\rm 850}-H_{\rm 160})+1.2),
\end{eqnarray*}
\begin{eqnarray*}
z\sim6: (i_{\rm 775}-z_{\rm 850}>1.0)\wedge (Y_{\rm 105}-H_{\rm 160}<1.0) \wedge \\
(i_{\rm 775}-z_{\rm 850} > 0.777(Y_{\rm 105}-H_{\rm 160})+1.0),
\end{eqnarray*}
\begin{eqnarray*}
z\sim7: (z_{\rm 850}-Y_{\rm 105}>0.7)\wedge (J_{\rm 125}-H_{\rm 160}<0.45) \wedge \\
(z_{\rm 850}-Y_{\rm 105} > 0.8(J_{\rm 125}-H_{\rm 160})+0.7),
\end{eqnarray*}
here, $\wedge$ and $\vee$ represent the logical AND and
OR symbols, respectively. The $\chi_{\rm UV_{\rm 225},UV_{\rm 275}}^2$ = $\Sigma_{\rm i}$ SGN$(f_{\rm i})(f_{\rm i}/\sigma_{\rm i})^2$, where $f_{\rm i}$ ($\sigma_{\rm i}$) is the flux (uncertainty) in the $i$-band and SGN$(f_{\rm i})$ is equal to 1 if $f_{\rm i}>0$ and -1 if $f_{\rm i}<0$.
As in \cite{Bouwens2015}, we use a 1$\sigma$ upper limit as the flux in the dropout band in the case of non-detection. The selected sources are required to be undetected (<2$\sigma$) in all bands blueward of the Lyman break and detected (>3$\sigma$) in all of the above bands redward of the break. We note that we do not include $U_{\rm 336}$, $UV_{\rm 225}$, and $UV_{\rm 275}$ in our work because our OFGs are undetected in these bands. Even if we only use the criteria of $z\sim3$: $(B_{\rm 435}-V_{\rm 606}<1.2) \wedge (i_{\rm 775}-Y_{\rm 105}<0.7)$, no galaxy at $z\sim3$ in our catalog is classified as LBG.
The color criteria for Lyman-break of \cite{Bouwens2020} are slightly different from those of \cite{Bouwens2012a} and \cite{Bouwens2015}, thus we also used the criteria of \cite{Bouwens2012a,Bouwens2015} to select LBGs. All three methods identified the same 5 LBGs in our OFG catalog: OFG9, OFG13, and OFG18  at $z\sim4$; OFG22 at $z\sim5$; and OFG24  at $z\sim6$.

\subsubsection{ $H$-dropouts}\label{Sect::Hdropout}
Galaxies bright in IRAC but not detected in the $H$-band are commonly referred to as $H$-dropouts. However, this definition can be confusing since different fields have been observed at different depths in the $H$-band. 
Here, we used the deepest $H$-band image to date in the GOODS-South field \citep[HLF;][]{Whitaker2019}, with a 5$\sigma$ point-source depth of approximately $27.0-29.8$ mag, to identify $H$-dropouts and to extend the sample to our more general definition of OFGs. In our sample, eight galaxies (OFG1-OFG7 and OFG27) are classified as $H$-dropouts, that is, there is no detection above 3$\sigma$ in the $H$-band (see Table \ref{table:properties}). We note that in \cite{Wang2019}, $H$-dropouts include all sources with no $H$-band flux above 5$\sigma$, that is, $H>27$ mag -- instead of 3$\sigma$ here -- and [4.5] < 24 mag. 
If we apply the same criterion, we find seven OFGs that meet this definition: OFG1, OFG2, OFG3, OFG5, OFG6, OFG11, and OFG27 (see Fig.~\ref{color_mag}).

In the GOODS-ALMA 2.0 catalog, \cite{Carlos2021} reported 13 OFGs out of 88 galaxies detected above 3.5$\sigma$ at 1.13 mm in the GOODS-ALMA field. Among them, 12 OFGs are included in our sample (see Table \ref{table:properties} with a "$\dagger$" exponent). The remaining one (A2GS2 in \citealt{Carlos2021} or AGS4 in \citealt{Franco2018,Franco2020a,Zhou2020}) does not meet our criterion of $H$ > 26.5 mag, since after applying our de-blending procedure ($\S$\ref{Sect:deblend}), we measured an $H$-band magnitude of $H$ = 24.76 mag. This value is consistent with the findings of \cite{Zhou2020}, who measured $H$ = 25.23 mag.


\section{Properties of individual galaxies}\label{individual}
In this section, we focus on a set of properties of individual galaxies: redshift, stellar mass, infrared luminosity, star formation rate, gas mass, dust mass, and dust temperature. The methodologies used to derive these properties are also used for the stacked samples, as described in $\S$\ref{stack}. 

\subsection{Redshifts and stellar masses}\label{Sect::redshift}
We fit the SED from the UV to MIR (rest-frame UV to NIR) to measure the photometric redshifts ($z_{\rm phot}$) and stellar masses ($M_{\star}$). The photometric redshifts were determined with the code \texttt{EAzY}\footnote{\texttt{EAzY}: \url{https://github.com/gbrammer/eazy-photoz}} \citep{Brammer2008}. Then we fixed the redshift and derived $M_{\star}$ with the code \texttt{FAST++}\footnote{\texttt{FAST++:} \url{https://github.com/cschreib/fastpp}, an updated version of the SED fitting code \texttt{FAST} \citep{Kriek2009}.}. The setup is described below.

Photometric redshifts were obtained with the galaxy template set \texttt{``eazy\_v1.3''}, which includes, in particular, a dusty starburst model to account for extremely dusty galaxies. We did not apply the redshift prior based on $K$-band magnitudes, as this prior is based on models that do not reproduce high-redshift mass functions \citep[see discussion in][]{Schreiber2018a}. 


Two sources, OFG15 and OFG27, have spectroscopic redshifts ($z_{\rm spec}$) confirmed by one-line detections. The galaxy OFG15 has a $z_{\rm spec}$ measured by the Lyman $\alpha$ line ($\sim$7$\sigma$) from the MUSE-Wide survey \citep[ID139013229 in][]{Herenz2017,Urrutia2019}. The galaxy OFG27 has a $z_{\rm spec}$ identified from the CO(6-5) line detection ($\sim$10$\sigma$) with ALMA \citep[AGS17 in][]{Zhou2020}. 
In addition, OFG1, 2, 25, and 27 were discovered in an overdensity region with a peak redshift of $z_{\rm overdensity}=3.47$ \citep[AGS24, 11, 15, 17 in][where they were studied in detail]{Zhou2020}. For the above five galaxies, we used their $z_{\rm spec}$ or $z_{\rm overdensity}$ in the following analysis.


Stellar masses were  then derived using the code \texttt{FAST++}, assuming a delayed, exponentially declining star formation history (SFH), with \cite{Bruzual2003} stellar population models and a \cite{Calzetti2000} dust attenuation law. The parameters used in \texttt{FAST++} are shown in Table \ref{table:fast}.

\begin{figure}
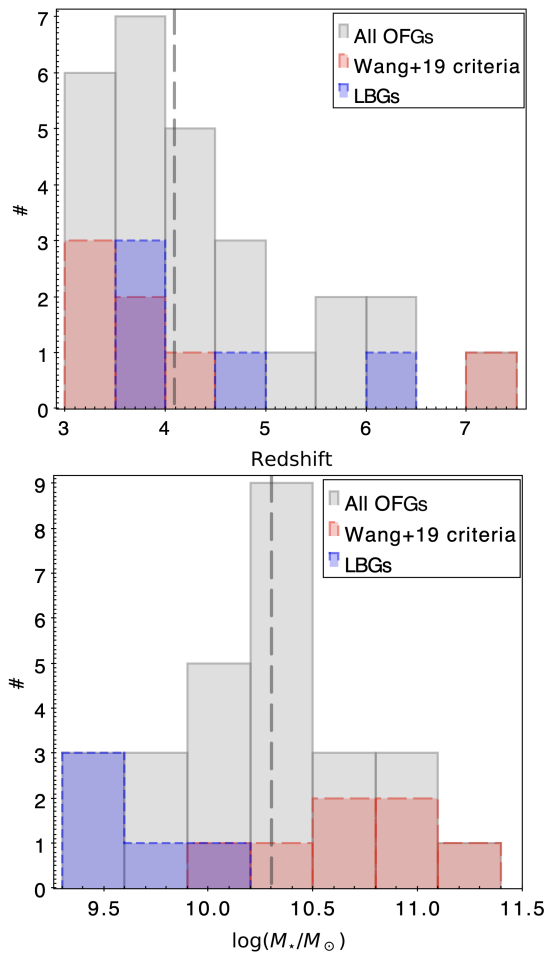

\centering
\includegraphics[scale=0.22]{img/hist_z3.pdf}
\hspace*{0.2cm}\includegraphics[scale=0.23]{img/hist_mass3.pdf}
\caption{Photometric redshift and stellar mass distributions for our total 27 OFGs (grey). Among them, five OFGs are identified as LBGs (blue; see $\S$\ref{Sect::lyman break}) and seven OFGs meet the criteria of $H$-dropouts (H > 27 mag \& [4.5] < 24 mag) in \cite{Wang2019} (red; see $\S$\ref{Sect::Hdropout}). 
$Top$: Photometric redshifts of our OFGs (including two sources with spectroscopic redshifts).  
$Bottom$: LBGs cover the low stellar mass end and $H$-dropouts cover the high stellar mass end. Our OFGs have a wide range of stellar masses with log($M_{\star}$/$M_{\odot}$) = $9.4-11.1$. The median redshift and stellar mass of the OFGs (dashed lines) are $z_{\rm med}=4.1$ and log($M_{\rm \star med}$/$M_{\odot}$) = 10.3, respectively.
\label{Fig:distribution}}
\end{figure}

In Fig.~\ref{Fig:distribution}, we show the distributions of derived redshifts and stellar masses of our OFGs. The redshift distribution confirms that our OFGs exhibit redshifts of $z_{\rm phot}$ > 3, which are consistent with the theoretical galaxy templates (see Fig.~\ref{track}). The median redshift of the distribution is $z_{\rm med}=4.1$. Compared to the LBGs covering the low stellar mass end and the $H$-dropouts in \cite{Wang2019} covering the high stellar mass end, 
our sample presents a broad distribution of stellar masses with log($M_{\star}$/$M_{\odot}$) = $9.4-11.1$ and a median value of log($M_{\rm \star med}$/$M_{\odot}$) = 10.3. The individual redshift and stellar mass values are listed in Table \ref{table:properties}, and the individual SEDs are presented in Figs.~\ref{Fig:SED1} and \ref{Fig:SED2}.

\begin{figure}
\centering
\includegraphics[scale=0.37]{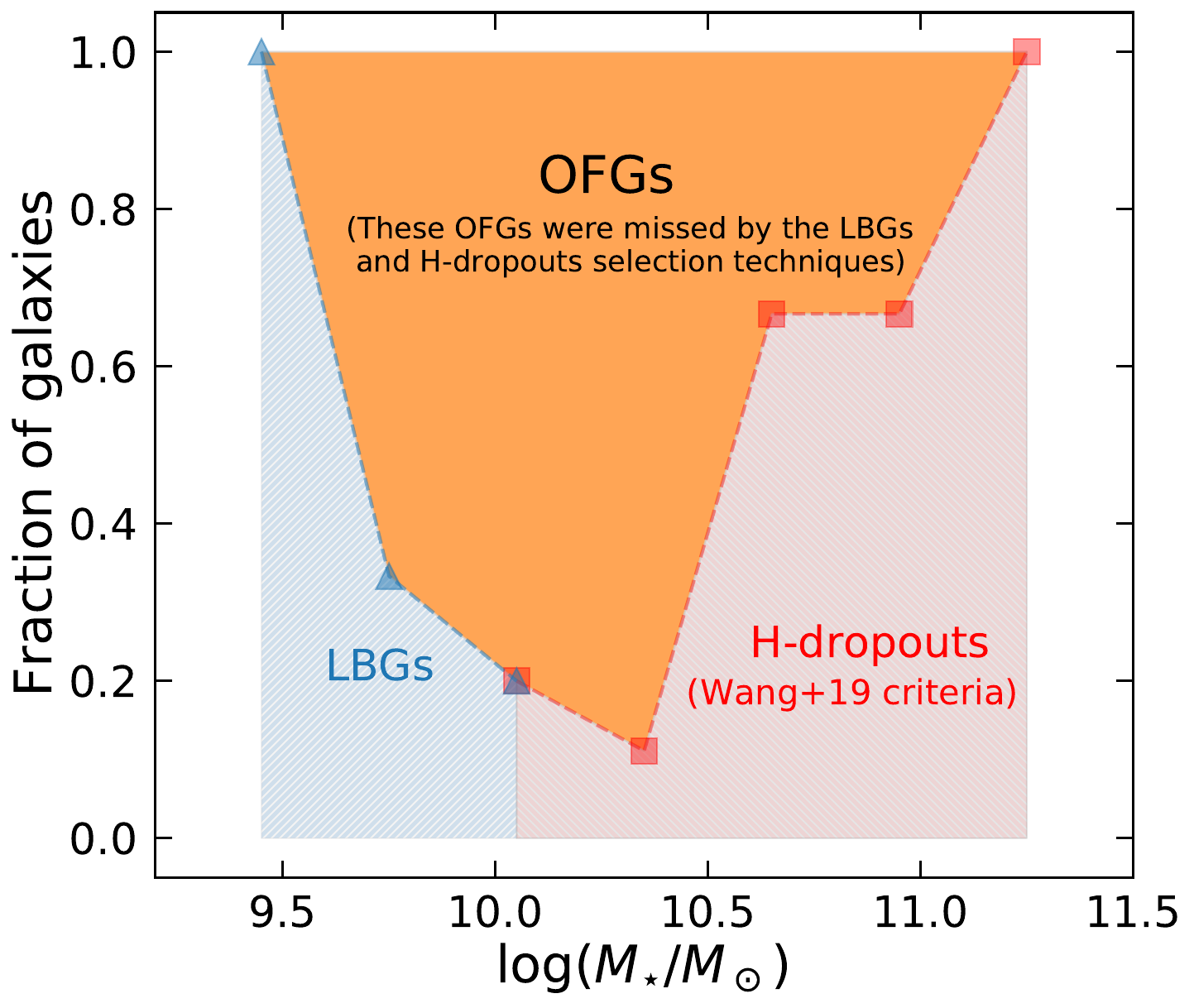}
\caption{Number fractions of galaxies in our OFG sample that are identified as LBGs (blue shaded region), $H$-dropouts (red shaded region), and the remaining OFGs (orange filled region) that are missed by the first two selection techniques. The OFGs undetected by the LBGs and $H$-dropouts criteria represent approximately 75\% at stellar masses between log($M_{\star}$/$M_{\odot}$) = $9.5-10.5$. 
\label{Fig:frac}}
\end{figure}

We investigate  in Fig.~\ref{Fig:frac}, the proportions of LBGs, $H$-dropouts, and remaining OFGs (after removing LBGs and $H$-dropouts) in our sample at different stellar masses. At stellar masses of log($M_{\star}$/$M_{\odot}$) = $9.5-10.5$, the fraction of OFGs is about three times the sum of LBGs and $H$-dropouts. In other words, up to 75\% of the galaxies with a stellar mass of  log($M_{\star}$/$M_{\odot}$) = $9.5-10.5$ at $z>3$ are missed by the previous LBG and $H$-dropout selection techniques.

\begin{table}
\caption{Input parameters used in the UV to 8$\mu m$ SED fitting procedures with \texttt{FAST++} to derive stellar masses.}            
\centering                          
\begin{tabularx}{0.4\textwidth}{l X}  
\hline\hline                
 Parameter      &  Value \\ 
\hline\hline                         
& Delayed SFH \\
\hline  
Age [log(yr$^{-1}$)]  & 6.0 - 10.2, step 0.1 \\
$\tau$ [log(yr$^{-1}$)]  & 6.5 - 11, step 0.1 \\
Metallicity & 0.02 (solar) \\
\hline
\hline
& Dust attenuation: \cite{Calzetti2000} \\
\hline
$A_v$ &  0 - 6, step 0.02\\
\hline                               
\end{tabularx}
\label{table:fast}     
\end{table}

\begin{table}
\caption{Input parameters used in the 24$\mu m$ to mm SED fitting procedures with \texttt{CIGALE}.}            
\centering                          
\begin{tabularx}{0.45\textwidth}{l X}  
\hline\hline                
 Parameter      &  Value \\ 
\hline\hline                         
& Dust emission: \cite{Draine2014} \\
\hline  
\textbf{$q_{\rm PAH}$}& 0.47, 1.12, 1.77, 2.50, 3.19, 3.90, 4.58, 5.26, 5.95, 6.63, 7.32\\
$U_{\rm min}$ & 0.1, 0.5, 1, 5, 12, 15, 20, 25, 30, 35, 40, 50 \\
$\alpha$  & 1, 1.5, 2, 2.5, 3.0 \\
$\gamma$ & 0.0001, 0.001, 0.01, 0.1, 0.5, 1 \\
\hline                               
\end{tabularx}
\label{table:cigale}     
\end{table}

\subsection{Infrared SED fitting}\label{Sect::LIR}
The infrared luminosities (8$-$1000 $\mu$m; $L_{\rm IR}$) and dust mass ($M_{\rm dust}$) are derived from the FIR SED-fitting. We used two methodologies in the fit, depending on whether the galaxies have a \textit{Herschel} counterpart or not (see Figs.~\ref{Fig:SED1} and \ref{Fig:SED2} for the entire sample).  

For galaxies with a \textit{Herschel} counterpart (3/27), we performed the FIR SED-fitting with \texttt{CIGALE}\footnote{\texttt{CIGALE:} \url{https://cigale.lam.fr}} \citep[Code Investigating GALaxy Emission;][]{Burgarella2005,Noll2009,Boquien2019}. We fit data from 24 $\mu$m up to millimeter wavelengths from the catalogs of Wang et al. (in prep.), \cite{Elbaz2011}, and GOODS-ALMA v2.0 1.13 mm \citep{Carlos2021}. The dust infrared emission model is the one of \cite{Draine2014}. 
The parameters used in \texttt{CIGALE} are shown in Table \ref{table:cigale}.

For galaxies without a \textit{Herschel} counterpart but with ALMA detections at 1.13 mm \citep[8/27;][]{Carlos2021}, applying the dust emission model \citep{Draine2014} of \texttt{CIGALE} would fit a single point in the FIR with four parameters, which would leave us with much less meaningful results. As a compromise, 
we used the IR template library\footnote{\texttt{S17} library: \url{http://cschreib.github.io/s17-irlib/}} 
described in \cite{Schreiber2018c}. In brief, it consists of two ingredients: $i$) dust continuum created by big dust grains (silicate + amorphous carbon grains) and $ii$) mid-infrared features contributed by polycyclic aromatic hydrocarbon (PAH) molecules. To form a full dust spectrum, the relative contribution of big grains and PAHs are characterized by the mid-to-total infrared color (IR8 = $L_{\rm IR}/L_{\rm 8}$), which is related to the redshift and starburstiness of a galaxy. The starburstiness is defined by $R_{\rm SB}$ $\equiv$ $\Delta$MS $\equiv$ SFR/SFR$_{\rm MS}$ \citep{Elbaz2011}, where SFR$_{\rm MS}$ is the average SFR of the main sequence galaxies presented in \cite{Schreiber2015}.

We fit the data iteratively with two templates: the star formation main sequence (MS) template and the starburst (SB) template, with fixed $R_{\rm SB}$ values of 1 and 5, respectively. For each template, given the known redshift and fixed $R_{\rm SB}$, we calculated the dust temperature ($T_{\rm dust}$) and IR8 using Eqs. (18) and (19) from \cite{Schreiber2018c}. The templates we used were normalized to $M_{\rm dust} = 1 M_\odot$. After re-normalizing the SED to the ALMA flux density at 1.13 mm, we obtained the total $M_{\rm dust}$ and total $L_{\rm IR}$ by integrating the SED in the 8$-$1000 $\mu$m rest-frame range. Then we computed $R_{\rm SB}$ using the output $L_{\rm IR}$ (\citealt{Kennicutt2012}; the contribution of UV to the SFR is negligible, as shown later in Table \ref{table:mean} and $\S$\ref{Sect::stacked_sfr}).
We computed two $R_{\rm SB}$ values for each galaxy derived from both templates (MS and SB).  If both $R_{\rm SB}$ values are less (greater) than 3, we consider a galaxy a MS (SB) galaxy. The best-fit SED was then generated with the typical value $R_{\rm SB}=1$ (MS) or $R_{\rm SB}=5$ (SB). Otherwise, that is, if both templates do not agree with each other for $R_{\rm SB}$ $<3$ or $>3$, we kept the two SEDs given by both templates as upper and lower limits and used the average template as the best SED. This approach is similar to the one used by \cite{Carlos2022} but slightly more conservative. 
Compared to the $L_{\rm IR}$ in \cite{Carlos2022} for the same sources, the results are generally consistent, 
with a median relative difference of ($L_{\rm IR}^{\rm This\,work}-L_{\rm IR}^{\rm G22})/L_{\rm IR}^{\rm G22}=0.04\pm0.18$. The relatively large dispersion was expected because the IR template fit is based on only one observed data point of ALMA 1.13 mm with large uncertainty in $z_{\rm phot}$. 

As a consistency test, for the three sources detected by Herschel and ALMA, we performed a SED fit using the IR template library normalized only to the ALMA point as if they had no Herschel values. We then derived $L_{\rm IR}$. The ratios of the IR luminosities, $L_{\rm IR}^{\rm CIGALE} / L_{\rm IR}^{\rm IR \, template}$ (where $L_{\rm IR}^{\rm IR \, template}$ is derived with only one photometric point), are 0.49, 2.95, and 1.95 for OFG2, OFG20, and OFG27, respectively. The sample is obviously statistically limited but we do not find a systematic offset when using only one photometric point.

One caveat for the $M_{\rm dust}$ estimates from \texttt{CIGALE} or the IR template library is that they are based on different dust models. Compared to the more standard dust models of \cite{Draine2014} \citep[an updated version of][]{Draine2007} that we adopted in \texttt{CIGALE}, the one used in the IR template library of \cite{Schreiber2018c} assumes that the carbonated grains are amorphous carbon grains rather than graphites. \cite{Schreiber2018c} stated that different dust grain species from the IR template library have different emissivities, systematically lowering the derived $M_{\rm dust}$ by a factor of about two. Therefore, to have comparable $M_{\rm dust}$ for galaxies with and without a \textit{Herschel} counterpart, we have corrected the differences in $M_{\rm dust}$ obtained using the IR template library in Table \ref{table:gas} and also in the following sections.


\subsection{SFRs}\label{Sect::sfr}

The total SFR was measured from the contributions of dust-obscured star formation (SFR$_{\rm IR}$) and unobscured star formation (SFR$_{\rm UV}$). 
The SFR$_{\rm IR}$ was calculated based on the total infrared luminosity ($L_{\rm IR}$), derived from integrating the best-fitted SED between 8 and 1000 $\mu$m in the rest frame, following \cite{Kennicutt2012}. The SFR$_{\rm UV}$ was derived from the luminosity emitted in the UV ($L_{\rm UV}$), which was not corrected for dust attenuation, following \cite{Daddi2004} (scaled to a \citealt{Chabrier:2003} IMF). We calculated the total SFR:
\begin{flalign}
\label{eq:SFRtot}
\rm{SFR}_{\rm tot}\ [M_{\odot}\,yr^{-1}]&= \rm{SFR}_{\rm IR}+\rm{SFR}_{\rm UV}&&\\\nonumber
&=\,1.49 \times 10^{-10} L_{\rm IR}\,+\,1.27 \times 10^{-10} L_{\rm UV},
\end{flalign}
both $L_{\rm IR}$ and $L_{\rm UV}$ in units of $L_{\rm \odot}$, and
\begin{eqnarray}
\label{eq:2}
L_{\rm IR}\ [L_{\rm \odot}]&=&4\pi D_L^2\int_{\rm 8\mu m}^{1000\mu m} F_{\rm \nu}(\lambda) \times \frac{c}{\lambda^2}d\lambda,,\\
\label{eq:3}
L_{\rm UV}\ [L_{\rm \odot}]&=&4\pi D_L^2\frac{\nu_{\rm 1500}}{(1+z)} \frac{10^{-0.4(48.6+m_{\rm 1500})}}{3.826 \times 10^{33}},
\end{eqnarray}
\noindent where $D_L$ is the luminosity distance (cm), $\nu_{\rm 1500}$ is the frequency (Hz) corresponding to the rest-frame wavelength 1500 $\AA$, and $m_{\rm 1500}$ is the AB magnitude at the rest-frame 1500 $\AA$.
Here, the value for $m_{\rm 1500}$ was derived from the best-fitting templates using \texttt{EAzY}, with a top-hat filter centered at 1500 $\AA$ and a width of 350 $\AA$. 

For individual OFGs with a \textit{Herschel} and/or ALMA counterpart (and therefore SFR$_{\rm IR}$), we present their SFR (SFR$_{\rm tot}=$ SFR$_{\rm UV}+$SFR$_{\rm IR}$) in Table \ref{table:properties}. Further discussion will be provided later in $\S$\ref{Sect::stacked_sfr} for the stacked samples.

\subsection{Molecular gas mass}\label{Sect::Mgas}
The gas mass, $M_{\rm gas}$, can be determined from $M_{\rm dust}$ by employing the gas-to-dust ratio ($\delta_{\rm GDR}$) with a metallicity dependency \citep[e.g.,][]{Magdis2012}:
\begin{flalign}
M_{\rm gas} = M_{\rm H_{\rm 2}} +  M_{\rm H_{\rm I}}  = \delta_{\rm GDR} M_{\rm dust},
\label{Mgas}
\end{flalign}
\begin{flalign}
\rm{log}(\delta_{\rm GDR}) = (10.54\pm1.0) - (0.99\pm0.12)\times(12 + \rm{log}(O/H)).
\label{GDR}
\end{flalign}
The metallicity was determined from the redshift-dependent mass-metallicity relation \citep[MZR;][]{Genzel2015}: 
\begin{flalign}
12 + \rm{log}(O/H) = a - 0.087[\rm{log}(M_{\rm *}) - b]^2, 
\label{Mgas}
\end{flalign}
where a = 8.74 and b = 10.4 + 4.46 $\times$ log($1+z$)$-$1.78 $\times$ log($1+z$)$^2$. We adopted an uncertainty of 0.2 dex in the metallicities \citep{Magdis2012}. 

With the estimates of $M_{\rm gas}$, we can calculate the gas fraction ($f_{\rm gas}$) and gas depletion time ($\tau_{\rm dep}$) as $f_{\rm gas} = M_{\mathrm{gas}}/(M_{\star} + M_{\mathrm{gas}})$ and $\tau_{\rm dep} = M_{\mathrm{gas}}/(\rm{SFR}_{\rm IR}+\rm{SFR}_{\rm UV})$. The $\tau_{\rm dep}$ is the inverse of the star formation efficiency (SFE $= 1/\tau_{\rm dep}$). The $M_{\rm gas}$, $f_{\rm gas}$, and $\tau_{\rm dep}$ for the individual sources are presented in Table \ref{table:gas}. We underline that only three OFGs have a \textit{Herschel} counterpart, and the rest have $M_{\rm dust}$ only based on ALMA 1.13 mm. Thus, there is a large uncertainty in the values of $M_{\rm dust}$ and, consequently, $M_{\rm gas}$, $f_{\rm gas}$, and $\tau_{\rm dep}$. Therefore, this paper does not go deeper into the $M_{\rm dust}$ and gas properties of individual galaxies. Instead, for the study of gas properties of the OFGs, we focus on the stacked sample,  described in $\S$\ref{figure_gas}.



\begin{table}
\caption{Dust and gas properties of the OFGs.}   
\tiny          
\centering
\begin{threeparttable} 
 
\begin{tabular}{l c c c cc}     
\hline\hline       
ID & log$(M_{\rm dust})$ & T$_{\rm dust}$&  log$(M_{\rm gas})$ & $f_{\rm gas}$ & $\tau_{\rm dep}$ \\  
& log$(M_{\odot})$ & (K)& log$(M_{\odot})$ & &(Myr)\\ 
(1)& (2) & (3) & (4)&(5)&(6)\\
\hline  
OFG1 & 8.49$_{-0.06}^{+0.05}$ &$(...)$& 10.64$_{-0.06}^{+0.05}$ & 0.42$_{-0.08}^{+0.08}$  & 115  $\pm$  20\\                
OFG2 & 8.80$_{-0.13}^{+0.10}$ & 37.0$_{-1.6}^{+1.7}$ & 10.96$_{-0.13}^{+0.10}$ & 0.61$_{-0.10}^{+0.09}$  & 249  $\pm$  67\\
OFG4 & 8.04$_{-0.09}^{+0.07}$ &$(...)$& 10.39$_{-0.09}^{+0.07}$ & 0.38$_{-0.14}^{+0.11}$  & 46  $\pm$  12 \\
OFG5 & 8.09$_{-0.07}^{+0.06}$ &$(...)$& 10.49$_{-0.07}^{+0.06}$ & 0.67$_{-0.11}^{+0.26}$  & 114  $\pm$  31 \\  
OFG7  & 8.37$_{-0.04}^{+0.04}$ &$(...)$& 10.75$_{-0.04}^{+0.04}$ & 0.72$_{-0.04}^{+0.05}$  & 52  $\pm$  7\\    
OFG11& 8.28$_{-0.07}^{+0.06}$ &$(...)$& 10.57$_{-0.07}^{+0.06}$ & 0.59$_{-0.05}^{+0.05}$  & 97  $\pm$  27\\    
OFG18& 7.96$_{-0.16}^{+0.11}$ &$(...)$& 10.71$_{-0.16}^{+0.11}$ & 0.95$_{-0.02}^{+0.02}$  & 180  $\pm$  77 \\
OFG19& 8.02$_{-0.06}^{+0.06}$ &$(...)$& 10.39$_{-0.06}^{+0.06}$ & 0.55$_{-0.06}^{+0.05}$  & 90  $\pm$  23\\
OFG20& 8.40$_{-0.05}^{+0.04}$& 68.5$_{-3.7}^{+4.2}$ & 10.59$_{-0.05}^{+0.04}$ & 0.34$_{-0.10}^{+0.10}$  & 43  $\pm$  6\\
OFG25& 8.66$_{-0.03}^{+0.03}$ &$(...)$&  11.11$_{-0.03}^{+0.03}$ & 0.93$_{-0.02}^{+0.12}$  & 105  $\pm$  9\\
OFG27& 9.01$_{-0.05}^{+0.04}$& 51.4$_{-2.1}^{+2.2}$ & 11.07$_{-0.05}^{+0.04}$ & 0.48$_{-0.09}^{+0.11}$  & 66  $\pm$  8\\   
\hline 
                 
\end{tabular}
\begin{tablenotes}
\item \textbf{Note:} (1) Source ID; (2) Dust mass obtained from \texttt{CIGALE} for the galaxies with a \textit{Herschel} counterpart or from the IR template library \citep{Schreiber2018c} for the galaxies without a \textit{Herschel} counterpart but with an ALMA counterpart (see $\S$\ref{Sect::LIR}). Since the dust emissivity used in \cite{Schreiber2018c} is different from the one of \cite{Draine2014} (used in \texttt{CIGALE}), resulting in a systematic twice lower $M_{\rm dust}$, here we have corrected the $M_{\rm dust}$ derived from the IR template library by multiplying by two. 
(3) Dust temperature obtained from a single temperature MBB model for the galaxies with a \textit{Herschel} counterpart (see $\S$\ref{Sect::Tdust}); (4) Gas mass obtained from the metallicity-dependent gas-to-dust mass ratio technique (see $\S$\ref{Sect::Mgas}); (5) Gas fraction: $f_{\rm gas} = M_{\mathrm{gas}}/(M_{\star} + M_{\mathrm{gas}})$; (6) Gas depletion time: $\tau_{\rm dep} = M_{\mathrm{gas}}/(\rm{SFR}_{\rm IR}+\rm{SFR}_{\rm UV})$, which is the inverse of the star formation efficiency (SFE $= 1/\tau_{\rm dep}$).

\end{tablenotes}
\label{table:gas} 
\end{threeparttable} 
\end{table}

\subsection{Dust temperatures}\label{Sect::Tdust}

For the comparison with previous studies, we measured the effective dust temperatures ($T_{\rm dust}$) by fitting single-temperature modified black-body (MBB) models to the FIR to mm photometry of the individual galaxies with a \textit{Herschel} counterpart, following:
\begin{equation}
  S_\nu \propto \frac{\nu^{3+\beta}}{\exp(\frac{h \nu}{k_B T_{dust}}) - 1},
\end{equation}
under the assumption of optically thin dust, where S$_\nu$ is the flux density, $k_B$ is Boltzmann constant, $h$ is the Planck constant, and $\beta$ is the dust emissivity index. We assumed $\beta$ = 1.5, a typical value for dusty star-forming galaxies \citep[e.g.,][]{Hildebrand1983,Kovacs2006,Gordon2010}. We note that changing  $\beta$ does not have a significant effect on $T_{\rm dust}$, as $\beta$ is affecting the slope of the Rayleigh-Jeans (RJ) tail of the dust emission at the rest-frame $\lambda \geq 200 \mu$m, while the peak of the dust SED is what determines $T_{\rm dust}$ \cite[e.g.,][]{Casey2012,Jin2019}.

Following the criteria used in \cite{Hwang2010} \citep[also used in][]{Franco2020b,Carlos2022}, we only fit the observed data points at $\lambda \geq 0.55 \lambda_{peak}$ to avoid contamination from small dust grains, polycyclic aromatic hydrocarbon (PAH) molecules, and/or AGNs in the MIR, where $\lambda_{peak}$ is the peak of IR SEDs of the \texttt{CIGALE} best fit. The fitted galaxies should satisfy  the following conditions: (i) at least one data point at $0.55\lambda_{peak} \leq \lambda < \lambda_{peak}$; and (ii) at least one data point at $\lambda >\lambda_{peak}$ (to exclude the synchrotron contribution from radio data). In our case, we finally fitted the photometry from \textit{Herschel}/SPIRE bands (250 $\mu$m, 350 $\mu$m, and 500 $\mu$m) and ALMA 1.13 mm. We note that we did not consider the CMB effect in the MBB fit because of the lack of data points in the RJ-tail where the CMB plays an important role \citep[e.g.,][]{Jin2019}. 

The MBB fit was performed using a Markov Chain Monte Carlo (MCMC) approach with 12000 iterations using the Python package \texttt{PyMC3}\footnote{\texttt{PyMC3} is available at: \url{https://docs.pymc.io/en/v3/}}. The derived $T_{\rm dust}$ for individual galaxies with a \textit{Herschel} counterpart are listed in Table \ref{table:gas}. For individual galaxies (three galaxies with \textit{Herschel} counterparts were fitted with MBB), their $T_{\rm dust}$ exhibit a large dispersion, that is, $T_{\rm dust}$ = $37-69$ K, and only one ($T_{\rm dust}$ = 51  $\pm$  2 K) is in agreement with the expected value from the redshift evolution in MS galaxies from the literature \citep{Schreiber2018c}. 
The fitting results are also shown in Figs.~\ref{Fig:SED1} and \ref{Fig:SED2}. Considering that only three OFGs have $T_{\rm dust}$ values, we do not go deeper into the discussion of individual galaxies. Instead, we further discuss the stacked optically dark/faint samples in $\S$\ref{figure_gas}.

\begin{figure*}
\centering
\includegraphics[scale=1.44]{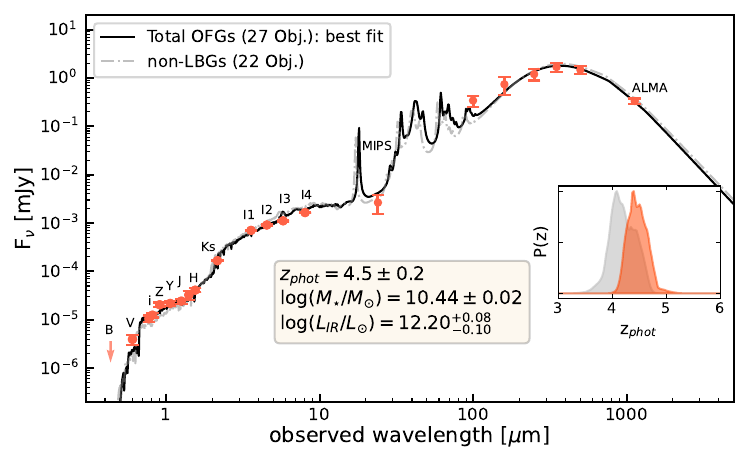}

\hspace*{1.7cm}\includegraphics[scale=0.44]{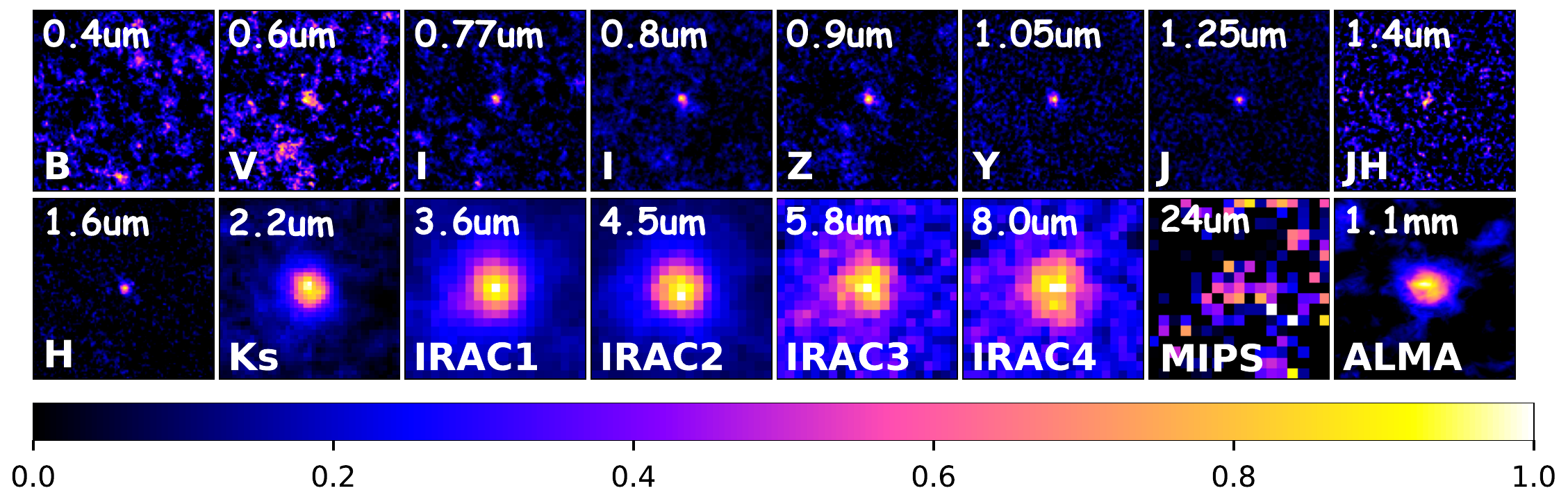}
\caption{Median stacked SED and images of the total sample of 27 OFGs in this work. 
$Top$:  Best-fit SED of the total sample (black line). The measured fluxes (red points) are derived from the stacked images. Error bars (1$\sigma$) and upper limits (3$\sigma$) are obtained from the Monte Carlo simulation (except $Herschel$) and bootstrap approach ($Herschel$; see $\S$\ref{Sect::stacked SED}). We also show the best-fit SED for 22 non-LBGs (grey line). These 22 non-LBGs will be used to calculate the cosmic SFRD. The inset shows the likelihood distributions of the photometric redshift of our samples (total sample in red, 22 non-LBGs in grey), based on the UV to MIR SED fitting from \texttt{EAzY}, which is normalized to the peak value.  The redshift obtained from the maximized likelihood is $z$ $\sim$ 4.5 for the total 27 OFGs and $z$ $\sim$ 4.2 for the 22 non-LBGs. 
$Bottom$: Stacked images of the total sample with peak fluxes normalized. Each panel is 6$^{\prime\prime}$ $\times$ 6$^{\prime\prime}$ except for the MIPS 24 $\mu$m, which is 24$^{\prime\prime}$ $\times$ 24$^{\prime\prime}$.
\label{Fig:stacked}}
\end{figure*}

\section{Properties of the stacked OFGs}\label{stack}
\subsection{Stacking analysis}\label{Sect::stacked data}
We performed a stacking analysis to study the global properties of our sample. Considering that OFGs are very faint in the optical/NIR ($H>26.5$ mag) and only 11/27 have \textit{Herschel} and/or ALMA counterparts, performing a stacking analysis helps improving the accuracy of the median photometric redshift and SFR measurements. To build the SED of our sample, we generated a median and mean stacked image in each filter, from the optical to 1.13 mm. Specifically, we used images from the HST/ACS (F435W, F606W, F775W, F814W, F850LP), HST/WFC3 (F105W, F125W, F140W, F160W), ZFOURGE $K_{\rm s}$-band, \textit{Spitzer}/IRAC (3.6, 4.5, 5.8, and 8\,$\mu$m), \textit{Spitzer}/MIPS (24\,$\mu$m), \textit{Herschel}/PACS\&SPIRE (100, 160, 250, 350 and 500\,$\mu$m), and ALMA 1.13 mm maps. 

The photometry was obtained mostly using aperture photometry techniques, except for the \textit{Herschel} bands, where appropriate aperture corrections were applied to account for flux losses outside the aperture. This procedure is very similar to that used previously in the deep surveys, which we summarize here. In the HST/ACS and HST/WFC3 bands, fluxes were extracted on the PSF-matched images (to the F160W) using the same aperture of 0.7$^{\prime\prime}$-diameter as in \cite{Whitaker2019}, which maximizes the signal-to-noise ratio (S/N) of the resulting aperture photometry. In the $K_{\rm s}$-band, we used a 1.2$^{\prime\prime}$ diameter circular aperture to measure flux on the ZFOURGE $K_{\rm s}$-band image following \cite{Straatman2016}, whose PSF was matched to a Moffat profile with FWHM=0.9$^{\prime\prime}$. In the IRAC bands, fluxes were extracted separately without PSF matching due to the broader PSFs. We adopted a 2.2$^{\prime\prime}$ diameter aperture to maximize the S/N of the resulting aperture photometry. In the MIPS 24\,$\mu$m band, we used a large aperture of 6$^{\prime\prime}$ in diameter corresponding to its full width at half maximum.  At 1.13 mm, we used a diameter of 1.6$^{\prime\prime}$ to measure the flux, which is the optimal trade-off between total flux and SNR. 

Uncertainties on the photometry were derived from the Monte Carlo simulations. For each band, we carried out the same stacking analysis as above, but at random positions, and measured the flux value on the stacked image. This was repeated 1000 times. We then calculated the 16th and 84th percentiles of the distribution of values as flux uncertainties.

For the \textit{Herschel}/PACS and SPIRE bands, we used the PSF fitting with a free background to fit the stacked image following \cite{Schreiber2015}. The uncertainties were obtained using the following methods: 1) a bootstrap approach; specifically, as an example, we generated a sample of 27 sources from 27 OFGs,  allowing the same galaxy to be picked repeatedly, and measured the stacked flux. This procedure was repeated 100 times, and we calculated their standard deviation as the flux uncertainty; and 2) a Monte Carlo simulation approach, which is the same as that used for lower wavelength images and 1.13 mm images. Here we adopted a 0.9 $\times$ FWHM diameter circular aperture. We note that the results given from the bootstrap approach include the uncertainties from i) the PSF fitting, ii) the clustering bias effect, and iii) background fluctuation. Thus, the derived values of uncertainties from bootstrap are larger than those from the Monte Carlo simulation. We conservatively take the former values as our uncertainties.

\begin{table}
\caption{Median physical properties of the total sample of 27 OFGs.}
\centering
\begin{threeparttable} 
 
\begin{tabular}{l ll}     
\hline\hline    
        \multicolumn{3}{l}{Derived from SED fitting with median stacked photometry$^{*}$}\\
        \hline
        $z_{\rm phot}$$^{a}$            &                       & 4.5  $\pm$  0.2                            \\
        $M_{\star}$$^{b}$               & $M_{\odot}$           & (2.8$^{+0.2}_{-0.1}$) $\times$ 10$^{10}$      \\
        $L_{\rm IR}$$^{c}$              & $L_{\odot}$           &  (1.6  $\pm$  0.3) $\times$ 10$^{12}$  \\
        $A_{\rm V}$$^{b}$         & $mag$               &  0.9$^{+0.3}_{-0.0}$ \\
        $M_{\rm dust}$$^{c}$ & $M_{\odot}$ &(1.2  $\pm$  0.2) $\times$ 10$^{8}$\\
        T$_{\rm dust}$$^{d}$            & K                     &       45.5$^{+2.1}_{-2.1}$                    \\
        
        \hline
        \multicolumn{3}{l}{} \\
        \multicolumn{3}{l}{Median stacked photometry}\\
        \hline
        H & mag & 27.4  $\pm$  0.1   \\
        $[4.5]$ & mag & 23.92  $\pm$  0.04 \\
        $S_{\rm 1.13 mm}$ & $\mu$Jy & 334 $\pm$ 24 \\
        $S_{\rm 3GHz}$          &$\mu$Jy                        & 4.1  $\pm$  0.7                            \\
        \hline
        
        \multicolumn{3}{l}{} \\
        \multicolumn{3}{l}{Derived quantities$^{**}$}\\
        \hline
        L$_{\rm 1.4 GHz}$$^{e}$  & erg s$^{-1}$ Hz$^{-1}$& (9.53  $\pm$  1.53) $\times$ 10$^{30}$        \\
        $q_{\rm TIR}$$^{f}$             &                       &        2.23  $\pm$  0.03                       \\
        SFR$_{\rm rad,med}$$^{g}$       & $M_{\odot}$ yr$^{-1}$ & 287.59  $\pm$  53.86 \\
        SFR$_{\rm IR,med}$$^{h}$        & $M_{\odot}$ yr$^{-1}$ & 235.33$_{-49.77}^{+47.37}$                                  \\
        SFR$_{\rm UV,med}$$^{i}$        & $M_{\odot}$ yr$^{-1}$ & 0.33  $\pm$  0.02                            \\
        $\Delta$MS$^{j}$ & &1.45$^{+0.29}_{-0.31}$ \\
         $M_{\rm gas}$$^{k}$ & $M_{\odot}$ & (2.6  $\pm$  0.4) $\times$ 10$^{10}$   \\
        $f_{\rm gas}$$^{l}$ &  & 0.48  $\pm$  0.05  \\
        $\tau_{\rm dep}$$^{m}$ & Myr & 110$^{+29}_{-30}$        \\

        \hline \hline
\end{tabular}
\begin{tablenotes}
\item \textbf{Note:} $^{*}$Uncertainties are the 16-84th percentile ranges of the probability distribution function given by the SED fitting. $^{**}$Uncertainties on derived quantities were calculated from the propagation of the errors in the parameter values. $^{a}$Photometric redshift, determined with the code \texttt{EAzY}. $^{b}$$M_{\star}$ and $A_{\rm V}$, derived from the UV to MIR SED fitting with the code \texttt{FAST++}. $^{c}$Given by IR SED fitting with \texttt{CIGALE}. $^{d}$Measured by MBB model fit (see $\S$\ref{Sect::Tdust}). $^{e}$Derived from $S_{\rm 3 GHz}$ assuming a radio spectral index $\alpha= -0.75$ (see Eq.~\ref{L1.4}). $^{f}$Calculated from the IR-radio correlation (see Eq.~\ref{qTIR}). $^{g}$Calculated following \cite{Delhaize2017} (see Eq.~\ref{SFRrad}), which was simply estimated from the radio emission without correction for AGN. $^{h}$Derived following \cite{Kennicutt2012} (see Eq.~\ref{eq:SFRtot}). $^{i}$Derived following \cite{Daddi2004}, scaled to a \cite{Chabrier:2003} IMF (see Eq.~\ref{eq:SFRtot}). $^{j}$Distance to the SFMS: $\Delta$MS = SFR/SFR$_{\rm MS}$, where SFR$_{\rm MS}$ is the average SFR of MS galaxies at fixed stellar mass and redshift \citep[][see Fig.~\ref{Fig:gas} and $\S$\ref{figure_gas}]{Schreiber2015}.  
$^{k}M_{\rm gas}$, computed based on gas-to-dust ratio. $^{l}$Gas fraction: $f_{\rm gas} = M_{\mathrm{gas}}/(M_{\star} + M_{\mathrm{gas}})$. $^{m}$Gas depletion time: $\tau_{\rm dep} = M_{\mathrm{gas}}/(\rm{SFR}_{\rm IR}+\rm{SFR}_{\rm UV})$, which is the inverse of the star formation efficiency (SFE $= 1/\tau_{\rm dep}$).
\end{tablenotes}
\label{table:stacked} 
\end{threeparttable} 
\end{table}


\begin{figure*}
\includegraphics[scale=0.46]{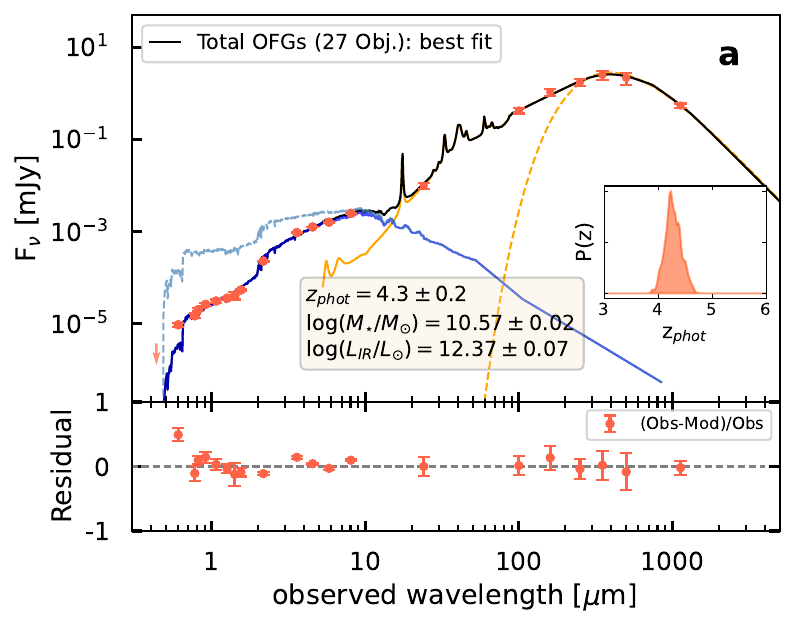}
\includegraphics[scale=0.46]{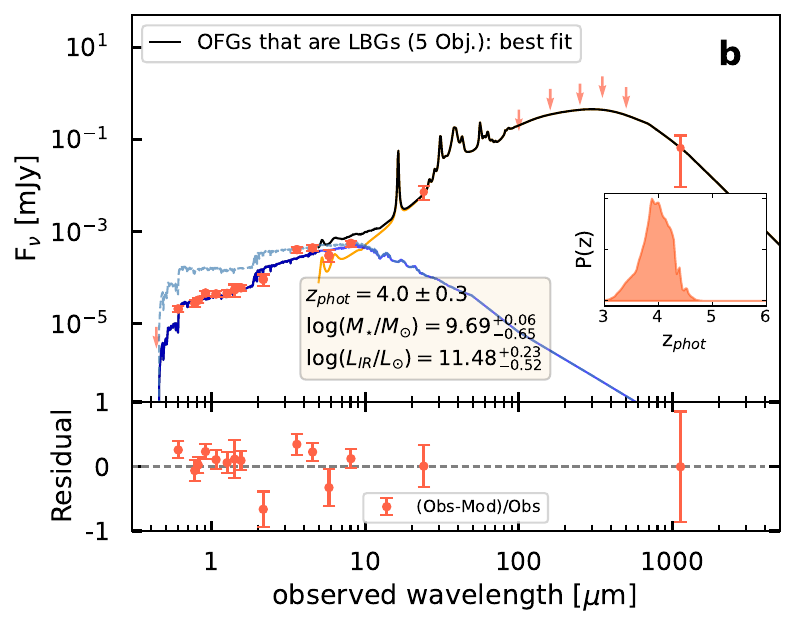}
\includegraphics[scale=0.46]{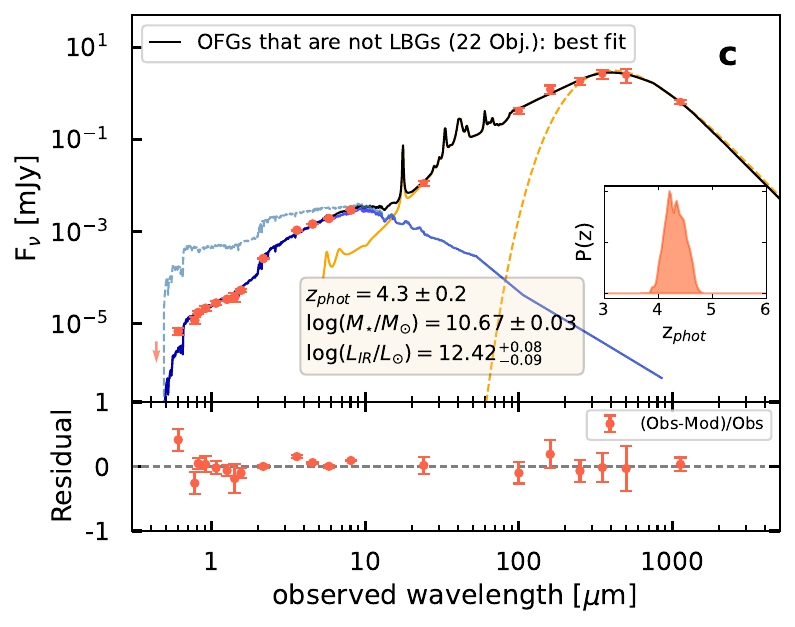}
\hspace*{3cm}\includegraphics[scale=0.46]{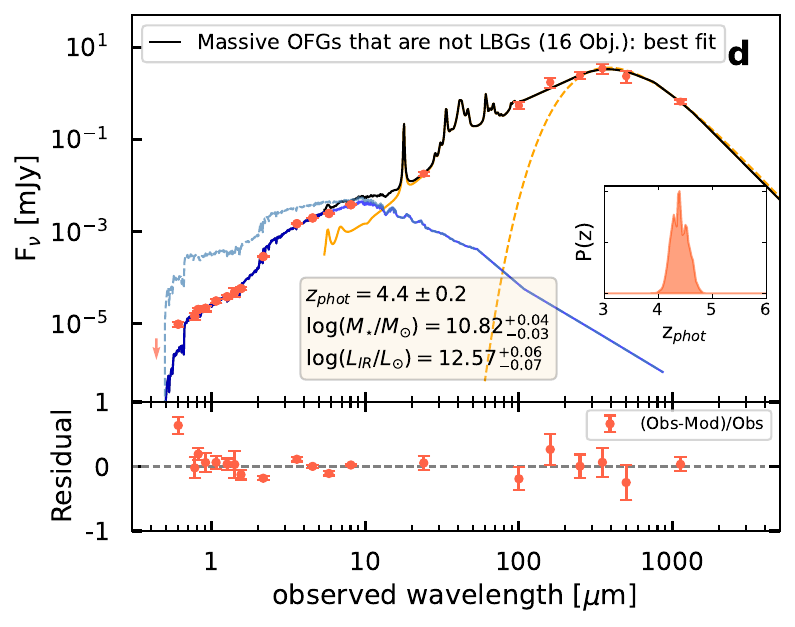}
\includegraphics[scale=0.46]{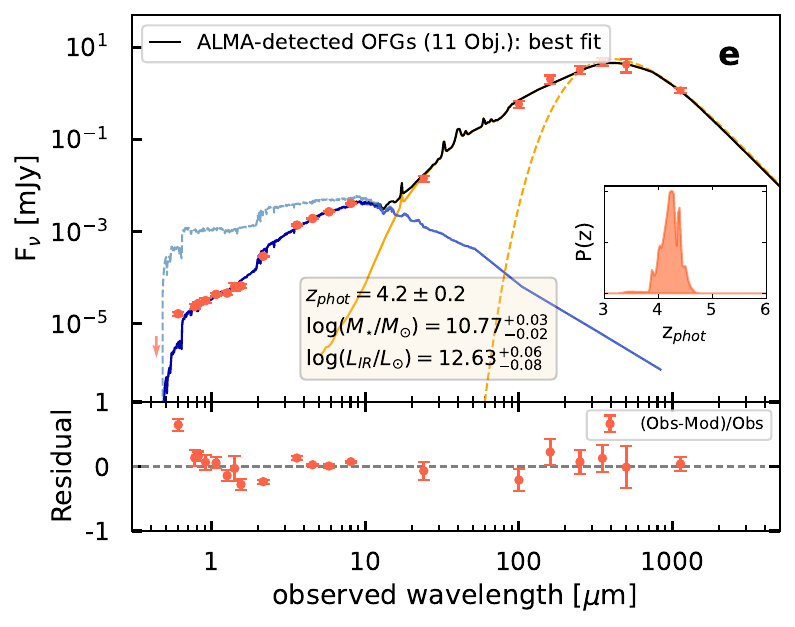}
\caption{Mean stacked SEDs for the total sample and four sub-samples of our OFGs. $Panels$ $a$: total sample of 27 OFGs; $b$, $c$,  $d$, and $e$: sub-samples of OFGs that are LBGs, OFGs that are not LBGs, massive OFGs (log($M_{\star}$/$M_{\odot}$) > 10.3) that are not LBGs, and OFGs with ALMA detections, respectively. We use the sub-samples in $panel$ $c$ and $panel$ $d$ to calculate the SFRD. 
$Top$:  Best-fit SED of the stacked OFGs (black line), which is composed of the uncorrected dust-attenuated stellar component (dark blue line) and the IR dust contribution (orange solid line). The corrected UV emission is shown as a faint blue line. We also plot the best fit of a MBB (orange dashed line) with a fixed dust emissivity index $\beta$ = 1.5. We note that there is no significant AGN contribution in our sample, except for the five OFGs that are LBGs displaying the presence of radio AGNs (see $\S$\ref{Sect::agn}).  The measured fluxes (red points) are derived from the stacked images. Error bars (1$\sigma$) and upper limits (3$\sigma$) are obtained from the Monte Carlo simulation (except $Herschel$) and bootstrap approach ($Herschel$; see $\S$\ref{Sect::stacked SED}). The inset shows the likelihood distribution of the photometric redshift based on the UV to MIR SED fitting from \texttt{EAzY} (see $\S$\ref{Sect::redshift}), which is normalized to the peak value. $Bottom$: Residuals calculated from (observation $-$ model)/observation. The SEDs for individual OFGs are given in the appendix (see Figs.~\ref{Fig:SED1} and \ref{Fig:SED2}).
\label{Fig:meanstack}}
\end{figure*}

\subsection{Fitting of the stacked SEDs}\label{Sect::stacked SED}
We obtained the stacked full-wavelength SEDs in the same way as for the individual galaxies (see $\S$\ref{Sect::redshift}, $\S$\ref{Sect::LIR}, Table \ref{table:fast}, \ref{table:cigale}). In brief, first, we fitted the broad photometry at OPT to MIR with the \texttt{EAzY} code to obtain photometric redshifts. Then, we independently performed the OPT to MIR SED fitting with \texttt{FAST++} and the MIR to mm SED fitting with \texttt{CIGALE}, respectively, at the previously obtained redshifts of the stacked sources. This approach helps to 1) disentangle the degeneracy between redshift and other parameters, such as stellar age and dust temperature; and 2) break the energy balance principle (the total energy emitted in the MIR and FIR is determined by the attenuation of observed starlight in the UV and optical) 
used in \texttt{CIGALE}. For dusty star-forming galaxies, especially for $H$-dropouts and $K_{\rm s}$-dropouts with strong dust obscuration, there could exist regions with strong UV extinction due to strong dust obscuration, which may not participate in the UV to optical part, but emit FIR light \citep[e.g.,][]{Simpson2015, Carlos2018, Elbaz2018}. Assuming an energy balance with a fixed redshift will lead to an underestimation of the $L_{\rm IR}$, hence, the SFR.

The best-fit SED is shown in Fig.~\ref{Fig:stacked}. The median redshift for the total sample is $z_{\rm med, stacked}$ = 4.5  $\pm$  0.2, which is consistent with $z_{\rm med}=4.1$ derived from the median value of individual OFGs with a wide distribution (see Fig.~\ref{Fig:distribution}, top). In addition, the median stacked SED peak (and the mean stacked SED peak; see Fig.~\ref{Fig:meanstack}) is between 350 and 500\,$\mu$m, also in agreement with being at $z \sim 4$. Thus, these agreements confirm that the bulk population of OFGs consists of dusty star-forming galaxies at $z \sim 4-5$. Remarkably, most fluxes in the stacked images are above the 3$\sigma$ confidence level, especially in the $H$, $K_{\rm s}$, and IRAC bands, helping to establish the position of the Balmer and 4000$\AA$ breaks very well, hence determining a robust redshift. The median properties derived from the stacked SED for the total sample are summarized in Table \ref{table:stacked}.

In addition, to further investigate the characteristics of different subpopulations of our OFGs, we performed median and mean stacked SED fitting for four sub-samples of OFGs.  The four subsamples and the purpose of our investigation are listed below.
 \begin{enumerate}[{1)}]
      \item OFGs that are LBGs: in our sample, five OFGs are classified as LBGs. Given that the traditional approach to estimating the cosmic SFRD at $z>3$ is mainly based on the LBGs \citep[e.g.,][]{Madau2014,Bouwens2020}, studying the differences in the properties of LBGs and  OFGs (after removing 5 LBGs) in our sample can help us understand the importance of OFGs in the cosmic SFRD.
      \item OFGs that are not LBGs: 22 OFGs are not classified as LBGs. To quantity the level of underestimation of the cosmic SFRD at $z>3$ \citep{Madau2014} based mainly on LBGs, we used the mean SFR of this sub-sample to calculate the SFRD to avoid contamination by LBGs (describe later in $\S$\ref{Sect::SFRD}).
      \item Massive OFGs that are not LBGs: there are 16 OFGs not classified as LBGs with log($M_{\star}$/$M_{\odot}$) > 10.3. To compare with the results of \cite{Wang2019} on the SFRD, here we used the same stellar mass cut for this sub-sample.
      \item ALMA-detected OFGs: 11 OFGs in our sample are detected by ALMA at 1.13 mm \citep[>3.5 $\sigma$;][]{Carlos2021}.  Since using ALMA detections to select OF sources is a very efficient method \citep[e.g.,][]{Franco2020a,Zhou2020,Carlos2021}, we study the properties of the ALMA-detected OFGs and compare them with other sub-samples to understand whether there is a selection bias using this approach.
 \end{enumerate}

\begin{table*}
\caption{Mean physical properties of sub-samples of OFGs.}
\tiny
\centering
\begin{threeparttable} 
 
\begin{tabular}{llllll}     
\hline\hline
Parameter               &       Unit            &OFGs that are LBGs & OFGs that are not LBGs               &  Massive OFGs that are not LBGs       & ALMA-detected OFGs      \\
&&  (5 Obj.) & (22 Obj.)&(16 Obj.) &(11 Obj.)   \\
\hline
        \multicolumn{4}{l}{} \\
        \multicolumn{4}{l}{Derived from SED fitting with mean stacked photometry}\\
        \hline
        $z_{\rm phot}$          & &              4.0  $\pm$  0.3        & 4.3  $\pm$  0.2                 &  4.4  $\pm$  0.2  & 4.2  $\pm$  0.2           \\
        $M_{\star}$             & $M_{\odot}$   &       (4.9$^{+0.7}_{-3.8}$) $\times$ 10$^{9}$ & (4.7  $\pm$  0.3) $\times$ 10$^{10}$ &  (6.6$^{+0.6}_{-0.4}$) $\times$ 10$^{10}$ &  (5.9$^{+0.4}_{-0.3}$) $\times$ 10$^{10}$  \\
        $L_{\rm IR}$            & $L_{\odot}$   & (2.9  $\pm$  2.1) $\times$ 10$^{11}$       & ($2.6^{+0.6}_{-0.5}$) $\times$ 10$^{12}$ & (3.7  $\pm$  0.6) $\times$ 10$^{12}$ & ($4.3^{+0.6}_{-0.7}$) $\times$ 10$^{12}$     \\
        $A_{\rm V}$        & $mag$              &0.7$^{+0.4}_{-0.3}$ &  1.4$^{+0.1}_{-0.3}$ & 1.2$^{+0.3}_{-0.1}$ & 1.7$^{+0.1}_{-0.2}$ \\
         $M_{\rm dust}$ & $M_{\odot}$ &(0.3  $\pm$  1.3) $\times$ 10$^{8}$&(2.9  $\pm$  0.7) $\times$ 10$^{8}$ &(2.8  $\pm$  0.6) $\times$ 10$^{8}$ &(5.3  $\pm$  1.1) $\times$ 10$^{8}$\\
        T$_{\rm dust}$          & K             & $(...)$       &  42.3$^{+1.2}_{-1.2}$  & 45.5$^{+1.4}_{-1.5}$ & 41.5$^{+1.2}_{-1.2}$   \\
        \hline
        \multicolumn{4}{l}{} \\
        \multicolumn{4}{l}{Mean stacked photometry}\\
        \hline
        H & mag &26.95  $\pm$  0.17& 27.09  $\pm$  0.08 & 26.99  $\pm$  0.08 & 26.85  $\pm$  0.10   \\
        $[4.5]$ & mag &24.74  $\pm$  0.31& 23.34  $\pm$  0.03 & 23.33  $\pm$  0.02 & 22.99  $\pm$  0.03 \\
      $S_{\rm 1.13 mm}$ & $\mu$Jy & 65 $\pm$ 56 & 650 $\pm$ 26 & 660 $\pm$ 30 & 1153 $\pm$ 40 \\
        $S_{\rm 3GHz}$          &$\mu$Jy        &       2.7  $\pm$  0.5 & 6.5  $\pm$  0.5 & 6.9  $\pm$  0.5 & 10.6  $\pm$  0.8                            \\
        \hline
        
        \multicolumn{4}{l}{} \\
        \multicolumn{4}{l}{Derived quantities}\\
        \hline
        L$_{\rm 1.4 GHz}$  & erg s$^{-1}$ Hz$^{-1}$& (4.81  $\pm$  0.92) $\times$ 10$^{30}$& (1.40  $\pm$  0.11) $\times$ 10$^{31}$ & (1.54  $\pm$  0.11) $\times$ 10$^{31}$  & (2.18  $\pm$  0.17) $\times$ 10$^{31}$ \\
        $q_{\rm TIR}$           &       &1.80$^{+0.15}_{-0.43}$ &2.28  $\pm$  0.03   &2.39  $\pm$  0.02 &    2.30  $\pm$  0.03               \\
        SFR$_{\rm rad,avg}$     & $M_{\odot}$ yr$^{-1}$ &189.95  $\pm$  58.40& 392.17  $\pm$  49.02 & 408.91  $\pm$  49.54 & 590.15  $\pm$  74.07 \\
        SFR$_{\rm IR,avg}$      & $M_{\odot}$ yr$^{-1}$ &44.52  $\pm$  33.08&         387.93$_{-72.08}^{+82.03}$&     556.32$_{-85.59}^{+84.94}$  &   635.15$_{-104.83}^{+96.61}$                      \\
        SFR$_{\rm UV,avg}$      & $M_{\odot}$ yr$^{-1}$ &0.63  $\pm$  0.07&  0.37  $\pm$  0.03&  0.45  $\pm$  0.03 &  0.59  $\pm$  0.04                      \\
        $\Delta$MS & &1.81  $\pm$  1.32&1.46$^{+0.31}_{-0.27}$ & 1.45  $\pm$  0.22 & 1.96$^{+0.30}_{-0.32}$  \\
        $M_{\rm gas}$ & $M_{\odot}$ & (1.3  $\pm$  5.7) $\times$ 10$^{10}$ & (5.0  $\pm$  1.2) $\times$ 10$^{10}$ & (4.4  $\pm$  0.9) $\times$ 10$^{10}$ & (8.4  $\pm$  1.8) $\times$ 10$^{10}$   \\
        $f_{\rm gas}$ &  & 0.73  $\pm$  0.87 & 0.52  $\pm$  0.06 & 0.40  $\pm$  0.06 & 0.59  $\pm$  0.05  \\
        $\tau_{\rm dep}$ & Myr & 291  $\pm$  1277 & 130$_{-40}^{+42}$ & 79  $\pm$  21 & 133$_{-35}^{+34}$\\
        $R_{\rm e(1.13 mm)}$ & kpc & $(...)$ & 1.09  $\pm$  0.05 & 1.05  $\pm$  0.06 & 0.80  $\pm$  0.03 \\
        ${\Sigma_{\rm SFR}}$$^{a}$ & $M_{\odot}$ yr$^{-1}$ kpc$^{-2}$ & $(...)$ & 52  $\pm$  11 & 80  $\pm$  15 & 158 $\pm$ 28  \\ 
        \hline
        
        \multicolumn{4}{l}{} \\
        \multicolumn{4}{l}{Cosmic SFRD$^{*}$}\\
        \hline
        V$^{**}$ &  Mpc$^3$ &$(...)$& 7.4 $\times$ 10$^{5}$ & 7.4 $\times$ 10$^{5}$ & 7.4 $\times$ 10$^{5}$\\
        SFRD & $M_{\odot}$ yr$^{-1}$Mpc$^{-3}$ &$(...)$& ($1.2\pm0.2$) $\times$ 10$^{-2}$ & ($1.2\pm0.2$) $\times$ 10$^{-2}$  & ($0.9\pm0.2$) $\times$ 10$^{-2}$\\
        \hline \hline
\end{tabular}
\begin{tablenotes}
\item \textbf{Note:} Same as Table \ref{table:stacked} but for the four sub-samples of OFGs. 
 $^{a}$SFR surface density: ${\Sigma_{\rm SFR}} = 0.5{\rm SFR}_{\rm tot}/(\pi R_{\rm e(1.13 mm)}^2)$, where $\rm{SFR}_{\rm tot} = \rm{SFR}_{\rm IR,avg}+\rm{SFR}_{\rm UV,avg}$.
$^{*}$The SFRD is discussed with details in $\S$\ref{SFRD}. $^{**}$Survey volume, calculated using Eq.~\ref{V} with a broad redshift range of $z=3.2-7.0$.
\\      
\end{tablenotes}
\label{table:mean} 
\end{threeparttable} 
\end{table*}

       
The best-fit mean SEDs of the total sample and the four sub-samples are shown in Fig.~\ref{Fig:meanstack}. For the OFGs that are LBGs (Fig.~\ref{Fig:meanstack}-$b$), there is a 3.3$\sigma$ detection at 24\,$\mu$m, no detection in all the \textit{Herschel} bands, and a 1.2$\sigma$ detection at 1.13 mm. To successfully perform the FIR SED fitting, we fit the fluxes at 24\,$\mu$m and 1.13 mm. We then compared the best-fit model with the 3$\sigma$ upper limits in the \textit{Herschel} bands (red arrows in Fig.~\ref{Fig:meanstack}-$b$). The best-fit model is below the red arrows, showing a good consistency with the \textit{Herschel} no detections. The mean properties derived from the mean stacked SEDs for the sub-samples are summarized in Table \ref{table:mean}.

For the mean stacked SED fit, one hypothesis here is that all OFGs have a similar SED shape. This is because the stacked SED we used to derive the SFR is a flux-weighted average in each band and if the brightest galaxy has a different SED shape, then the fitting results will be biased towards the properties of the brightest galaxy. For example, it has been shown that the dust temperature derived from the mean stacked SED is biased by $1.5$ K to a higher temperature than the true value since the starburst galaxies in the sample are warmer and brighter \citep{Schreiber2018c}. However, we do not yet know the true IR SED shapes of most OFGs in our sample because of the lack of the \textit{Herschel} detections. It is also unclear whether the brightest OFGs have a different SED shape compared to the remaining OFGs, therefore causing a bias. Hence, we cannot correct this potential bias here. Instead, we performed a median stacked SED fitting as a comparison. Although the median stacked SED fitting exhibits a lower confidence level (because it is less influenced by the brightest sources) compared to the mean one, their properties are more robust against outliers and are representative of the vast majority of galaxies in the sample. 
On the other hand, and most importantly, we need the SFR derived from the mean stacked SED fitting to calculate the cosmic SFRD (described later in $\S$\ref{Sect::SFRD}).


\subsection{SFRs and AGN}
\subsubsection{SFRs}\label{Sect::stacked_sfr}
We obtained the SFR$_{\rm tot}$ of the stacked optically dark/faint (sub)samples using the same method as for individual galaxies (see $\S$\ref{Sect::sfr}), following Eq.~\ref{eq:SFRtot}. With the 3 GHz VLA observations in the GOODS-South, we can also calculate the radio-based SFR (SFR$_{\rm rad}$; assuming a \citealt{Chabrier:2003} IMF) following \cite{Delhaize2017}:
\begin{flalign}
\rm{SFR}_{\rm rad}\ [M_{\rm \odot}\,yr^{-1}]=\,10^{-24} \times 10^{q_{\rm TIR}(z)}\,L_{\rm 1.4GHz}\ [W Hz^{-1}],
\label{SFRrad}
\end{flalign}
\noindent where $L_{\rm 1.4GHz}$ is the rest-frame 1.4 GHz luminosity converted from the 3 GHz flux density ($S_{\rm 3GHz}$ at observed-frame; $W m^{-2} Hz^{-1}$) using:
\begin{flalign}
L_{\rm 1.4GHz}\ [W Hz^{-1}]=\frac{4\pi D_L^2}{(1+z)^{\alpha + 1}}\, \left(\frac{1.4\ [GHz]}{3\ [GHz]}\right)^{\alpha}\,S_{\rm 3GHz},
\label{L1.4}
\end{flalign}
\noindent  here, the radio spectral index $\alpha^{\rm 3GHz}_{\rm 1.4GHz}$ is assumed to be $\alpha=-0.75$. 
The $q_{\rm TIR}(z)$ in Eq.~\ref{SFRrad} is the IR-to-radio luminosity ratio, which was recently found to evolve primarily with the stellar mass and depend secondarily on the redshift \citep{Delvecchio2021}:
\begin{flalign}
q_{\rm TIR}(M_\star,z)& = (2.646 \pm 0.024) \times (1+z)^{-0.023 \pm 0.008}&& \label{qTIR(mz)}\\
&-(0.148 \pm 0.013) \times (\rm{log}\,M_\star/M_\odot - 10). \nonumber
\end{flalign}

The derived IR-based and radio-based SFR values (SFR$_{\rm IR}$ and SFR$_{\rm rad}$) are in good agreement (except for the sub-sample of OFGs that are LBGs), as shown in Tables \ref{table:stacked} and \ref{table:mean}. For the sub-sample of OFGs that are LBGs, the SFR$_{\rm rad}$ is about four times higher than the SFR$_{\rm IR}$, although with large uncertainty, hinting at the existence of radio AGNs in the sub-sample of OFGs that are LBGs. 
The median SFRs for our total 27 OFGs are given in Table \ref{table:stacked}, while the mean SFRs for our OF sub-samples are summarized in Table \ref{table:mean}. For our entire sample, the median contribution from SFR$_{\rm UV}$ to SFR$_{\rm tot}$ is only 0.1$\%$, which is negligible.

\subsubsection{AGN}\label{Sect::agn}
As our selection criterion has been designed to avoid selecting passive galaxies, the OFGs in our sample are mainly dusty star-forming galaxies. Studying the presence of AGN in our sample can help us understand the co-evolution between AGN and star formation activity in the early Universe. It is also crucial for ensuring that our calculations of the SFR and, eventually, the cosmic SFRD are correct (uncontaminated by AGN).
Here, we examine our sample for AGN contributions using three different methods, that is, studying their IR, radio, and X-ray excesses.

First, we fit the stacked IR SED with an additional AGN template using \texttt{CIGALE} \citep[][see $\S$\ref{Sect::stacked SED} for the infrared SED fitting]{Fritz2006}. The contribution of the IR-bright AGN to the total IR luminosity ($f_{\rm AGN}$) can lead to an overestimation of the dust IR emission and thus of the total SFR. We found that the SED fitting yields a $f_{\rm AGN} < 0.01$, indicating the absence of IR-bright AGN in our sample.

Secondly, the $q_{\rm TIR}$ is defined as the IR-to-radio luminosity ratio \citep[e.g.,][]{Helou1985,Yun2001}:
\begin{flalign}
q_{\rm TIR}\equiv\, \rm{log}\, \left(\frac{L_{\rm IR}\,[W]}{3.75\times10^{12}\,[Hz]}\right) - \rm{log}\, \left(L_{\rm 1.4GHz}\,[W Hz^{-1}]\right).
\label{qTIR}
\end{flalign}
\noindent The derived $q_{\rm TIR}$ from the IR-radio correlation (Eq.~\ref{qTIR}) is presented in Tables \ref{table:stacked} and \ref{table:mean}. Except for the sub-sample of OFGs that are LBGs, $q_{\rm TIR}$ values of the OFGs are consistent with those in \cite{Delvecchio2021} for star-forming galaxies at the same redshift and stellar mass. 
These agreements suggest that for the OFGs that are not LBGs, there is a lack of strong AGN activity in the radio band. 
On the other hand, the sub-sample of OFGs that are LBGs present a mean $q_{\rm TIR}$ = 1.8, much smaller than the typical $q_{\rm TIR}$ of 2.6 for star-forming galaxies at the same redshift and stellar mass, and would thus be classified as radio AGNs \cite[Fig. 12 in][]{Delvecchio2021}. 

We also searched for X-ray-bright AGN in the CDF-S 7 Ms catalog \citep{Luo2017}. Among 1008 sources in the main catalog and 47 lower-significance sources in the supplementary catalog, we did not find any X-ray counterpart for the individual OFGs within a 0.6$^{\prime\prime}$ radius. 
None of the sources in our catalog exhibit a total X-ray luminosity integrated over the entire 0.5-7 keV range larger than $L_{X} = 10^{42.5} erg$ $s^{-1}$ \citep[AGN definition in][]{Luo2017}. Hence, we find no evidence for any bright X-ray AGN in our catalog. We also performed mean and median stacking for the 27 OFGs in 0.5-7 keV images and did not find any significant detections ($\ll 3\sigma$) in either of the stacked images.

In addition, we considered the MIR-AGN selection criterion developed by \cite{Donley2012} to diagnose the presence of a power-law AGN based on IRAC colors. However, this criterion does not apply to our high-$z$ OFGs. At $z>3$, the IRAC bands mainly collect emission from stars below 2 $\mu$m in the rest frame, outside the typical domain where power-law AGNs contribute. In summary, we do not find evidence for significant contamination by AGNs in our OFG sample, except for the five OFGs that are LBGs displaying the presence of radio AGNs. 

\begin{figure}
\centering
\includegraphics[scale=0.37]{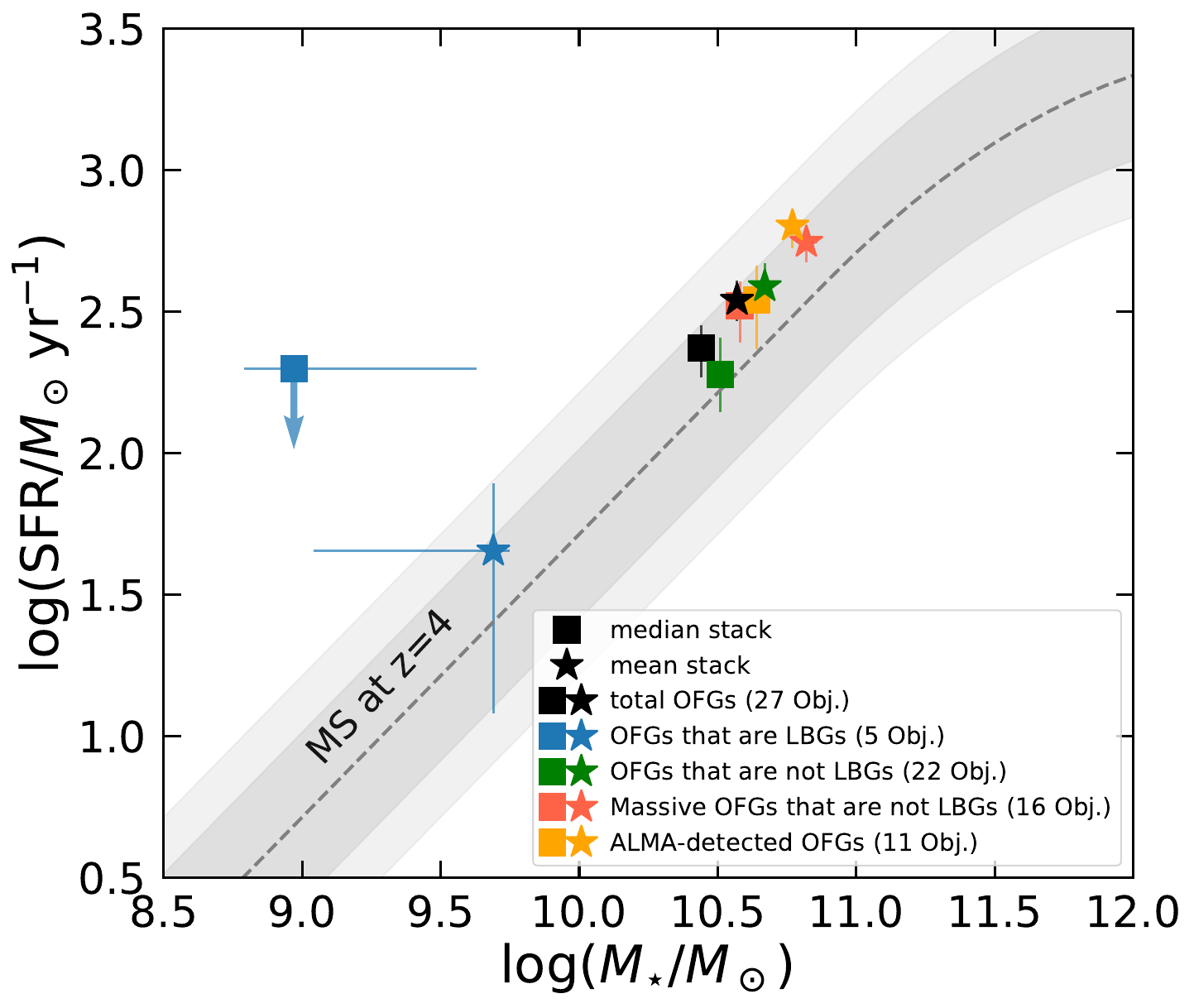}
\caption{Locations of OFGs compared to the SFMS in the SFR-$M_{\star}$ plane. The SFMS at $z=4$ \citep{Schreiber2015}, 1$\sigma$ scatter (0.5 < $\Delta$MS < 2, i.e., $\sim$0.3 dex), and $\pm 3 \times \Delta$MS region (0.33 < $\Delta$MS < 3, i.e., $\sim0.5$ dex) are highlighted with a grey dashed line, a grey shaded area, and a light grey shaded area, respectively. $\Delta$MS > 3 is commonly used to separate MS and SB galaxies. 
 Squares and stars respectively represent the median and mean stacking results for our total sample (black) and four sub-samples.  The four sub-samples are the OFGs that are LBGs (blue), the OFGs that are not LBGs(green), the massive OFGs that are not LBGs(red), and the OFGs with ALMA detections (orange). 
 When necessary, data from the literature have been converted to a \cite{Chabrier:2003} IMF.
\label{Fig:MS}}
\end{figure}

\subsection{The main sequence of star-forming galaxies and the properties of gas and dust for the stacked OFGs}\label{figure_gas}

\begin{figure*}
\centering
\includegraphics[scale=0.227]{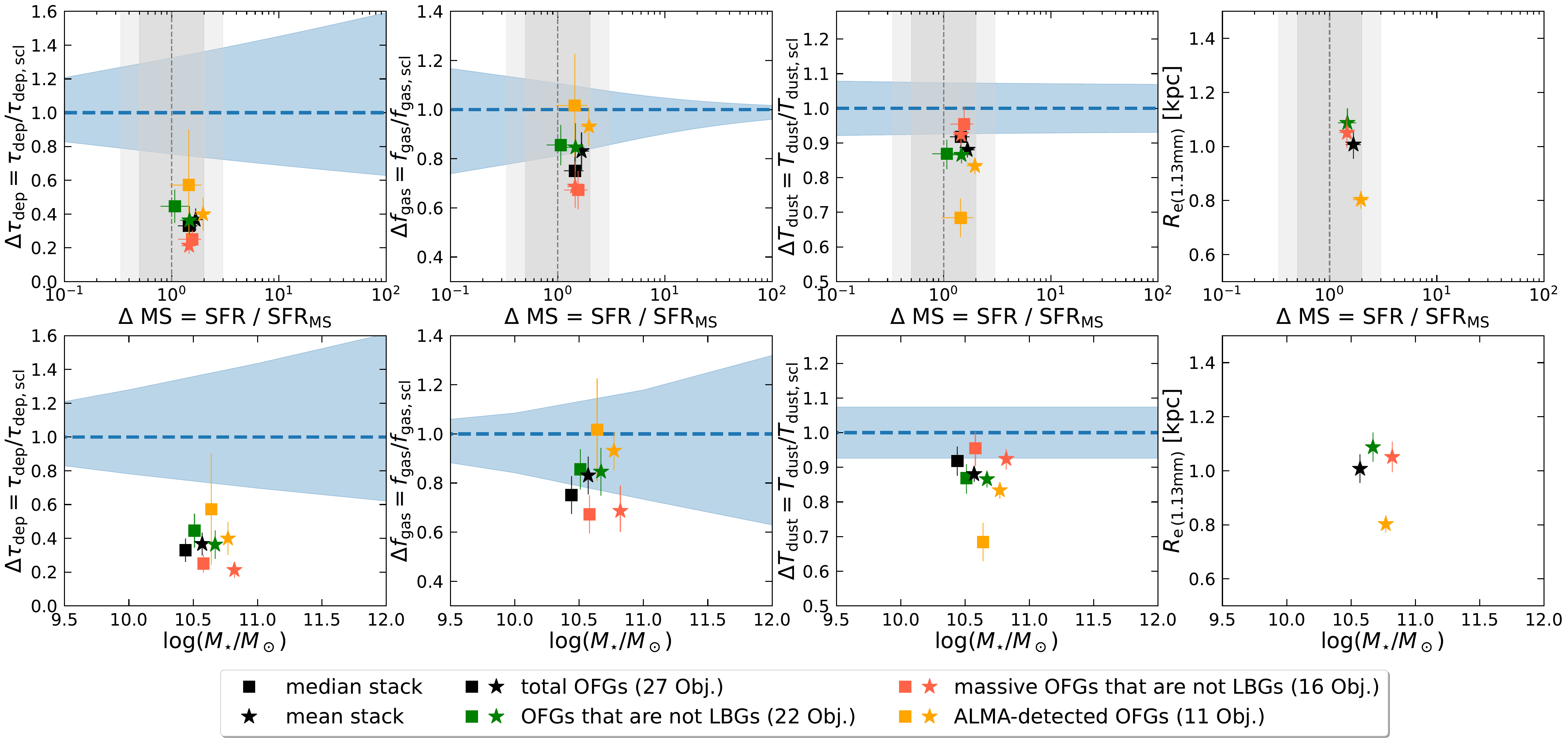}
\caption{Normalized $\tau_{\rm dep}$ (left), $f_{\rm gas}$ (mid-left), and $T_{\rm dust}$ (mid-right) by the scaling relation of $\tau_{\rm dep,scl}(z, M_{\star}, \Delta$MS) and $f_{\rm gas,scl}(z, M_{\star}, \Delta$MS) from \cite{Tacconi2018} and of $T_{\rm dust, scl}(z, \Delta$MS) from \cite{Schreiber2018c} as well as the half-light radius $R_{\rm e}$ (right) at ALMA 1.13 mm as a function of $\Delta$MS (first row) and $M_{\star}$ (second row). Here, $\Delta$MS is the distance to the SFMS \citep{Schreiber2015}, $\Delta$MS = SFR/SFR$_{\rm MS}$, at its own stellar mass and redshift for each sample. 
The blue dashed line and shaded area show the scaling relation and scatter. For each data point, the $\tau_{\rm dep,scl}$, $f_{\rm gas,scl}$, and $T_{\rm dust,scl}$ are calculated at a fixed redshift, stellar mass, and $\Delta$MS.
Squares, stars, and shaded grey areas are the same as in Fig.~\ref{Fig:MS}. 
\label{Fig:gas}}
\end{figure*}

In this section, we investigate the properties of the stacked samples derived from the SED fitting. We examine their locations in the star-formation main sequence, their gas depletion timescales, gas fractions, and dust temperatures in the framework of the scaling relations for galaxy evolution, and their dust sizes.

In Fig.~\ref{Fig:MS}, we place the stacked total OFG sample and the four sub-samples in the SFR-$M_{\star}$ plane, showing the locations compared to the SFMS. In the SFR-$M_{\star}$ plane, it is well known that the SFMS as a whole moves to higher SFRs with increasing redshift \citep[e.g.,][]{Elbaz2007,Elbaz2011,Noeske2007, Magdis2010, Whitaker2012,Whitaker2014,Speagle2014,Schreiber2015,Lee2015,Leslie2020}. We adopted a fixed $z=4$ for the SFMS \citep{Schreiber2015} as a comparison since it is close to $z_{\rm med}=4.1$ from the median value of the individual OFGs and $z_{\rm med,stacked}=4.5$ from the median stacked total OFGs. This figure shows that all the (sub-)samples are located within the SFMS region ($\sim$0.5 dex) at $z=4$, and most of them 
lie within the 1$\sigma$ scatter of the SFMS (0.5 < $\Delta$MS < 2, i.e., $\sim$0.3 dex), consistent with being normal star-forming galaxies at the same redshift. It suggests that unlike studies limited to a rare population of extreme starburst galaxies \citep[e.g.,][]{Riechers2013,Strandet2017,Marrone2018,Dudzeviciute2020,Riechers2020}, our OFGs represent a less extreme population of dusty star-forming galaxies at $z > 3$.

Furthermore, we study the gas and dust properties of the OFGs by focusing on their gas depletion timescales, gas fractions, dust temperatures, and dust sizes. In Fig.~\ref{Fig:gas}, we show the normalized $\tau_{\rm dep}$, $f_{\rm gas}$, and $T_{\rm dust}$ by scaling them to the observed relation (scl; which is the median of the MS) of $\tau_{\rm dep,scl}(z, M_{\star}, \Delta$MS) and $f_{\rm gas,scl}(z, M_{\star}, \Delta$MS) from \cite{Tacconi2018} and of $T_{\rm dust,scl}(z, \Delta$MS) from \cite{Schreiber2018c} as a function of $\Delta$MS and $M_{\star}$. The $\Delta$MS is the SFR of each stacked (sub)sample normalized by the SFR of the SFMS \citep[$\Delta$MS = SFR/SFR$_{\rm MS}$;][]{Schreiber2015} at its own redshift and stellar mass. The $\tau_{\rm dep,scl}$, $f_{\rm gas,scl}$, and $T_{\rm dust,scl}$ are calculated for each data point at fixed redshift, stellar mass, and $\Delta$MS. We also present dust continuum sizes at 1.13 mm ($R_{\rm e(1.13 mm)}$) of the mean stacked optically dark/faint (sub)samples as a function of $\Delta$MS and $M_{\star}$. Here, the half-light radius $R_{\rm e(1.13 mm)}$ was measured in the $uv$  plane by fitting a circular Gaussian (task $uvmodelfit$) after performing $uv$ plane stacking according to the method described by \cite{Carlos2021}. We note that we did not scale $R_{\rm e(1.13 mm)}$ to the observed relations \citep[e.g.,][]{van2014} because the redshifts ($z>3$) of our OFGs exceed the limits of these relations.

In Fig.~\ref{Fig:gas}, there is no global offset between mean and median results of normalized $\tau_{\rm dep}$, $f_{\rm gas}$, and $T_{\rm dust}$ for the stacked (sub)samples. It indicates no significant differences in the SED shapes of the brightest OFGs compared to the remaining ones, which would otherwise cause a strong bias (as discussed in $\S$\ref{Sect::stacked SED}) and further show a global offset even for the different stacked samples. The mean SFR is larger than the median SFR (see Fig.~\ref{Fig:MS}), which is expected since the former is influenced by the brightest sources in the flux-weighted average in each band (as discussed in $\S$\ref{Sect::stacked SED}).

In the first and second columns of Fig.~\ref{Fig:gas}, the stacked (sub)samples show $\tau_{\rm dep}$ values below the scatter of the scaling relation, while $f_{\rm gas}$ is at the lower boundary of the scaling relation. That is to say, the OFGs have shorter $\tau_{\rm dep}$ and slightly lower $f_{\rm gas}$ values compared to normal star-forming galaxies. This indicates that galaxies with stronger dust obscuration tend to have lower gas fractions and shorter gas depletion times. Their gas is consumed more rapidly, hence, they form their stars with a high efficiency, which sets them in the so-called class of starbursts in the main sequence \citep{Elbaz2018,Carlos2022}.

Among all the stacked (sub)samples, the ALMA-detected OFGs have the longest gas depletion timescale and the highest gas fraction. We believe this is due to a selection effect, as galaxies with higher dust content are more easily detected by ALMA at 1.13 mm. The $M_{\rm gas}$ was derived from $M_{\rm dust}$ in our study by employing the gas-to-dust ratio (see $\S$\ref{Sect::Mgas}). Thus, the ALMA-detected galaxies tend to have higher $M_{\rm gas}$  and, consequently, higher values of $\tau_{\rm dep}$ and $f_{\rm gas}$ as well. Furthermore, the SFR is positively correlated with $M_{\rm dust}$ for star-forming galaxies at fixed $T_{\rm dust}$ \citep[e.g.,][]{Genzel2015,Orellana2017,Donevski2020}. It explains why they show a higher SFR in the SFR-$M_{\star}$ plane compared to the total stacked OFGs (in Fig.~\ref{Fig:MS}). Notably, it raises the caveat that the approach of selecting only ALMA-detected galaxies in studies of  OFGs will end up biasing the sample toward larger SFRs, longer $\tau_{\rm dep}$, and larger $f_{\rm gas}$.

In addition, the massive OFGs (excluding LBGs) present the lowest gas fraction and the shortest gas depletion timescale of all stacked (sub)samples in Fig.~\ref{Fig:gas}. Yet, we did not see any significant difference in the $\Delta$MS of the massive OFGs compared with the other stacked (sub)samples. 
This suggests that these galaxies are observed just before becoming passive.

The median $T_{\rm dust}$ = $46 \pm 2$ K for the stacked total OFGs (see Table \ref{table:stacked}) is consistent with the scaling relation of $T_{\rm dust}(z, \Delta$MS) \citep[][black squares in third column of Fig.~\ref{Fig:gas}]{Schreiber2018c}. However, surprisingly, most of the stacked (sub-)samples show slightly colder dust temperatures compared to the scaling relation. In particular, the ALMA-detected OFGs have the most abundant dust but show the lowest $T_{\rm dust}$, indicating that the dust is colder in the more obscured sources ($A_{\rm V}$ = 1.7$^{+0.1}_{-0.2}$ in Table \ref{table:mean}). The mean $T_{\rm dust}$ =  $42 \pm 1$ K of the ALMA-detected OFGs (see Table \ref{table:mean}) is consistent with the $T_{\rm dust}$ = $40\pm2$ K\footnote{$T_{\rm dust}$ is derived using the IR template library \citep{Schreiber2018c}. To compare with our results, it has been scaled to the light-weighted dust temperature by applying Equation (6) in \cite{Schreiber2018c}.} of the median stacked ALMA-detected massive $H$-dropouts at $z=4$ \citep{Wang2019}. The median $T_{\rm dust}$ of the ALMA-detected OFGs is much lower, with $T_{\rm dust}$ = $33\pm3$ K. 
The low $T_{\rm dust}$ of the ALMA-detected OFGs cannot be explained by current studies \citep[see, e.g.,][]{Magnelli2014,Schreiber2018c}, which suggest that an increasing $T_{\rm dust}$ is correlated to an enhanced specific star formation rate. Furthermore, this is contrary to the findings of \cite{Sommovigo2022}, for instance, where the authors conclude that dust is warmer in obscured sources because a larger obscuration leads to more efficient dust heating. However, quite intriguingly, cases of cold dusty star-forming galaxies at high redshifts have already been reported in the literature, such as GN20 at $z=4.05$ with $T_{\rm dust}$ = 33 K \citep{Magdis2012, Cortzen2020} and four ALMA-detected sources at 3mm at $z\sim5$ \citep{Jin2019}. 
A possible reason for the cold dust temperature is that the dust emission in the FIR of the dust-obscured sources may be optically thick rather than optically thin, where a warm and compact dust core is hidden \citep{Jin2019,Jin2022}. 
 Indeed, the compact dust core is shown in the last column of Fig.~\ref{Fig:gas}. Among (sub-)samples of our OFGs, the ALMA-detected OFGs with the highest dust obscuration (largest $A_{\rm V}$) present the most compact dust core with a half-light radius $R_{\rm e(1.13 mm)}=0.80\pm0.03$ kpc. The SFR surface density ($\Sigma_{\rm SFR}$) of the ALMA-detected OFGs is about two to three times higher than the others. This would imply that the measured dust temperature underestimates the actual dust temperature, that would be higher after correcting for the attenuation in the shorter FIR bands. Making this correction is out of the scope of this paper due to the limited information that we have on those galaxies.

\subsection{The hidden side of the dust region} \label{SFRuvcorr}


\begin{table}
\caption{UV-corrected SFR vs. total SFR for the mean stacked OFGs.}   
\tiny      
\centering
\begin{threeparttable} 
 
\begin{tabular}{l l l c}
\hline\hline       
Mean stacked OFGs &  SFR$_{\rm UV}^{\rm corr}$ &  SFR$_{\rm tot}$ & SFR$_{\rm tot}$/SFR$_{\rm UV}^{\rm corr}$ \\  
& ($M_{\odot}$ yr$^{-1}$)&($M_{\odot}$ yr$^{-1}$)&\\ 
(1)& (2) & (3) & (4)\\
\hline  
Total OFGs & 46$_{-4}^{+3}$ & 348 $\pm$  59 & 8 $\pm$ 1\\                
OFGs that are LBGs & 5$_{-1}^{+1}$ & 45 $\pm$  33 & 9 $\pm$ 7\\
OFGs that are not LBGs& 57$_{-4}^{+6}$  & 388 $\pm$  82 & 7 $\pm$ 2\\            
Massive OFGs that are not LBGs& 88$_{-8}^{+9}$& 557 $\pm$ 86& 6 $\pm$ 1  \\
ALMA-detected OFGs & 83$_{-7}^{+10}$ & 636 $\pm$ 105 & 8$\pm$ 2\\
\hline 
                 
\end{tabular}
\begin{tablenotes}
\item \textbf{Note:} (1) Mean stacked total sample and four sub-samples of OFGs; (2) SFR$_{\rm UV}$ corrected for dust extinction (see $\S$\ref{SFRuvcorr}); (3) $\rm{SFR}_{\rm tot} = \rm{SFR}_{\rm IR}+\rm{SFR}_{\rm UV}$; (4) Ratio of total SFR to SFR$_{\rm UV}^{\rm corr}$.

\end{tablenotes}
\label{table:sfr} 
\end{threeparttable} 
\end{table}

An important check is to test whether UV continuum emission alone (after correcting for dust extinction) provides a robust estimate of the total SFR, especially for those highly dust-obscured galaxies. We again used the stacked total sample and four sub-samples of OFGs.

We derived the SFR$_{\rm UV}$ corrected for dust extinction (i.e., SFR$_{\rm UV}^{\rm corr}$) using the \cite{Calzetti2000} reddening law and assuming a constant star formation history 
from the UV to MIR SED fitting with the code \texttt{FAST++}. Specifically, similar to $\S$\ref{Sect::sfr}, the SFR$_{\rm UV}^{\rm corr}$ was obtained from the $L_{\rm UV}^{\rm corr}$ following \cite{Daddi2004}, which was calculated based on the AB magnitude at the rest-frame 1500 $\AA$ (see Eq.~\ref{eq:3}). The intrinsic flux was derived with
\begin{flalign}
f_{\rm int}(\lambda) = f_{\rm obs}(\lambda) \,10^{0.4\,A_{\lambda}}, 
\label{fint}
\end{flalign}
where $f_{\rm int}$ and $f_{\rm obs}$ are the intrinsic and observed fluxes, respectively. The extinction $A_{\lambda}$ is related to the reddening curve $k(\lambda)$:
\begin{flalign}
A_{\lambda} = \frac{k(\lambda)\,A_{\rm V}}{R_{\rm V}}.
\label{fint}
\end{flalign}
From the \cite{Calzetti2000} reddening law, we have $R_{\rm V}$ = 4.05 and
\begin{flalign}
K_{\lambda} = 2.659\,(-2.156+\frac{1.509}{\lambda}-\frac{0.198}{\lambda^{2}}+\frac{0.011}{\lambda^{3}})+R_{\rm V}, 
\label{fint}
\end{flalign}
with $0.12\,\mu m\leq\lambda\leq0.63\,\mu m$.

We then compared the derived SFR$_{\rm UV}^{\rm corr}$ with the total SFR ($\rm{SFR}_{\rm tot} = \rm{SFR}_{\rm IR}+\rm{SFR}_{\rm UV}$) and calculated their ratios. The results are presented in Table \ref{table:sfr}. We find that all the stacked (sub-)samples have SFR$_{\rm tot}$ much larger than SFR$_{\rm UV}^{\rm corr}$. 
In addition, they have similar SFR$_{\rm tot}$/SFR$_{\rm UV}^{\rm corr}$ ratios within uncertainties, with SFR$_{\rm tot}$/SFR$_{\rm UV}^{\rm corr}$ = $8\pm1$ for the mean stacked total OFGs.  
It suggests: ($i$) the existence of a hidden dust region in the OFGs (even for the LBGs) that absorbs all the UV photons, which cannot be reproduced with a dust extinction correction, indicating that the dust emission in these OFGs might be optically thick; and ($ii$) that it is fundamental to include IR/mm band observations when studying extremely dusty star-forming galaxies. Otherwise, the total SFR and, therefore, the cosmic SFRD will be strongly underestimated.

\begin{figure}
\centering
\includegraphics[scale=0.37]{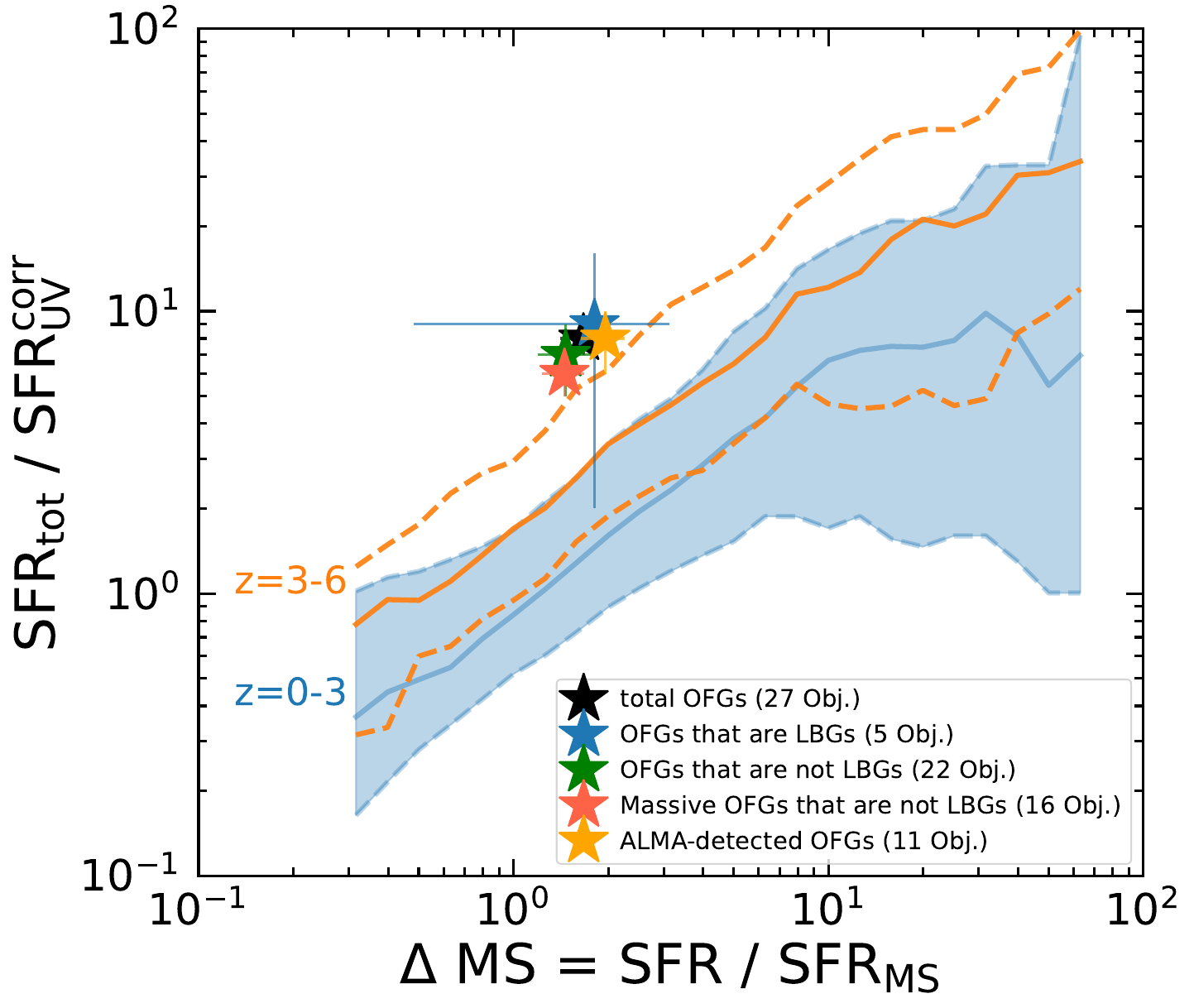}
\caption{Ratio of the total SFR ($\rm{SFR}_{\rm tot} = \rm{SFR}_{\rm IR}+\rm{SFR}_{\rm UV}$) and the SFR$_{\rm UV}$ corrected for the dust extinction (SFR$_{\rm UV}^{\rm corr}$) as a function of the starburstiness ($\Delta$MS = SFR$_{\rm tot}$/SFR$_{\rm MS}$). Stars represent the mean stacked optically dark/faint (sub)samples as in Fig.~\ref{Fig:MS}. The solid and dashed lines show the sliding median and 16-84th percentile range of star-forming galaxies from the CDF-S field with a \textit{Herschel} detection at $0\leq z \leq 3$ (blue) and $3\leq z \leq 6$ (orange) from the ZFOURGE catalog \citep{Straatman2016}.
\label{compare}}
\end{figure}

Furthermore, we compared the stacked optically dark/faint (sub)samples with the star-forming galaxies from the ZFOURGE catalog \citep{Straatman2016}. These star-forming galaxies are from the CDF-S field with a \textit{Herschel} detection, and are split into two redshift bins ($0 \leq z \leq 3$ and $3\leq z \leq 6$). 
We calculated their SFR$_{\rm tot}$/SFR$_{\rm UV}^{\rm corr}$ ratios using the same method as for the stacked optically dark/faint (sub)samples. 
As shown in Fig.~\ref{compare}, the SFR$_{\rm tot}$/SFR$_{\rm UV}^{\rm corr}$ ratio increases with increasing starburstiness, indicating the presence of more hidden dust regions in galaxies that are likely to be optically thick. It suggests that using the UV emission alone to determine the total SFR of starburst galaxies, even after dust attenuation correction, could result in strong underestimates, consistent with the findings of \cite{Elbaz2018} and \cite{Puglisi2017}. We further find that the strong underestimations appear at both redshift bins, suggesting that this may be a general phenomenon for starburst galaxies, regardless of the redshift. In addition, for MS galaxies ($\Delta$MS $\sim$ 1) with $0 \leq z \leq 3$, their SFR$_{\rm tot}$ and SFR$_{\rm UV}^{\rm corr}$ are very similar, showing that both SFR estimators agree with each other for typical MS galaxies at low redshifts. However, for MS galaxies with $3 \leq z \leq 6$, their SFR$_{\rm tot}$ is about twice ($\sim$0.3 dex) larger than the SFR$_{\rm UV}^{\rm corr}$. Generally, the median SFR$_{\rm tot}$/SFR$_{\rm UV}^{\rm corr}$ ratio is about twice higher for star-forming galaxies with $3 \leq z \leq 6$ than those with $0 \leq z \leq 3$. We note that in Fig.~\ref{compare}, we did not perform a stellar-mass cut for the star-forming galaxies from the ZFOURGE catalog due to the small number of galaxies. As a test, we selected galaxies with log($M_{\star}/M_{\odot}) > 9$ and obtained similar results but with a larger dispersion of the galaxy distribution because of their small number. A more detailed study of the stellar masses, and the SFR$_{\rm tot}$/SFR$_{\rm UV}^{\rm corr}$ ratio is beyond the scope of this paper. 

In Fig.~\ref{compare}, the stacked optically dark/faint (sub)samples, with $z_{\rm med,stacked}=4.5$, lie above the 16-84th percentile range of the star-forming galaxies at $3 \leq z \leq 6$. This is consistent with the fact that these dusty star-forming galaxies are the more extreme cases (more dust-obscured), with lower dust temperatures compared to typical star-forming galaxies at similar redshifts and with relatively compact dust sizes ($R_{\rm e(1.13 mm)}=1.01\pm0.05$ kpc; black star in Fig.~\ref{Fig:gas}).

\section{Cosmic star formation rate density}\label{SFRD}

\subsection{Star formation rate density}\label{Sect::SFRD}
In this section, we calculate the SFRD in a very conservative way. In our calculation, the derived value would be a lower limit for the OFGs.

As our OFGs were discovered randomly within a blind GOODS-ALMA survey area, we can simply calculate their SFRD by using their total SFR divided by the survey volume. The survey volume is the volume between the shells defined by the redshift range of the sources and within a solid angle:
\begin{equation}
\label{V}
V = \frac{\Omega}{4 \pi} \,(V_{z_{1}} - V_{z_{0}}) = \frac{\Omega}{4 \pi} \,[\frac{4 \pi}{3}(d^3_{z_{1}} - d^3_{z_{0}})] = \frac{\Omega}{3} \,(d^3_{z_{1}} - d^3_{z_{0}}),
\end{equation}
where the solid angle $\Omega$ corresponding to the effective area of 72.42 arcmin$^2$, in units of steradian ($6.1\times10^{-6}$ sr), of the GOODS-ALMA survey; d$_{z_{0}}$ and d$_{z_{1}}$ are the comoving distances at given redshifts of $z_{0}$ and $z_{1}$.  
To be conservative, here we use  a broad redshift range of $z=3.2-7.0$ from the whole OFG sample (see Fig.~\ref{Fig:distribution}) for our total 22 OFGs that are not LBGs, instead of using the redshift and 1$\sigma$ confidence interval given by the stacked SED. 

\begin{figure*}
\centering
\includegraphics[scale=0.85]{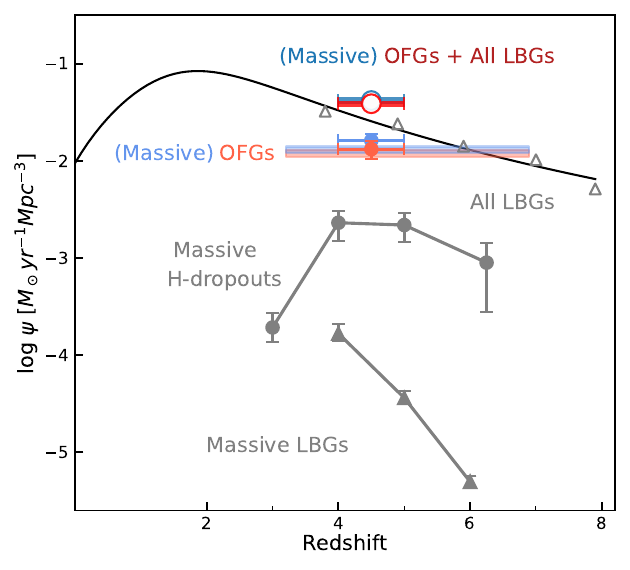}
\includegraphics[scale=0.85]{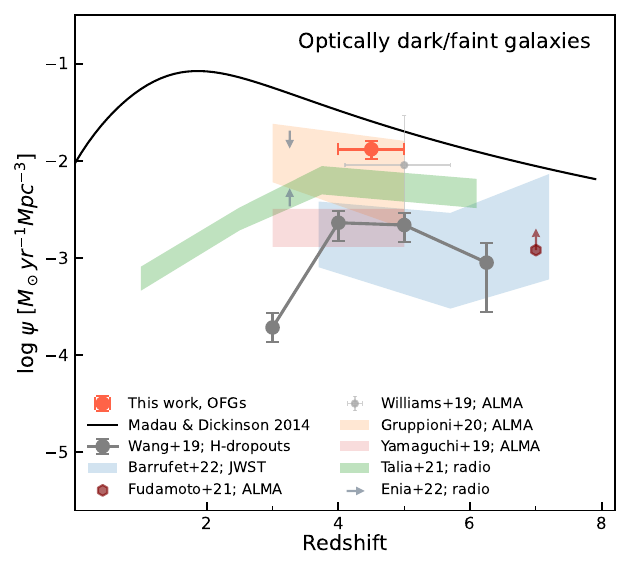}
\caption{Cosmic star formation history of the Universe. $Left$: Contribution of our OFGs to the cosmic SFRD. The black line is the cosmic SFRD, $\psi$, as a function of redshift from \cite{Madau2014}, which is based on LBGs at $z\gtrsim4$ with a dust correction applied \citep[grey open triangles;][]{Bouwens2012a,Bouwens2012b}. The 22 OFGs that are not LBGs are shaded in red and the 16 massive OFGs that are not LBGs (log($M_{\star}$/$M_{\odot}$) > 10.3) are shaded in blue, where the sources at $z=4-5$ are shown in faint red and blue filled circles, respectively.  Grey filled circles are ALMA-detected massive $H$-dropouts with log($M_{\star}$/$M_{\odot}$) > 10.3 \citep{Wang2019}. Grey filled triangles are massive LBGs with log($M_{\star}$/$M_{\odot}$) > 10.3 whose SFRD is based on the dust-corrected UV \citep{Wang2019}. 
We also show the combined contribution of OFGs and LBGs to the cosmic SFRD. The red (blue) open circle indicates the combined contribution of OFGs (massive OFGs) and all LBGs \citep{Madau2014} at $z=4-5$. 
$Right$: Summary of the contribution of optically dark/faint galaxies to the cosmic SFRD from the literature. These optically dark/faint galaxies were selected in different methods, including our OFGs (red point; this work),  ALMA-detected massive $H$-dropouts \citep[grey filled circles;][]{Wang2019}, JWST-selected HST-dark galaxies from CEERS \citep[blue area;][]{Barrufet2022}, ALMA-detected optically dark sources from REBELS \citep[brown point and arrow;][]{Fudamoto2021}, an ALMA-detected NIR-dark galaxy in COSMOS \citep[light grey point;][]{Williams2019}, ALMA-detected HST+near-IR dark galaxies from ALPINE \citep[orange area;][]{Gruppioni2020}, ALMA-detected $K$-dropouts from ASAGAO \citep[red area;][]{Yamaguchi2019}, radio 3 GHz-selected UV-dark galaxies in COSMOS \citep[green area;][]{Talia2021}, and radio 1.4 GHz-selected HST-dark galaxies in GOODS-N \citep[grey arrows;][]{Enia2022}.
All data from the literature have been converted to a \cite{Chabrier:2003} IMF. 
\label{Fig:SFRD}}
\end{figure*}

The cosmic star-formation-rate density for our total OFG sample is:
 \begin{equation}
\psi \,[\rm{M}_{\odot} \rm{yr}^{-1}\rm{Mpc}^{-3}] = \frac{\rm{SFR}_{\rm mean}\times22}{V},
\end{equation}
where the SFR$_{\rm mean}$ is the UV+IR-based SFR given from the mean stacked SED fitting of the 22 OFGs that are not LBGs (Table \ref{table:mean}).  Assuming that the intrinsic infrared SED of our OFGs is the same as the SED derived from mean stacking, the SFRD of our 22 OFGs (excluding LBGs) reaches approximately 1.2 $\times$ 10$^{-2}$ $M_{\odot}$ yr$^{-1}$Mpc$^{-3}$ with the mean redshift of $z_{\rm mean}=4.3$ (and $z_{\rm median}=4.1$). The survey volume, $V$, is 7.4 $\times$ 10$^{5}$ Mpc$^3$. The number density is $n \sim 3 \times 10^{-5}$ Mpc$^{-3}$, which is slightly higher than the one of massive $H$-dropouts.


\subsection{Incompleteness of our understanding of the cosmic star-formation history}
The reference cosmic star-formation history at $z\gtrsim4$ \citep{Madau2014} is based on measurements of the LBGs corrected for dust attenuation in the rest-frame UV \citep{Bouwens2012a,Bouwens2012b}. The study of OFGs (excluding LBGs) can help us quantify the incompleteness of our understanding of the cosmic star-formation history at high redshifts.

The left panel of Fig.~\ref{Fig:SFRD} shows the contribution of OFGs ($z_{\rm mean}=4.3$) to the cosmic SFRD. At $z=4-5$, 
the contribution of OFGs (red-filled circles) reaches about 52\% of the reference SFRD from LBGs \citep[][black solid line with grey open triangles]{Madau2014}. If we combine the contributions of OFGs (excluding LBGs) and LBGs \citep{Madau2014} as total SFRD, the OFGs contribute about 34\% and the LBGs contribute about 66\% of the total SFRD. It shows that the LBGs dominate the total SFRD, while the OFGs (excluding LBGs) make up about a third of the total SFRD.  

In addition, we  investigate the contribution of massive OFGs to the cosmic SFRD. Out of a total of 22 OFGs (excluding LBGs), 16 massive OFGs have log($M_{\star}$/$M_{\odot}$) > 10.3, the same stellar mass cut as the so-called ALMA-detected massive $H$-dropouts \citep[][]{Wang2019}. The SFRD of massive OFGs is approximately 1.2 $\times$ 10$^{-2}$ $M_{\odot}$ yr$^{-1}$Mpc$^{-3}$ with $z_{\rm mean}=4.4$ (and $z_{\rm median}=4.1$), shown as the blue-shaded area in the left panel of Fig.~\ref{Fig:SFRD} (see also Table \ref{table:mean}). If we limit the redshift range to $z=4-5$, the SFRD of massive OFGs is approximately 1.6 $\times$ 10$^{-2}$ $M_{\odot}$ yr$^{-1}$Mpc$^{-3}$, as illustrated by the blue-filled circles. This value is consistent with the SFRD of the total 22 OFGs within errors (red-filled circles in Fig.~\ref{Fig:SFRD} ). 
Since the massive and total OFGs have similar SFRD values, it indicates that among the OFGs, the massive ones dominant the SFRD compared to the remaining less massive ones.

We further compare the massive OFGs with equivalently massive LBGs and $H$-dropouts.
Firstly, comparing the massive OFGs with equivalently massive LBGs, the SFRD of massive OFGs is at least two orders of magnitude higher. Thus, we conclude that for massive galaxies (log($M_{\star}$/$M_{\odot}$) > 10.3), the cosmic SFRD is dominated by massive OFGs (99\%) rather than massive LBGs (1\%) at high redshifts. Secondly, the SFRD of massive OFGs is more than four times higher than that of massive $H$-dropouts detected by ALMA \citep{Wang2019}. We recall that the selection criteria for $H$-dropouts are $H$ > 27 mag \& [4.5] < 24 mag \citep{Wang2019}, whereas our OFGs are selected with $H>$ 26.5 mag \& [4.5] $<$ 25 mag. 
It implies that the optically faint galaxies that contribute significantly to the SFRD of massive galaxies have been neglected in previous studies of LBGs as well as $H$-dropouts.  
Therefore, we emphasize the importance of considering moderately obscured, non-LBG sources in complementing the high-$z$ SFRD measurements, rather than focusing only on those extremely dusty star-forming galaxies.

Table \ref{table:mean} presents the SFRD of the total OFGs, the massive OFGs, and the ALMA-detected OFGs, with LBGs excluded. Table \ref{table:mean} shows that these three (sub)-samples have the same SFRD values within errors. A comparison of the  total OFGs and the massive OFGs has already been discussed above and shown in the left panel of Fig.~\ref{Fig:SFRD}. For the ALMA-detected OFGs, their SFRD value is lower than the total OFGs but still within errors. This implies that the investigation of the ALMA-detected sources alone might be sufficient to account for the contribution of OFGs in the SFRD.
However, limited by the number of sources, this result needs to be further confirmed with a larger OFG sample in future studies. Besides, we would like to remind that the dust and gas properties of the ALMA-detected OFGs are not representative of all OFGs, as described in $\S$\ref{figure_gas}.

In our work, we find that the OFGs contribute $\sim$ 52\% to the reference SFRD at $z=4-5$. Many works have also studied optically dark/faint galaxies selected or originally detected with different methodologies. 
The right panel of Fig.~\ref{Fig:SFRD} shows a summary of the cosmic SFRD of optically dark/faint galaxies from the literature.
Specifically, in \cite{Wang2019}, the contribution of the ALMA-detected massive $H$-dropouts accounts for $\sim$10\% of the reference SFRD from LBGs at $z\sim4-6$. At $z\sim5$, this value is broadly consistent with that from the ALMA-detected NIR-dark galaxy in COSMOS \citep{Williams2019} and the ALMA-detected HST+near-IR dark galaxies from ALPINE \citep{Gruppioni2020}. Similar results have also been found for the ALMA-detected $K$-dropouts from ASAGAO \citep[$\sim$10\% at $z\sim3-5$;][]{Yamaguchi2019} and the JWST-selected HST-dark galaxies from CEERS \citep[$\sim$10\% at $z\sim4-6$, but increasing to $\sim$ 36\% at $z\sim7$;][]{Barrufet2022}. 
At z$\sim$7, \cite{Fudamoto2021} identified two optically dark galaxies in two separate REBELS pointings that contribute $\sim 10-25$\% to the reference SFRD. 
In radio, the 3 GHz-selected UV-dark galaxies from VLA-COSMOS \citep{Talia2021} contribute 10-25\% at $z\sim3-4.5$ and 25-40\% at $z>4.5$ to the reference SFRD. More recently, \cite{Enia2022} found a 7-58\% contribution from 1.4 GHz radio-selected HST-dark galaxies in GOODS-N at $z\sim3$. 
Despite the broad range of contributions from these galaxies, the high values all agree that the contribution of optically dark/faint galaxies to the cosmic SFRD cannot been neglected, suggesting that highly dust-obscured star formation is relatively common in the $z>3$ Universe.

Finally, we calculated the combined contribution of the OFGs and LBGs \citep[values given from][]{Madau2014} to the cosmic SFRD. As shown by the red open circle in the left panel of  Fig.~\ref{Fig:SFRD}, the total SFRD is 4.0 $\times$ 10$^{-2}$ $M_{\odot}$ yr$^{-1}$Mpc$^{-3}$ at $z=4-5$, about 0.15 dex higher (43\%) than the SFRD at the same redshift from \cite{Madau2014}. As we mention previously, the calculation of the SFRD values was very conservative and the true total SFRD of the OFGs and LBGs could be greater.




\section{Conclusions}\label{Sec:summary}
This work aims to obtain a more complete picture of the cosmic star formation history in the $z>3$ Universe, that is, to bridge the extreme population of optically dark/faint galaxies (or $H$-dropouts) with the most common population of lower mass, less attenuated galaxies, such as LBGs. We use a more permissive criterion ($H>$ 26.5 mag \& [4.5] $<$ 25 mag) to select optically dark/faint galaxies (i.e., OFGs) at high redshifts, which avoids limiting the sample to the most extreme cases. This criterion selects extremely dust-obscured massive galaxies that are normal star-forming galaxies, with dust obscuration typically at E(B-V) $>$ 0.4, with lower stellar masses at high redshifts than $H$-dropouts. In addition, our selection method has the capacity to select OFGs without contamination from passive or old galaxies. In the GOODS-ALMA region, we have a total of 28 OFGs (including a candidate IRAC 4.5$\mu$m dropout). We analyzed the properties of individual and stacked OFGs, respectively. We calculated their SFRD and quantified the incompleteness of our understanding of the cosmic star-formation history in the $z>3$ Universe.

 Here are the main results of this work:
   \begin{enumerate}
 \item After performing SED analyses with the code \texttt{EAzY} and \texttt{FAST++}, we find that the OFGs cover a redshift of $z_{\rm phot}$ > 3, as indicated by the theoretical galaxy templates. The median redshift of individual OFGs is $z_{\rm med}=4.1$, with a wide distribution ($z=3.2-7.0$), consistent with $z_{\rm med,stacked}=4.5\pm0.2$ derived from the median stacked SED. The OFGs have a broad stellar mass distribution with log($M_{\star}$/$M_{\odot}$) = $9.4-11.1$, with a median of log($M_{\rm \star med}$/$M_{\odot}$) = 10.3. 
 \item We investigate the proportions of LBGs, $H$-dropouts, and remaining OFGs (after removing LBGs and $H$-dropouts) in our sample. Remarkably, at stellar masses of log($M_{\star}$/$M_{\odot}$) = $9.5-10.5$, the fraction of remaining OFGs is about three times the sum of LBGs and $H$-dropouts. In other words, up to 75\% of the OFGs with log($M_{\star}$/$M_{\odot}$) = $9.5-10.5$ at $z>3$ are neglected by the previous LBGs and $H$-dropout selection techniques. 
    \item All stacked OFGs are located within the SFMS region (0.33 < $\Delta$MS < 3, i.e., $\sim$0.5 dex), which is consistent with being normal star-forming galaxies at the same redshift. It suggests that rather than being limited to a rare population of extreme starburst galaxies, our OFGs represent a normal population of dusty star-forming galaxies at $z > 3$.
    \item The gas properties of the OFGs imply that the OFGs have shorter $\tau_{\rm dep}$ and slightly lower $f_{\rm gas}$ values compared to the scaling relation followed by typical main sequence galaxies. Their gas is consumed more rapidly, hence they form their stars with a high efficiency, setting them in the so-called class of starbursts in the main sequence. In addition, the massive OFGs have the shortest $\tau_{\rm dep}$ and lowest $f_{\rm gas}$ of all stacked (sub)samples, indicating that they are in the process of becoming passive. Finally, we point out that the approach of selecting only ALMA-detected galaxies in studies of OFGs will end up biasing the sample toward larger SFRs, longer $\tau_{\rm dep}$, and larger $f_{\rm gas}$.
    \item Studying dust temperatures in the OFGs, we find that, surprisingly, most of the stacked (sub)samples show colder $T_{\rm dust}$ than the scaling relation. In particular, the ALMA-detected OFGs have the most abundant dust but show the lowest $T_{\rm dust}$, indicating that the dust is colder in more obscured sources. A possible reason for the cold dust temperature is that the dust emission in the FIR of the dust-obscured sources may be optically thick rather than optically thin, where a warm and compact dust core is hidden.
    \item In the comparison of SFR$_{\rm UV}^{\rm corr}$ and SFR$_{\rm tot}$, we find that ($i$) all the stacked (sub-)samples have SFR$_{\rm tot}$ much larger than SFR$_{\rm UV}^{\rm corr}$. There could be a hidden dust region of OFGs (even for LBGs as well) that absorbs all the UV photons, which cannot be reproduced with a dust extinction correction; and ($ii$) 
    it is fundamental to include IR/mm band observations when studying extremely dusty star-forming galaxies; otherwise, the total SFR and, therefore, the cosmic SFRD will be underestimated.
      \item After excluding five LBGs in the OFG sample, we study the hidden side of cosmic SFRD at high redshift ($z\gtrsim3$) and find that: ($i$) among all galaxies, the total SFRD is dominated by LBGs, followed by OFGs. At $z=4-5$, the contribution of OFGs reaches about 52\% of the SFRD \citep[][]{Madau2014}, which is calculated mainly based on the LBGs \citep[][]{Bouwens2012a,Bouwens2012b};   ($ii$) for the OFGs, the massive and total OFGs have similar SFRD values, indicating that the massive OFGs make a major contribution to the SFRD compared to the remaining low-mass OFGs; ($iii$) for massive galaxies, the SFRD is dominated by massive OFGs rather than massive LBGs. The SFRD contributed by massive OFGs is at least two orders of magnitude higher than the one contributed by massive LBGs; ($iv$) the contribution of massive OFGs to the SFRD is more than four times higher than that of $H$-dropouts \citep{Wang2019}. It implies that optically faint galaxies also contribute significantly to the SFRD, which has been neglected in previous studies of LBGs and $H$-dropouts; and ($v$) the ALMA-detected OFGs and the total OFGs have similar SFRD values, but with different gas and dust properties (as we mentioned in point 4), such as $\tau_{\rm dep}$, $f_{\rm gas}$, $T_{\rm dust}$, and $R_{\rm e(1.13 mm)}$, which require attention when studying the OFGs selected by ALMA detection alone.

      \item Finally, we calculate the combined contribution of the OFGs and LBGs to the cosmic SFRD at $z=4-5$, which is 4.0 $\times$ 10$^{-2}$ $M_{\odot}$ yr$^{-1}$Mpc$^{-3}$, about 0.15 dex higher (43\%) than the SFRD derived from LBGs alone \citep{Madau2014} at the same redshift. This value could be even larger as our calculation was very conservative.

                     \end{enumerate}

\begin{acknowledgements}
We are very grateful to the anonymous referee for instructive comments, which helped to improve the overall quality and strengthened the analyses of this work. 
We thank Alain Omont, Boris Sindhu Kalita, Dangning Hu, Shuowen Jin, Pascal Oesch, and Bing Luo for valuable discussions and suggestions that improved this paper. 
This paper makes use of the following ALMA data: ADS/JAO.ALMA \#2015.1.00543.S and ADS/JAO.ALMA \#2017.1.00755.S. ALMA is a partnership of ESO (representing its member states), NSF (USA), and NINS (Japan), together with NRC (Canada), NSC, ASIAA (Taiwan), and KASI (Republic of Korea), in cooperation with the Republic of Chile. The Joint ALMA Observatory is operated by ESO, AUI/NRAO, and NAOJ. 
This work is supported by the National Key Research and Development Program of China (No. 2017YFA0402703), and by the National Natural Science Foundation of China (No. 11733002, 12121003, 12192220, 12192222, 12173017, and 12141301). 
This work is also supported by the Programme National Cosmology et Galaxies (PNCG) of CNRS/INSU with INP and IN2P3, co-funded by CEA and CNES.
M.-Y. X. acknowledges the support by China Scholarship Council (CSC). 
M.F. acknowledges NSF grant AST-2009577 and NASA JWST GO Program 1727. 
G.E.M. acknowledges the Villum Fonden research grants 13160 and 37440 and the Cosmic Dawn Center of Excellence funded by the Danish National Research Foundation under the grant No. 140. 
R.D. gratefully acknowledges support by the ANID BASAL projects ACE210002 and FB210003. 
     
\end{acknowledgements}

%
%

\begin{appendix}
\section{IRAC catalog compilation}\label{IRAC catalog}
We constructed the IRAC catalog using the deepest IRAC 3.6 and 4.5\,$\mu$m images from the GREATS program \citep[][]{Stefanon2021}. Source detection was performed using \texttt{Source Extractor} \citep[SE version 2.25.0;][]{Bertin1996} on the background-subtracted 3.6 and 4.5\,$\mu$m images, respectively. This is to help us exclude as many false detections as possible and get pure sources after cross-matching two catalogs from 3.6 and 4.5\,$\mu$m images. Each detected source was required to have a minimum area
of 4 pixels, 
with each pixel achieving the threshold of 0.25$\sigma$. The deblending parameters were optimized by setting  {\tt DEBLEND\_THRESH $=64$} and {\tt DEBLEND\_MINCONT $=0.0001$}, which were a compromise between deblending neighboring sources and minimizing a potential division of a larger object into multiple components. The sky subtraction was performed with SE, using a bicubic interpolation of the background with an adopted mesh size of 512 pixels and a median filter size of 5 pixels. A Mexhat filter was used to smooth the images before detection, to help detect faint and extended objects. The Mexhat filter can very well de-blend faint sources around the bright source, but it also extracts artifacts around the bright source caused by diffraction, which needs to be cleaned. We set {\tt CLEAN\_PARAM} $=$ {\tt 1}  to clean spurious detections with SE. Then, we visually inspected images to further remove artifacts around the bight sources. For the 3.6 and 4.5\,$\mu$m images, we obtained their source catalogs respectively. The total sources in the catalogs are 125,338 for 3.6 \,$\mu$m and 154,234 for 4.5\,$\mu$m  in the entire GOODS-S field.

To ensure the purity of detections, we then cross-matched two catalogs from 3.6 and 4.5\,$\mu$m images with a radius of 1.0$^{\prime\prime}$ ($\sim$0.5$\times$FWHM). For sources that are simultaneously in the GOODS-ALMA 2.0 catalog \citep{Carlos2021} and detected in at least one IRAC band, we considered them as real sources and kept them in the final catalog. 
Finally, we have 5,127    
pure sources detected by both 3.6 and 4.5\,$\mu$m bands and/or ALMA 1.13 mm in the GOODS-ALMA field. 


To estimate the completeness of our detection strategy, we employed a Monte Carlo approach where we simulated 10,000 artificial
model sources with random magnitudes between 14.5 mag and 29 mag. They are point sources with the same PSF profiles as 3.6 and 4.5\,$\mu$m images, respectively. The simulated sources were allowed to fall at random positions on the real image, including on top of other sources, to account for the impact of source blending in the real image. We injected ten sources each time on the IRAC 3.6um and 4.5um images to avoid excessively artificial source confusion caused by bright fake sources. After each injection, we performed the same blind source detection procedure using SE. 

\begin{figure}[h!]
\centering
\includegraphics[scale=0.21]{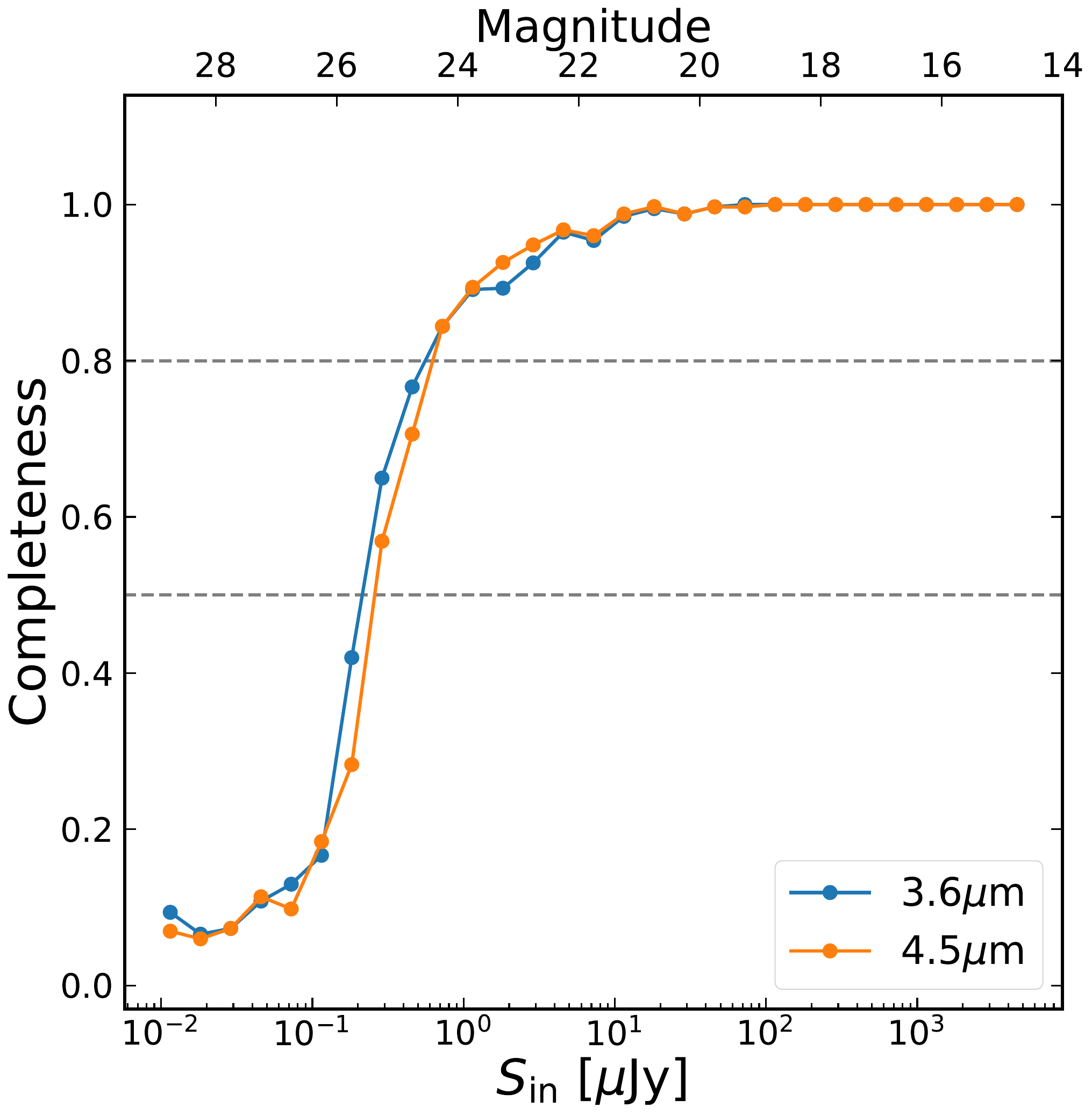}
\includegraphics[scale=0.21]{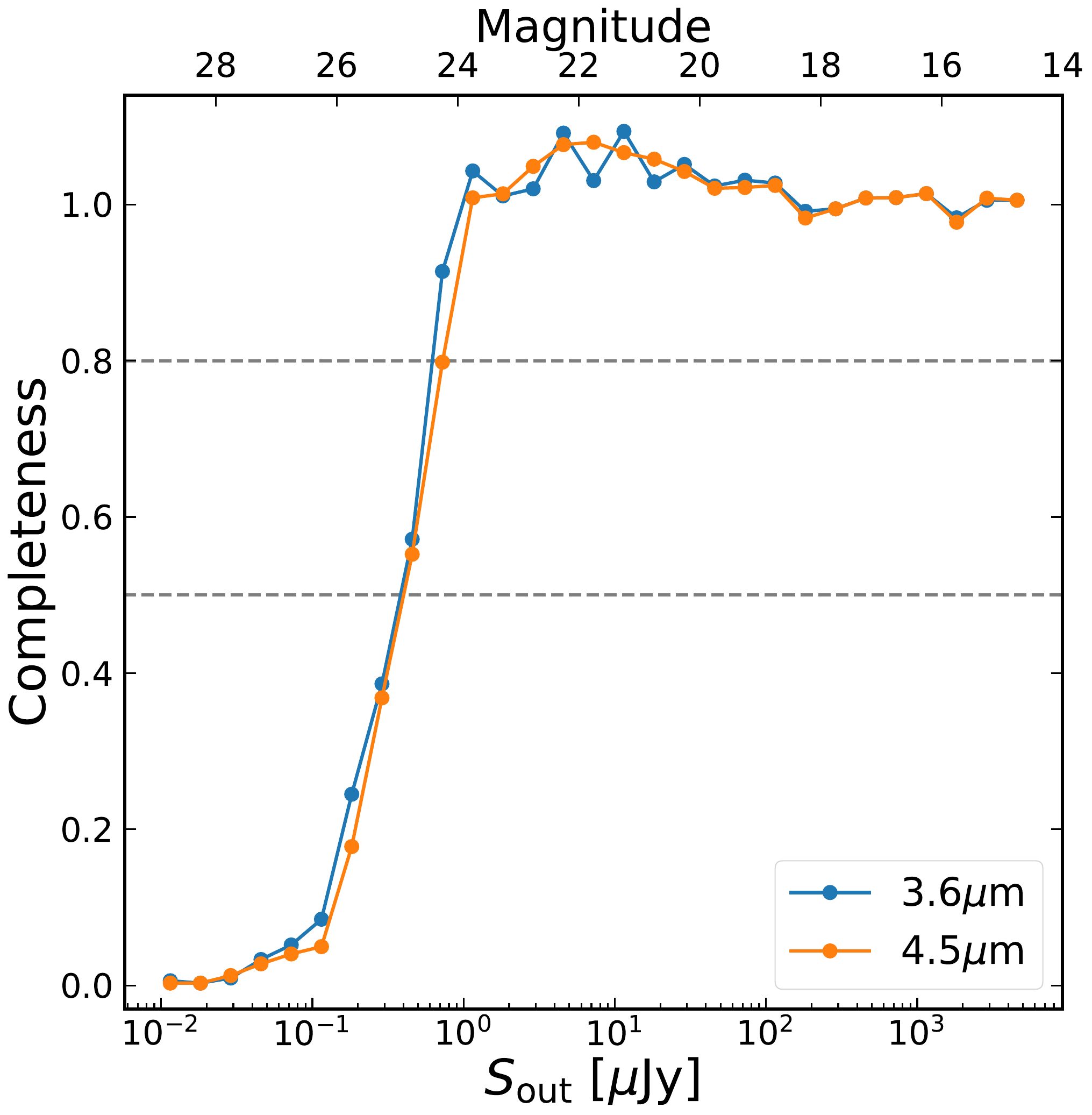}
\caption{Completeness as a function of the input flux density ($S_{\rm in}$; top) and the output flux density ($S_{\rm out}$; bottom). The dashed lines represent the 50\% and 80\% completeness limits as a reference.
\label{Fig:completeness_in}}
\end{figure}

In Fig.~\ref{Fig:completeness_in} (left), we show the completeness as a function of the input flux density ($S_{\rm in}$) for the simulated sources in 3.6 and 4.5\,$\mu$m images. The survey reaches a 100\% completeness for all simulated sources for flux densities $S_{\rm in} \gtrsim$ 20 $\mu$Jy at both 3.6 and 4.5\,$\mu$m wavelength. To understand the incompleteness of our OFG sample, we need to know the behavior of completeness as a function of output flux density ($S_{\rm out}$) shown in Fig.~\ref{Fig:completeness_in} (right). Here, $S_{\rm out}$ is the parameter being measured directly.

\onecolumn
\section{Multiwavelength Postage-stamp}

\begin{figure*}[h!]
\centering
\includegraphics[scale=0.33]{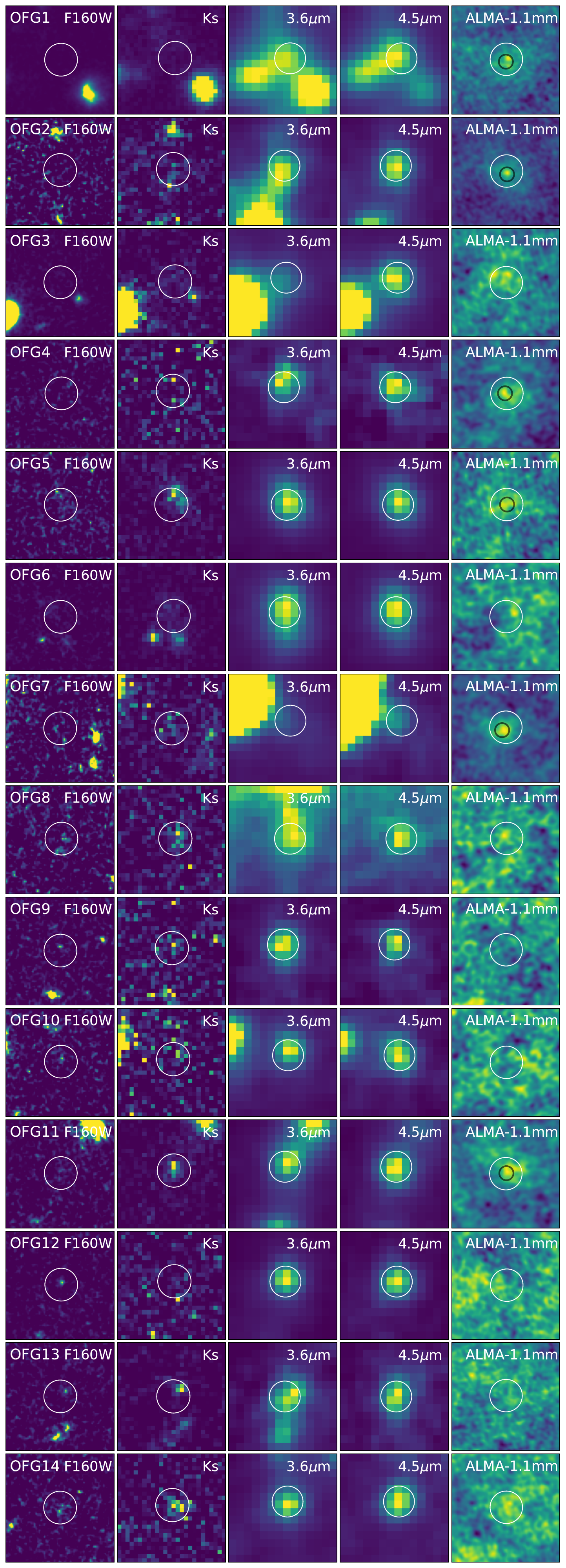}
\includegraphics[scale=0.33]{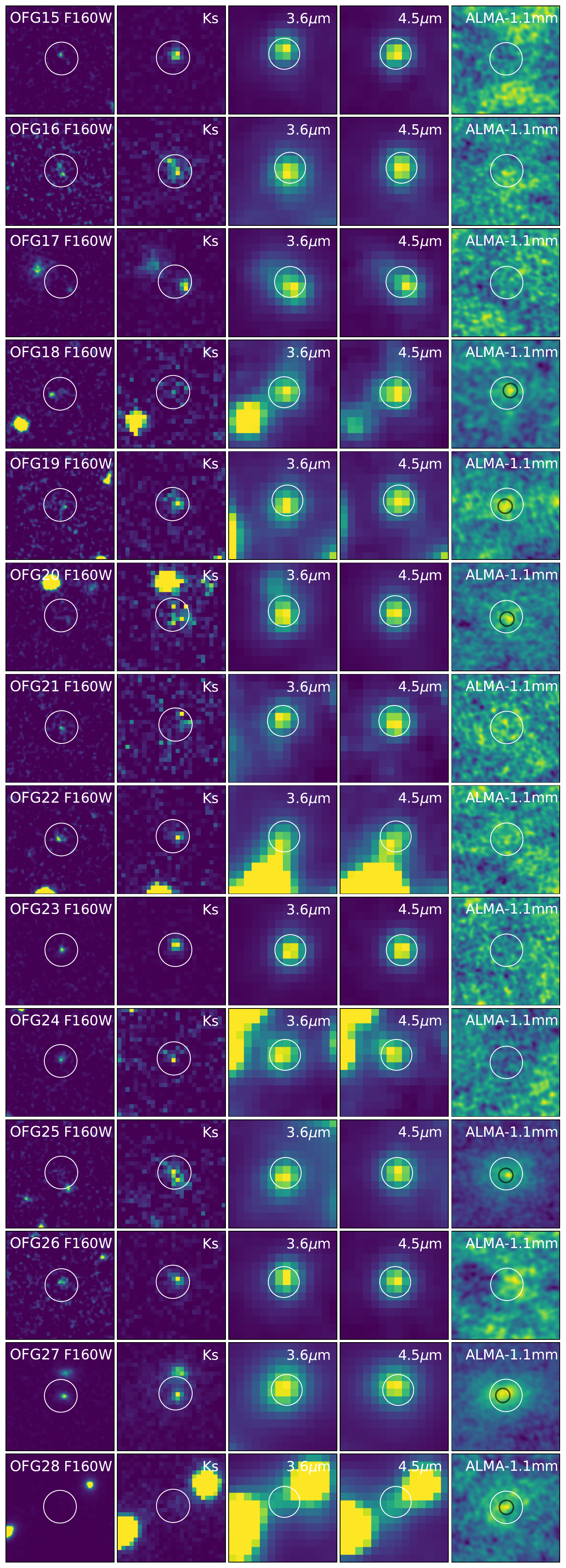}
\caption{Postage-stamps (4$^{\prime\prime}$ $\times$ 4$^{\prime\prime}$) of our OFGs. From left to right: HST/WFC3 (F160W), ZFOURGE ($K_{\rm s}$), \textit{Spitzer}/IRAC (3.6 $\mu$m and 4.5 $\mu$m), and ALMA band 6 (1.13 mm). The white circles mark our target galaxies with 0.6$^{\prime\prime}$ in radius. The black circles indicate the positions of the ALMA detections. North is up and east is to the left.
\label{Fig:stamp}}
\end{figure*}

\onecolumn
\section{SEDs for the individual OFGs}\label{individual SEDs}
\begin{figure*}[h!]
\centering
\includegraphics[scale=0.4]{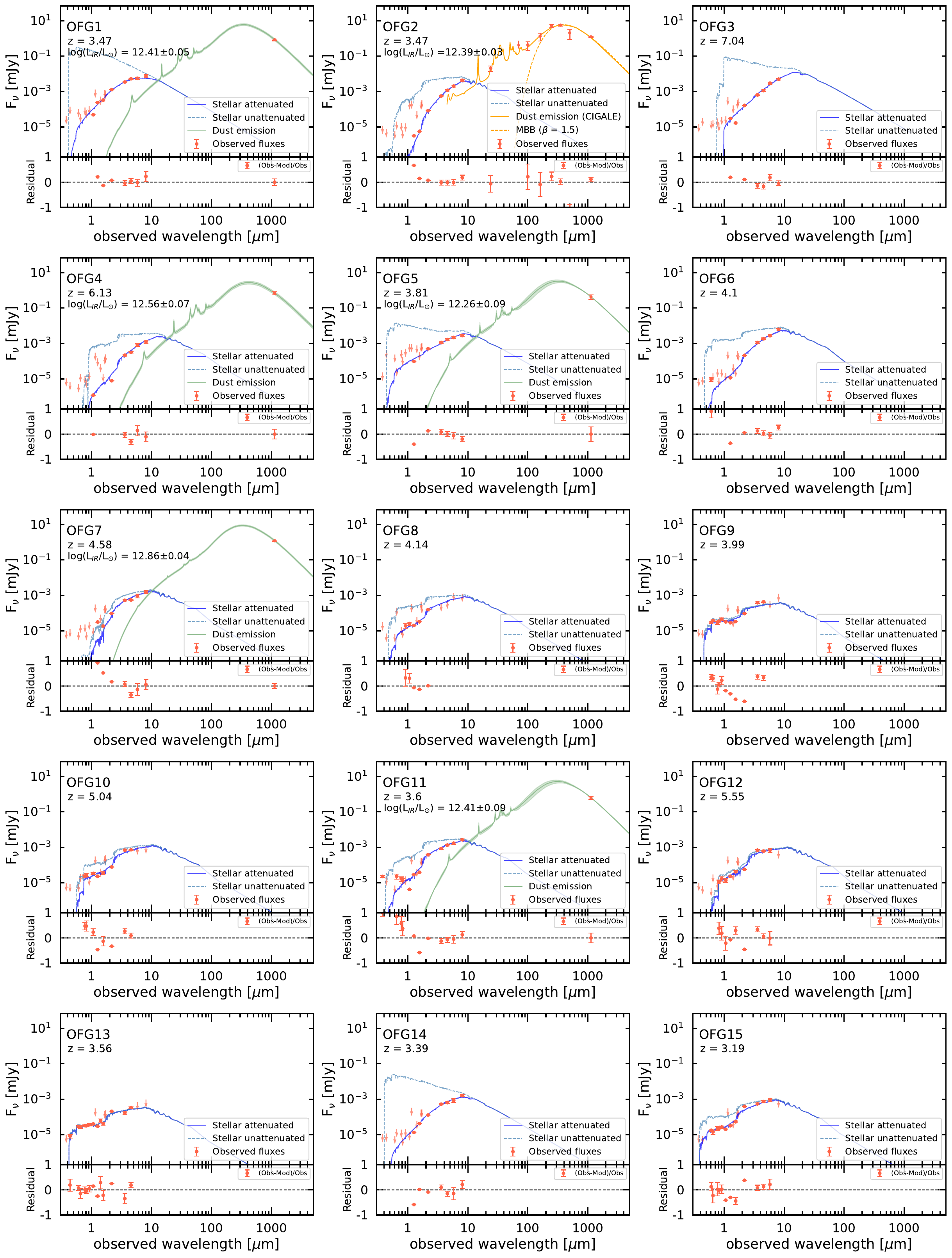}
\caption{UV to millimeter SEDs for 27 OFGs. The symbol convention and lines are the same as those in Figs.~\ref{Fig:meanstack}. The data from the UV to MIR (16 $\mu$m) bands are fitted with the \texttt{FAST++} code (see $\S$\ref{Sect::redshift}). From 24 $\mu$m up to millimeter wavelengths, we fitted galaxies with a \textit{Herschel} counterpart using the \texttt{CIGALE} code (orange solid line) and galaxies without a \textit{Herschel} counterpart but with the ALMA 1.13 mm detection using the dust template libraries \citep[green line;][see $\S$\ref{Sect::LIR}]{Schreiber2018c}.
\label{Fig:SED1}}
\end{figure*}

\begin{figure*}[h!]
\centering
\includegraphics[scale=0.41]{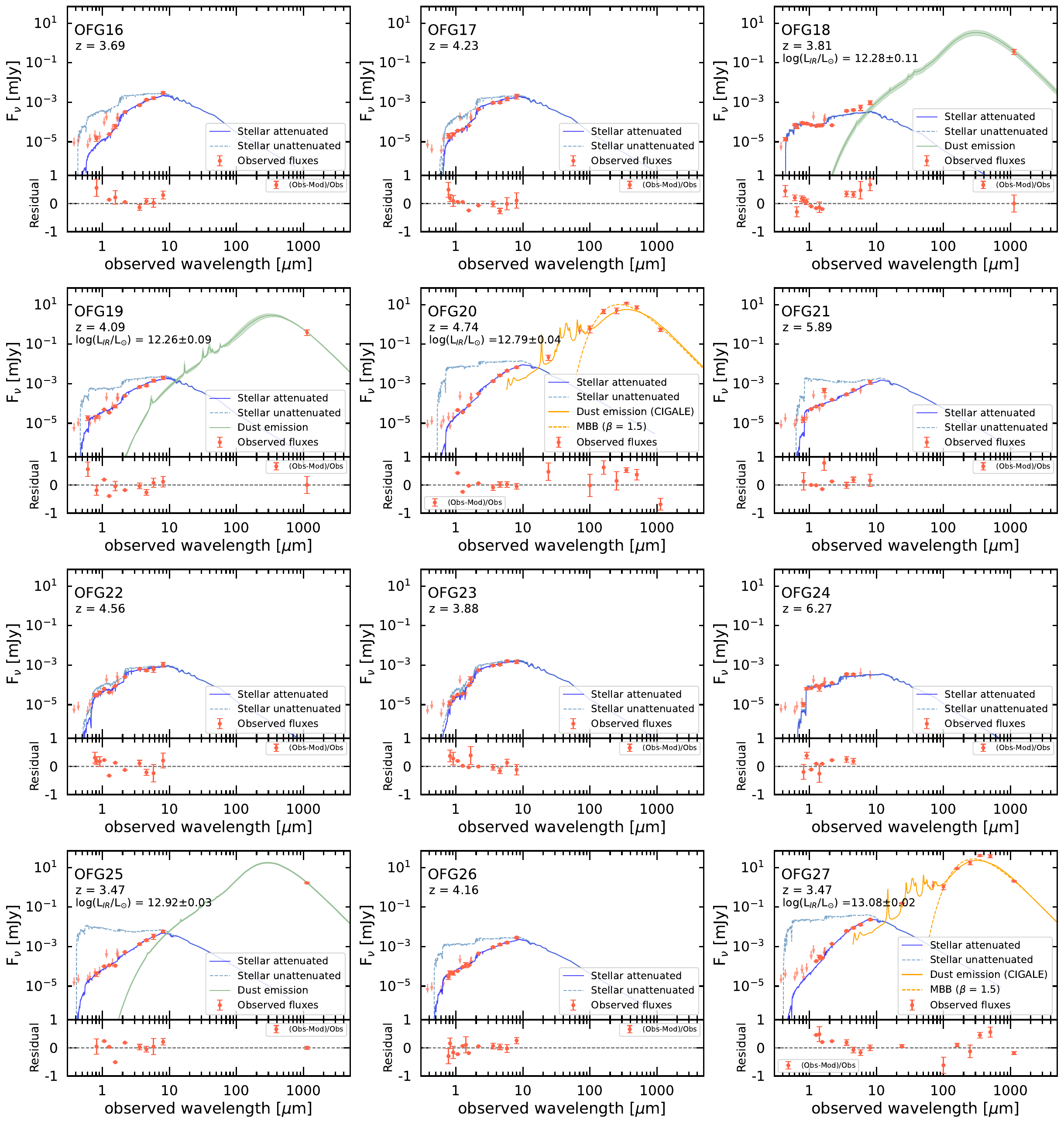}
\caption{(continued).
\label{Fig:SED2}}
\end{figure*}

\end{appendix}
\end{document}